\documentclass[12pt]{article}
\usepackage{latexsym}
\usepackage{amsmath}
\usepackage{amssymb}
\usepackage{epsfig,graphics}
\usepackage{graphicx}
\usepackage{booktabs}
\usepackage{multirow}
\usepackage{caption}
\usepackage{subcaption}
\usepackage{tikz}
\usetikzlibrary{matrix,snakes,arrows,shapes,decorations.pathmorphing,decorations.markings,calc}

\newcommand{\be}{\begin{equation}}
\newcommand{\ee}{\end{equation}}

\begin{document}

\begin{titlepage}

\vspace*{0.6in}

\begin{center}
{\large\bf SU(N) gauge theories in 2+1 dimensions:\\
glueball spectra and k-string tensions}\\
\vspace*{0.75in}
{Andreas Athenodorou$^{a,b}$ and Michael Teper$^{c}$\\
\vspace*{.25in}
$^{a}$Department of Physics, University of Cyprus, POB 20537, 1678 Nicosia, Cyprus\\
\vspace*{.1in}
$^{b}$Computation-based Science and Technology Research Center, The Cyprus Institute, 20 Kavafi Str., Nicosia 2121, Cyprus \\
\vspace*{.1in}
$^{c}$Rudolf Peierls Centre for Theoretical Physics, University of Oxford,\\
1 Keble Road, Oxford OX1 3NP, UK}
\end{center}

\vspace*{0.6in}

\begin{center}
{\bf Abstract}
\end{center}

We calculate the low-lying glueball spectrum and various string tensions 
in $SU(N)$ lattice gauge theories in $2+1$ dimensions, and extrapolate the results
to the  continuum limit. We do so for for the range $N\in[2,16]$ so 
as to control the $N$-dependence with a useful precision.
We observe a number of striking near-degeneracies in the various $J^{PC}$ 
sectors of the glueball spectrum, in particular between $C=+$ and $C=-$ states.
We calculate the string tensions of flux tubes in a number of representations,
and provide evidence that the leading correction to the $N$-dependence of the 
$k$-string tensions is $\propto 1/N$ rather than $\propto 1/N^2$, and that the 
dominant binding of $k$ fundamental flux tubes into a $k$-string is via pairwise
interactions. We comment on the possible implications of our results for the
dynamics of these gauge theories.

\vspace*{0.95in}

\leftline{{\it E-mail:} athenodorou.andreas@ucy.ac.cy, mike.teper@physics.ox.ac.uk}

\end{titlepage}

\setcounter{page}{1}
\newpage
\pagestyle{plain}

\tableofcontents

\section{Introduction}
\label{section_intro}

In this paper on $SU(N)$ gauge theories in $2+1$ dimensions we provide a 
calculation of the light glueball spectrum and of the confining string tension
for fluxes in various representations. This work improves significantly upon earlier work~\cite{MT98_SUN,HMMT_spin,BLMT02_glue}; 
in particular, we cover a wider range of $N$, from $N=2$ to $N=16$, and we calculate 
the continuum limit of more excited states, by using a much larger basis of 
operators. And we do all this with greater accuracy than before. Since one of 
the motivations of such calculations is to look for interesting regularities 
in the spectrum, one needs accuracy and many states. Another motivation is to 
provide a testbed for analytic methods or models, e.g. closed flux tube
\cite{NIJP}
or constituent gluon
\cite{FB13}
models, and this makes similar demands.

In Section~\ref{section_lattice_energies} we begin with a brief description of
the lattice setup and outline how we calculate the masses of glueballs as
well as the energies of flux tubes from which we extract string tensions.
We also list the main systematic errors one needs to be aware of.
We then move on, in Section~\ref{section_tensions}, to the results of
our string tension calculations and, in  Section~\ref{section_glueballs},
to our results for the glueball spectrum. Our calculations point to a number
of interesting regularities which we discuss in the concluding section
of the paper.

Our results for the glueball spectrum in this paper are largely consistent
with the earlier work in
\cite{MT98_SUN,HMMT_spin,BLMT02_glue}
but are much more precise and extensive. Some of our preliminary results 
appeared in
\cite{AARLMT_SONSUN}
where it was shown how a comparison of $SO(N)$ and $SU(N)$ gauge theories
can provide an explanation for the very weak $N$-dependence of a number of
glueball mass ratios. While the spectrum in this paper supersedes that quoted in 
\cite{AARLMT_SONSUN},
none of the conclusions of that paper are affected by the changes.

\section{Calculating energies on a lattice}
\label{section_lattice_energies} 

Here we outline how our lattice calculations are performed. The methods we use
are standard and so we shall be brief.

\subsection{lattice setup}
\label{section_latticesetup}

Our lattice field variables are $SU(N)$ matrices, $U_l$, residing on the links $l$
of the periodic $L^2_s L_t$ lattice, with lattice spacing $a$. The Euclidean path 
integral is 
\begin{equation}
Z=\int {\cal{D}}U \exp\{- \beta S[U]\}\,,
\label{eqn_Z}
\end{equation}
where ${\cal{D}}U$ is the Haar measure over all the lattice link matrices and we use
the standard plaquette action,
\begin{equation}
\beta S = \beta \sum_p \left\{1-\frac{1}{N} {\text{ReTr}} U_p\right\}  
\quad ; \quad \beta=\frac{2N}{ag^2}.
\label{eqn_S}
\end{equation}
Here $U_p$ is the ordered product of link matrices around the plaquette $p$. 
We write  $\beta=2N/ag^2$, where $g^2$ has dimensions of mass and this becomes 
the continuum coupling when $a\to 0$. Monte Carlo calculations are performed 
using a standard Cabibbo-Marinari heat bath plus over-relaxation algorithm.
(See e.g.
\cite{MT98_SUN}
for more details.)
\subsection{calculating glueball masses}
\label{subsection_glue_calc}

The quantum numbers of our particle (`glueball') states are as follows
\cite{MT98_SUN,HMMT_spin}
(We ignore momentum since we always take $p=0$). In continuum $D=2+1$
the rotation group is Abelian and the spins are $J=0, \pm1,\pm2,...$.
(More exotic values --- anyons -- are possible in two spatial dimensions,
but we shall assume that they do not occur in $SU(N)$ gauge theories.)
Parity, $P$, does not commute with rotations: 
$J \stackrel{P}{\to} -J$. So particle (`glueball') states can be labelled 
by parity $P=\pm$ and $|J|$ as well as charge conjugation $C$.
(For notational simplicity we shall use the label $J^{PC}$ 
rather than the more correct $|J|^{PC}$ in this paper.) 
Now, for $J\neq 0$ the state $|J\rangle$ will have an orthogonal partner 
$|-J\rangle$ that is degenerate with it. So the states 
$\propto \{|J\rangle \pm |-J\rangle|\}$, which are degenerate and non-null,
have parities $P=\pm$ respectively. That is to say we have the  well known
parity doubling for states with $J\neq 0$ in 2 space dimensions
\begin{equation}
M_{JC}^{P=+} = M_{JC}^{P=-}\,,   \qquad J\neq 0\,.
\label{eqn_Pdoubling}
\end{equation}
For $J=0$ the argument breaks down because the $P=-$ state can clearly be null. 
We also recall
\cite{MT98_SUN,HMMT_spin}.
that on a square lattice, where the full rotation group is 
broken down to the group of rotations by $\pi/2$,  states with $J^P= 1^+, 1^-$ 
(and odd spins in general) will still be degenerate, while states with $J^P= 2^+, 2^-$ 
(and even non-zero spins in general) need not be -- although any 
non-degeneracy is due to lattice spacing correction and so will usually 
be very small. Note also that the finite volume periodic boundary conditions 
may also break the $2^{P=\pm}$ degeneracy, even in the continuum limit, but
not that of odd spins. 

The fact that on a square lattice we have exact rotational invariance only
under rotations of $\pi/2$ implies that states that tend to continuum
states with $J=0$ or $J=4$ or $J=8$,... will all fall into the same 
representation of the lattice rotation group, and similarly for 
$J=2,6,...$ and  for $J=1,3,...$. For simplicity, we shall 
label states belonging to these representations as $J=0$, $J=2$ and $J=1$.
What are the real spins of these states has been investigated in
\cite{HMMT_spin}
and we shall refer to the detailed implications of that work for our results
later on. For now we merely remark that in some channels the actual ground
states are not those with the lowest spin $J$.

Ground state masses $M$ are calculated from the asymptotic time
dependence of correlators, i.e.
\begin{equation}
\langle \phi^\dagger(t) \phi(0) \rangle 
= \sum_n |\langle vac|\phi^\dagger|n \rangle|^2 e^{-E_n t} 
\stackrel{t\to\infty}{\propto} e^{-Mt}\,,
\label{eqn_M}
\end{equation}
where $M$ is the mass of the lightest state with the quantum numbers
of the operator $\phi$. (If  $\phi$ has the quantum numbers of the vacuum
then we use vacuum-subtracted operator $\phi - \langle \phi \rangle$,
so as to remove the contribution of the vacuum in eqn(\ref{eqn_M}).)
The operator $\phi$ will be the product of $SU(N)$ 
link matrices around some contractible closed path, with the trace then 
taken. We will use zero momentum operators, $p=0$, so that there is 
no momentum integral on the right side of eqn(\ref{eqn_M}).
To calculate the excited states $E_n$ in eqn(\ref{eqn_M}), one
starts with a number of operators, $\phi_i$, and calculates all their 
(cross-)correlators $\langle \phi_i^\dagger(t) \phi_j(0) \rangle$.
One uses this as the basis for a systematic variational calculation 
in $e^{-Ht_0}$ where $H$ is the Hamiltonian (corresponding to our lattice 
transfer matrix) and $t_0$ is some convenient distance. Typically we 
choose $t_0=a$. The procedure produces a set of orthonormal
operators  $\psi_i$ that are our best variational estimates for
the true eigenstates. We then improve the energy estimate by
using eqn(\ref{eqn_M}) with the correlators 
$C_i(t)=\langle \psi_i^\dagger(t) \psi_i(0) \rangle$. With these 
operators we can hope to have good overlaps onto the
desired states, so that one can evaluate their masses at values of $t$
where the signal has not yet disappeared into the statistical noise.
To achieve this in practice, we use blocked and smeared operators
(for details see e.g.
\cite{MT98_SUN,BLMTUW_ops})
and a large number of such operators -- in our case $O(100)$ for
each set of quantum numbers. Some examples of the basic
loops we employ are displayed in Table~\ref{table_glueloops}.

An important aside. The accuracy of such a calculation is
constrained by the fact that the statistical errors are roughly
independent of $t$ (for pure gauge theories) while the desired `signal'
is decreasing  exponentially with $t$. So if the overlap of the desired
state onto our basis is small, the relevant correlator will disappear
into the statistical `noise' before we get to large enough $t$ for the correlator
to be dominated by a simple exponential. Similarly the correlator of a 
more massive state will disappear into the `noise' at smaller $t=an_t$
and this may make ambiguous the judgement of whether it is dominated by
a single exponential. All this may provide an important source of
systematic error.

\begin{figure}[htb]
\begin	{center}
\leavevmode
\input	{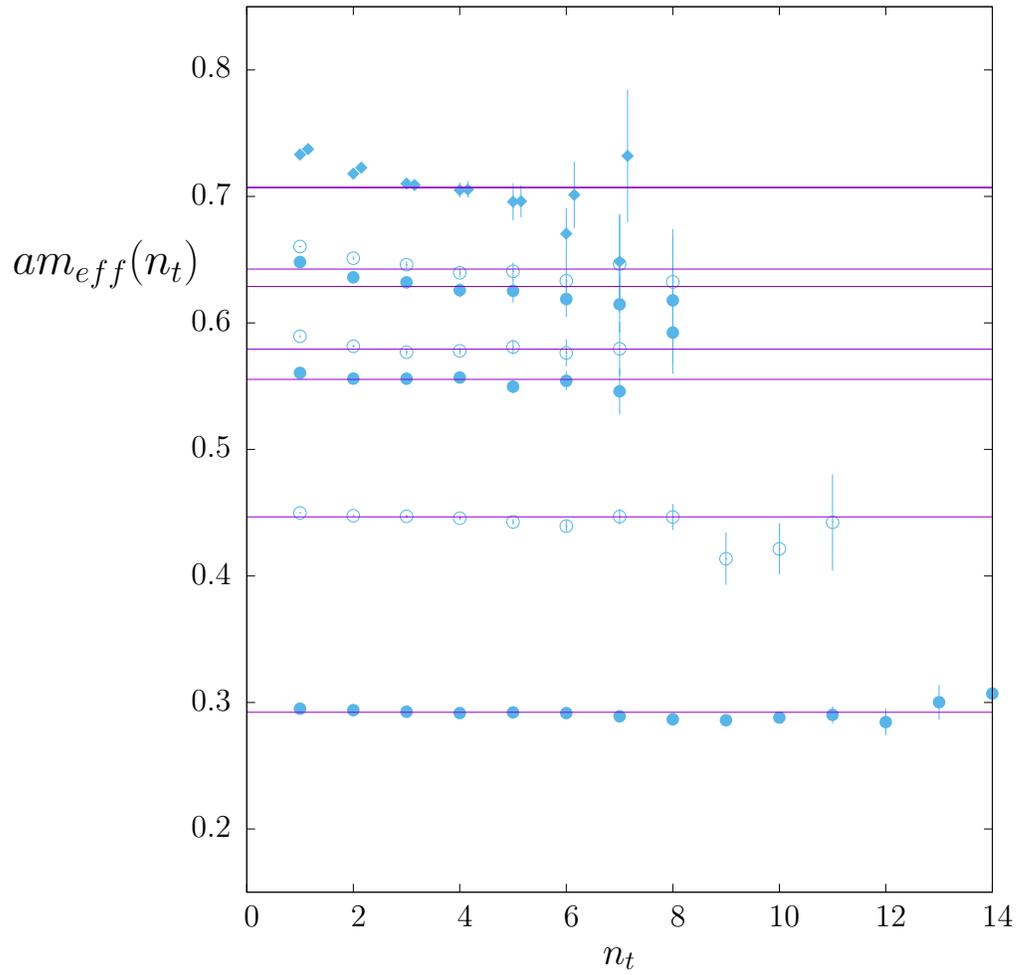}
\end	{center}
\caption{Effective masses of the lightest six $0^{++}$ glueballs ($\bullet$,$\circ$), 
  in $SU(6)$ at $\beta=206.84$. Also the lightest two $1^{++}$ glueballs ($\blacklozenge$),
  shifted in $n_t$ for clarity. Lines are our energy estimates.}
\label{fig_meff_JPC_su6}
\end{figure}

\begin{figure}[htb]
\begin	{center}
\leavevmode
\input	{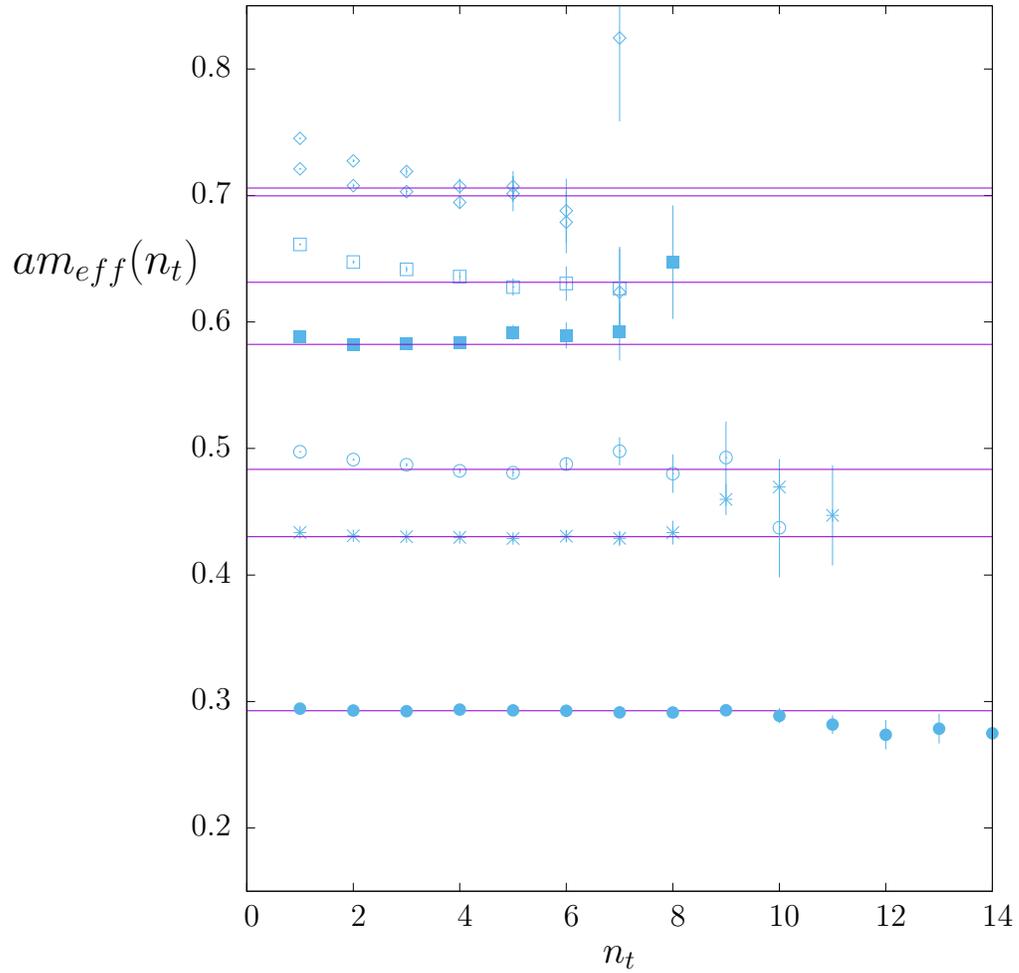}
\end	{center}
\caption{Effective masses of lightest glueballs with $J^{PC}=0^{++}$ ($\bullet$), 
$0^{--}$ ($\ast$), $2^{-+}$ ($\circ$), $2^{--}$ ($\blacksquare$), $0^{-+}$ ($\square$), 
and $1^{-+}$ ($\Diamond$). In $SU(8)$ at $\beta=370.0$. Lines are our energy estimates.}
\label{fig_meff_JPC_su8}
\end{figure}

We extract the energy of the state by going to vale of $t=an_t$ that is large
enough that $C_i(t=an_t) \propto \exp\{-aE_in_t\}$ within our errors. For
illustrative purposes it is convenient to define the effective energy
\begin{equation}
E_{eff,i}(n_t) = -\ln \frac{C_i(n_t)}{C_i(n_t-1)}.
\label{eqn_Eeff}
\end{equation}
If $\tilde{t}=a\tilde{n}_t$ is the lowest value of $t$ for which 
$E_{eff,i}(n_t)$ is independent of $t\geq\tilde{t}$ within our errors,
then we can say that a single exponential dominates and $E_{eff,i}(\tilde{t})$
provides an estimate of the energy $E_i$ of the state.
(In practice we extract our $E_i$ from a fit over a range of $t$.)
That is to say, we search for a `plateau' in the values of $E_{eff}(n_t)$ 
against $n_t$. As a concrete example we show in Fig.\ref{fig_meff_JPC_su6}
the values of $E_{eff,i}(n_t)$ for our six lightest $0^{++}$ glueballs in 
a calculation in $SU(6)$ on a $60^3$ lattice at $\beta=206.84$, which
corresponds to our smallest value of $a$. As we can see, the overlaps
are not far from $\sim 100\%$ so that it is easy to identify a plateau 
in $E_{eff}$ even for the highly excited states. In  Fig.\ref{fig_meff_JPC_su8} 
we show the ground states for various quantum numbers,
as obtained in $SU(8)$ at a comparably small value of $a$. Here 
the overlaps of some higher excited states are not quite so good,
but an $E_{eff}$ plateau can still be plausibly identified. All this to
indicate that while the mass estimates given in this paper should
be largely reliable, this is less certain for the most massive states.

\subsection{calculating flux tube energies}
\label{subsection_flux_calc} 

We calculate the string tension by calculating the energy of a flux
tube of length $l$ that closes on itself by winding once around
a spatial torus of size $l$. We use exactly the same technique as for
the glueballs, except that the operator $\phi$ is now the product of
$SU(N)$ link matrices taken around a non-contractible closed path that
winds once around the spatial torus.
The simplest such operator is the spatial Polyakov loop
\begin{equation}
\phi(n_t) = l_p(n_t) =  \sum_{n_y} \mathrm{Tr} 
\left\{\prod^{L_x}_{n_x=1} U_x(n_x,n_y,n_t)\right\} \,,
\label{eqn_poly}
\end{equation}
where we take the product of the link matrices in the $x$-direction 
around the $x$-torus of length $l=aL_x$.
Here $(x,y,t)=(an_x,an_y,an_t)$, and we sum over $n_y$ to produce
an operator with zero transverse momentum, $p_\perp = p_y = 0$.
(We only consider flux tubes with zero transverse momentum, since we
do not expect to learn anything qualitatively new from $p_\perp \neq 0$.)
In addition to this simple straight-line operator we construct other
winding operators with a variety of kinks and loops extending from the
original straight line and we also use smeared and blocked operators
just as for the glueballs. Using this large basis of operators
we apply a variational procedure as for the glueballs, and this allows
us to calculate not only the ground state energy but also excitation energies
of the flux tube. We can label the flux tubes states by their transverse parities
and their longitudinal momenta, and by their longitudinal  parities if 
the momentum is zero. We can do all this for flux tubes carrying flux
in any representation ${\cal{R}}$ by taking the trace in eqn(\ref{eqn_poly})
in that representation, i.e. by using $\mathrm{Tr}_{\cal{R}}$.
All this is described in detail in for example
\cite{BLMT01_strings,AABBMT11_strings}.

Since we will focus in this paper on the string tension rather than on
the full flux tube spectrum
\cite{AABBMT11_strings}
we only need to calculate the ground state energy, $E_{gs}(l)$, of
the flux tube. We extract the string tension $\sigma$
from $E_{gs}(l)$ using the `Nambu-Goto' formula
\begin{equation}
E_{gs}(l)
\stackrel{NG}{=}
\sigma l \left(1-\frac{\pi}{3\sigma l^2}\right)^{1/2}\,,
\label{eqn_gsNG}
\end{equation}
which arises from the light-cone quantisation of the bosonic string
and is known to provide an excellent approximation to the lattice
calculations
\cite{AABBMT11_strings,AAMT16_strings}
for reasons that have now become well understood
\cite{OA}
(see also
\cite{SD_long}).
We calculate in this way not only the fundamental string tension, $\sigma_f$,
but also string tensions, $\sigma_{\cal{R}}$, for the flux in a number of
other representations  ${\cal{R}}$ as discussed
in Section~\ref{section_tensions}.

\begin{figure}[htb]
\begin	{center}
\leavevmode
\input	{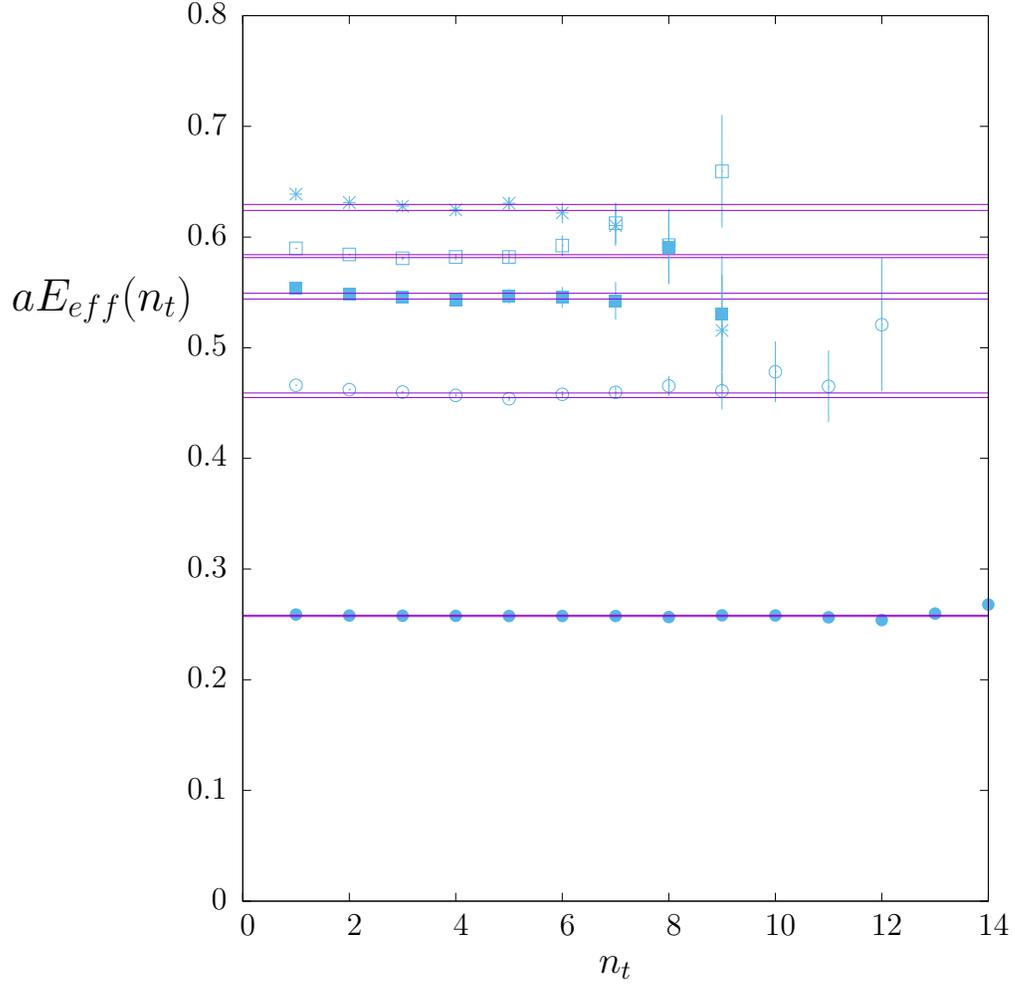}
\end	{center}
\caption{Effective energies of ground state flux tube correlators in various
representations: fundamental $k=1$, $\bullet$; $k=2$, $\circ$; adjoint $k=0$, 
$\blacksquare$; $k=3$, $\square$; $k=4$, $\ast$. In $SU(8)$ at $\beta=370.0$.
Pairs of lines are our energy estimates, embracing the statistical errors.}
\label{fig_Ekeff_su8}
\end{figure}

The reliability of such calculations depends on how well we can
calculate  $E_{gs}(l)$. As  ${\cal{R}}$ becomes larger, so does the
corresponding  ground state energy $E_{gs,{\cal{R}}}(l)$ which makes the
calculation less reliable. To give some indication of the uncertainties
involved we show in Fig.\ref{fig_Ekeff_su8} the effective energies, as
defined in eqn(\ref{eqn_Eeff}), for the representations of interest in
this paper. We do this for the case of $SU(8)$ at our smallest lattice
spacing. We also show the estimates of $E_{gs,{\cal{R}}}(l)$ that we
obtain from these effective energies. We see that the identification
of the effective energy `plateau' appears to be unambiguous in all cases.

\subsection{ambiguities and systematic errors}
\label{subsection_sys_errors}

A calculation of the glueball mass spectrum is subject to a
number of systematic errors that one needs to control. (For
flux tubes this is also the case, but because we are only interested 
in their ground states here, they are less important.) 
We discuss in this section our control over some of these systematic
errors.

\subsubsection{missing states}
\label{subsubsection_missing_states}

Our variational bases are, of course, incomplete and as we calculate
ever higher excited states we will, at some point, begin to miss states.
As it happens we have two sets of operators that were independently
produced and which differ substantially, and comparing the spectra
obtained using these bases provides some check on whether we are
in fact missing any states in the range of interest to us here.
One set of operators was used for our $N=3,6$ calculations, while the
other was used for $N=2,4,8,12,16$. However we have also performed
some extra calculations to compare the results  obtained
using the two bases. One comparison was in $SU(4)$ at $\beta=51$
on a $40^248$ lattice. Another was in $SU(6)$, comparing the spectrum
on a $54^260$ lattice at $\beta=206.0$ with that from the second basis on a
$60^3$ lattice at $\beta=206.84$. This latter comparison has a much smaller
lattice spacing and hence a more reliable estimate of the masses of the
heavier states, although the differing lattice sizes and values of $\beta$
make it less than ideal. (But since to leading order $aM \propto 1/\beta$,
the expected shift in the masses should be a negligible $\approx 0.4\%$.)
We compare the spectra obtained with the two bases of operators in
Tables~\ref{table_msu4_opcomp},~\ref{table_msu6_opcomp}.  We include
for each set of quantum numbers as many states as we will eventually
include in our continuum extrapolations. We see that the spectra are
broadly consistent within errors. (This provides us with
some confidence that we are not missing any states in the range of energies
being considered. (The consistency of the spectra obtained with the two
bases is of course a prerequisite for the large $N$ extrapolations we
perform later in this paper.)

\subsubsection{multi-glueball and unstable states}
\label{subsubsection_multiglueballs}

The full glueball spectrum will contain multi-glueball states. Such states
are represented by multi-trace operators and so as $N$ increases the overlap
of such a state onto our basis of single trace operators will vanish by
standard large-$N$ counting. However at modest values of $N$ such states
may be present. Here two $J^{PC}=0^{++}$ glueballs with angular momentum $J$
can produce $|J|^{\pm +}$ states (with $P=-$ for $J\neq 0$). Similarly
$|J|^{\pm -}$ states can arise from a $0^{++}$ and $0^{--}$ glueball with angular
momentum $J$. Some of our more massive states certainly fall into
this `dangerous' mass range. However the overlap suppression tells us that
such a state will manifest itself through an effective energy that is much
larger at small $t$ and then decreases, asymptoting to $2m_{0^{++}}$ or
$m_{0^{++}}+m_{0^{--}}$ at large $t$. Since we order our states by the value
of $E_{eff}(n_t)$ at $t=a$, it is unlikely that any such states would appear
in our energy range, and indeed we see no obvious sign of any such states.

\begin{figure}[htb]
\begin	{center}
\leavevmode
\input	{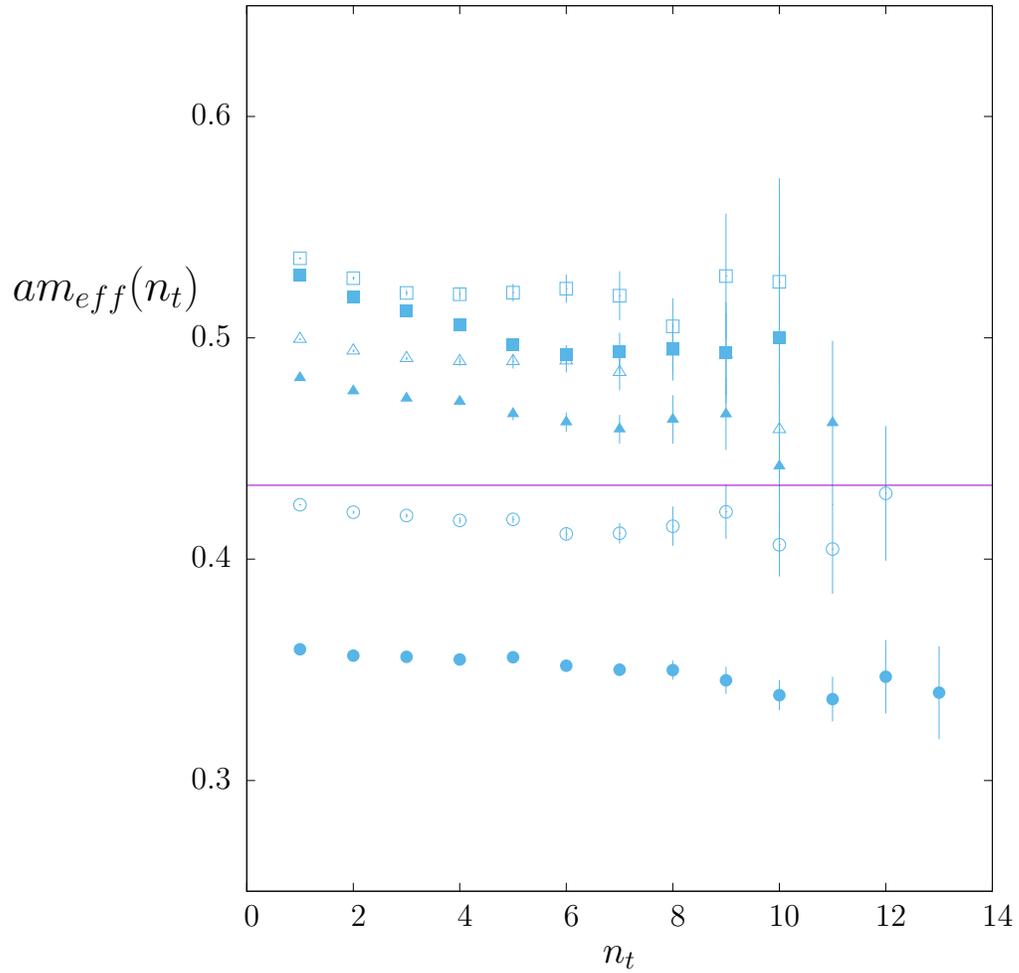}
\end	{center}
\caption{Effective masses of the lightest six $2^{++}$ glueballs 
  in $SU(2)$ at $\beta=30$. Horizontal line is twice the scalar glueball mass.}
\label{fig_meff_2pp_su2}
\end{figure}

A closely related issue concerns  genuine single glueball states that are heavy
enough to decay into lighter glueballs. Again the decay width will vanish
as $N\to\infty$ and at moderate $N$ we expect the state to resemble a narrow
resonance. So the effective energy should have something like an approximate
plateau at small $t$, but at large $t$ it will drift down to the threshold value
of its decay products. However because the error/signal ratio grows
exponentially with $t$, this large-$t$ behaviour is unlikely to be
visible and the state will appear just like any stable state. Of course
things may be different at small $N$. To see what happens at smaller $N$
we turn to our $SU(2)$ calculation, where the decay widths should be
largest, and to our smallest lattice spacing, which corresponds to 
$\beta=30.0$, performed on a $120^290$ lattice,
We show in Fig.~\ref{fig_meff_2pp_su2} the effective energies of the
lightest few $2^{++}$ states. The horizontal line corresponds to the
threshold energy of the potential decay products, i.e.
$aE_{th}=2am_{0^{++}} \simeq 0.43$. We see that while several of the
states are heavier than this theshold energy, there is no sign, within
our errors, of a drift in the effective mass towards this threshold energy
as $t$ increases. There is of course a decrease at small $t$, as the
higher excited state contributions to the correlator die away,
but that does not persist to larger $t$, as one would expect it to do
if the decay width was substantial.

All this encourages us to conclude that any systemtic errors due to
multiglueball states or the instability of the states we consider
are unlikely to be substantial in the calculations of this paper.

\subsubsection{finite V corrections}
\label{subsubsection_finiteV}

The leading finite volume correction, at large volumes, to the mass of a
glueball comes from the emission of the lightest glueball, of mass $m$, which
then winds around the spatial torus of length $l$, before being reabsorbed
by the glueball. So these contributions are typically $\propto \exp(-ml)$
and they are suppressed at large $N$ because the triple glueball coupling
is $g^2_{GGG} \sim 1/N$. Now for our calculations in this paper $ml$ varies
between $\sim 25$ at small $N$ and $\sim 12$ at our largest values of $N$,
so we can assume that these leading finite volume effects are completely
insignificant.

In practice, as is well known, the important finite volume corrections
are quite different and arise from the presence of finite volume states
composed of, for example, a winding flux tube
together with a conjugate winding flux tube. These states are described by
double-trace operators so their overlap onto our basis of single-trace
glueball operators will vanish at large $N$ but is known to be substantial
at small values of $N$. Now, when both flux tubes are in their
ground states, the energy of such a `winding' state is $E_{T} \sim 2\sigma_f l$,
and this can contribute to both $0^{++}$ and $2^{++}$. For $J=1$ or $P=-$ or
$C=-$, one needs to have one or both of the flux tubes in a suitable excited
state and this has a much larger energy. Our strategy to control these
finite volume corrections is therefore the very simple one of making $l$ large
enough at small $N$ that all the  $0^{++}$ and $2^{++}$ glueball states that
we consider have energies no greater than $E_T$. We then rely on the large-$N$
suppression of the overlaps to allow ourselves smaller values of $l$ at larger $N$.
(A decreasing overlap means higher effective masses at small $t$, pushing the
state out of our range of masses, which is determined by the effective mass at
$t=a$.) Our specific choices are listed in Table~\ref{table_sizeN}. Of course
one needs to ask whether the large-$N$ suppression is actually sufficient to
eliminate such extra finite volume states at our `large' values of $N$. This
can only be answered convincingly by explicit finite volume studies. So in
Table~\ref{table_msu8_Vcomp} we compare the spectrum obtained at $\beta=306.25$
on a $44^248$ lattice (our standard size) with the spectrum on a much larger
$60^248$ lattice. Any finite volume state should be apparent from a large upward
shift in its mass as we go to the larger lattice. As we see in 
Table~\ref{table_msu8_Vcomp} the spectra are consistent, with no sign of any
finite volume states. The smallest volumes we use are in $SU(16)$ so we perform
another comparison in that case. We list in Table~\ref{table_msu16_Vcomp}
the spectrum on a $22^230$ lattice (our standard volume) and that on a
larger $26^230$ lattice, both at $\beta=800$. We see that these two spectra
are entirely consistent at the level of 2 standard deviations.
All this strongly suggests that these finite volume corrections are insignificant
in the calculations presented in this paper.

\subsubsection{nearly degenerate states}
\label{subsubsection_degeneracy}

One of the interesting features of our results is that there are a number of
nearly degenerate states. When these have the same quantum numbers and
therefore arise from a single variational calculation (as described above)
the variational procedure can induce an extra splitting, which is driven
by statistical fluctuations and is therefore on the order of the errors.
Moreover this may also lead to the two states being ordered differently
in different bins, leading possibly to biases in the estimated errors
and eventually to unsatisfactory fits when performing continuum or large $N$
extrapolations. Any ambiguities here are small as long as the errors are small, 
but some of the most interesting near-degeneracies occur in the
$J=1$ sector where all the states are massive and the statistical errors are
substantial.

\section{String tensions}
\label{section_tensions} 

\subsection{fundamental string tension and $N$-dependence}
\label{section_ftension} 

The ground state of the fundamental flux tube of length $l$ that winds 
once around our spatial torus is usually the lightest of all the states
(on the volumes we typically use) and hence it is
the state whose energy, $E_{gs,f}(l)$, we can calculate most accurately 
and most reliably. We extract the string tension $\sigma_f$ from the 
energy using the `Nambu-Goto' formula in eqn(\ref{eqn_gsNG}).

\begin{figure}[htb]
\begin	{center}
\leavevmode
\input	{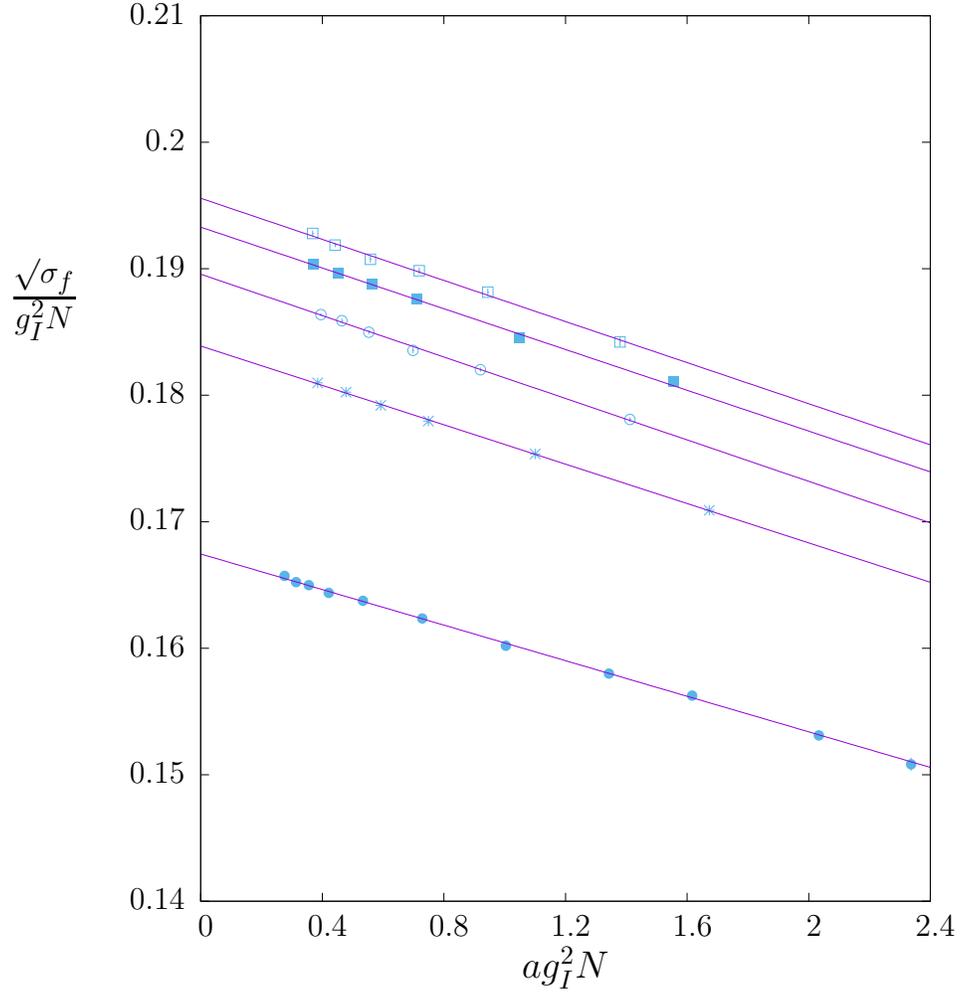}
\end	{center}
\caption{Linear continuum extrapolations of string tensions in units of the (lattice) 't Hooft coupling.
For $SU(2)$ ($\bullet$), $SU(3)$ ($\ast$), $SU(4)$ ($\circ$), $SU(6)$ ($\blacksquare$), and $SU(12)$ ($\Box$).}
\label{fig_k1g_cont}
\end{figure}

We list in Tables~\ref{table_param_su2}-\ref{table_param_su16} 
the resulting values of $a\surd\sigma_f$ 
for our various lattice calculations. To obtain values in the
continuum limit we need to express $\surd\sigma_f$ in units
of a quantity with dimensions of mass, and an obvious choice 
is the coupling $g^2$ which in $2+1$ dimensions has dimensions $[m^1]$. 
The lattice coupling scheme that we choose to use is the
`mean-field improved' coupling $g^2_I$ defined by
\cite{b_imp}
\begin{equation}
\beta_I \equiv \beta \times \langle \frac{1}{N} {\text{Tr}}U_p \rangle
=
\frac{2N}{ag^2_I}\,.
\label{eqn_gMFI}
\end{equation}
Since different choices of lattice coupling differ at $O(g^2)$,
the leading correction to the continuum limit will be $O(a)$ rather 
than $O(a^2)$. We therefore extrapolate our string tensions to the 
continuum limit, at each value of $N$, using
\begin{equation}
\frac{\beta_I}{2N^2}a\surd\sigma_f
=
\left.\frac{\surd\sigma_f}{g^2_IN}\right|_a
=
\left.\frac{\surd\sigma_f}{g^2N}\right|_{a=0} + c_1 ag^2_IN +  c_2 (ag^2_IN)^2 + \ldots \,.
\label{eqn_sigf_cont}
\end{equation}
We have used the 't Hooft coupling, $g^2N$, since that is what 
needs to keep fixed for a smooth $N\to\infty$ limit. We provide some 
examples of such continuum extrapolations in Fig.\ref{fig_k1g_cont}. As is 
apparent from the figure, we can get good fits with just the leading $O(a)$
correction. We list in Table~\ref{table_sigf_cont} the continuum limit 
for each $N$, together with the coefficients of the linear correction, and
the goodness of fit as measured by the total $\chi^2$ and the
number of degrees of freedom, $n_{dof}$, of the fit.

\begin{figure}[htb]
\begin	{center}
\leavevmode
\input	{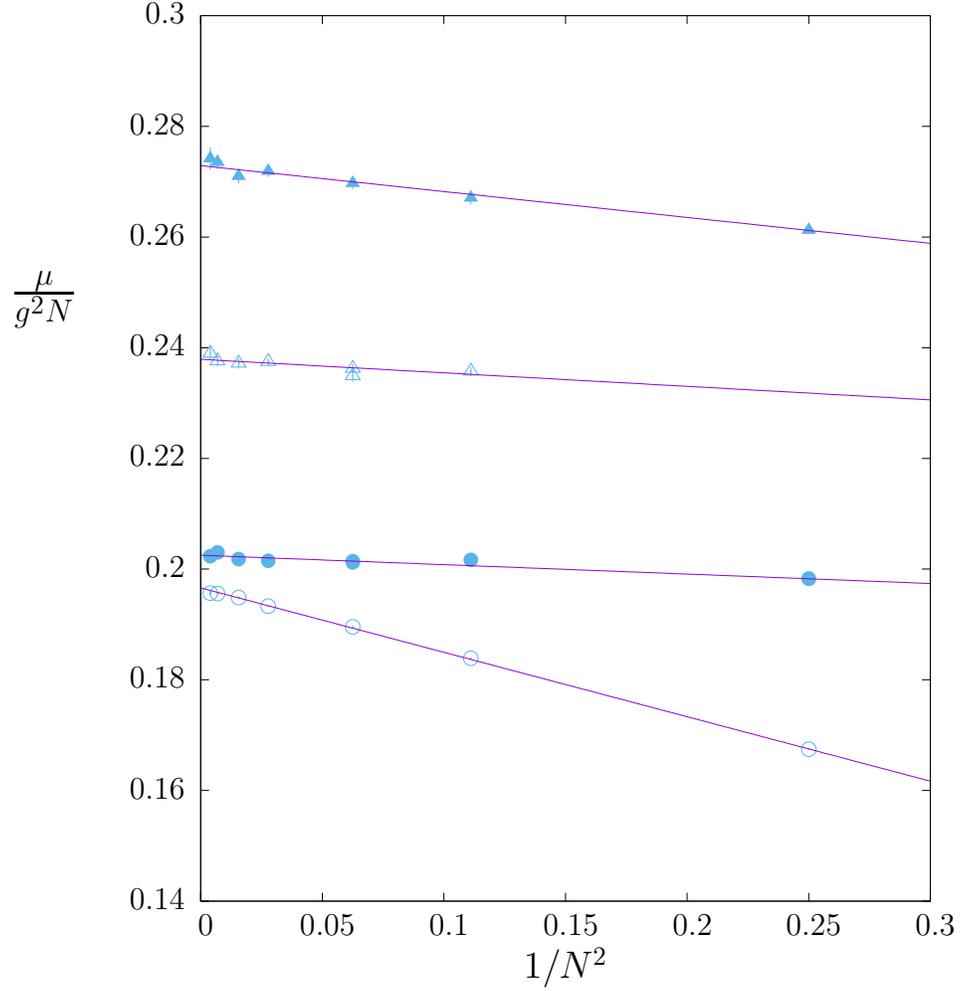}
\end	{center}
\caption{String tension $\mu=\surd\sigma_f$ ($\circ$), and rescaled lightest glueballs:
$\mu = 0.25\times m_{0^{++}}$ ($\bullet$), 
$\mu = 0.2 \times m_{0^{--}}$ ($\triangle$), and
$\mu = 0.2 \times m_{2^{++}}$ ($\blacktriangle$). Versus $1/N^2$, with linear fits 
to the large-$N$ limit.}
\label{fig_k1g_m02g_N}
\end{figure}

We can now extrapolate our results to $N=\infty$. (For $N=2$ and $N=4$ we use the
values on lattices of medium size, where the flux tube energies are small enough
that one expects to avoid the systematic error associated with the large energies
one obtains on large lattices.) We expect the leading correction to be $O(1/N^2)$
\cite{tHooft_N}
and indeed we find that we get a marginally acceptable fit to all our 
calculated values with just this correction,
\begin{equation}
\frac{\surd\sigma_f}{g^2N}
=
0.196573(81) -\frac{0.1162(9)}{N^2} \qquad N\geq 2 \quad \chi^2/n_{dof}=11.6/5\,,
\label{eqn_sigf_N}
\end{equation}
which is displayed in Fig.\ref{fig_k1g_m02g_N}.
In order to see how robust this result is to the inclusion of higher 
order corrections in $1/N^2$, or to dropping the lowest $N$ data point,
we perform the corresponding fits, giving
\begin{equation}
\frac{\surd\sigma_f}{g^2N}
= 0.19636(12) -\frac{0.1085(32)}{N^2}-\frac{0.029(12)}{N^4} \qquad N\geq 2 \quad \chi^2/n_{dof}=5.0/4\,,
\label{eqn_sigf_Nb}
\end{equation}
\begin{equation}
\frac{\surd\sigma_f}{g^2N}
=
0.19642(11) -\frac{0.1118(26)}{N^2} \qquad \qquad \qquad N\geq 3 \quad \chi^2/n_{dof}=6.0/4\,.
\label{eqn_sigf_Nc}
\end{equation}
In the rest of the paper we shall use the result obtained in eqn(\ref{eqn_sigf_Nb}).


\subsection{$k$-string tensions and $N$-dependence}
\label{section_ktensions} 

We calculate string tensions in other representations in the same way as for the
fundamental. A flux tube carrying a flux that transforms under the $Z_N$ centre
like the product of $k$ fundamental fluxes is called a $k$-string. For $N\geq 2k$
the lightest flux tube in each $k$-sector is stable
against decay as one can infer for $k=2,3,4$ from the energies of the various
ground states listed in Tables~\ref{table_Egs_su4}-\ref{table_Egs_su16}
and from the continuum string tension ratios listed in
Table~\ref{table_sigksigf_cont}. So there is no ambiguity in extracting the flux tube 
energy, and hence the string tension using eqn(\ref{eqn_gsNG}). The ground
states of these flux tubes fall into the totally antisymmetric
representation to an excellent approximation
\cite{BLMT01_strings,AAMT16_strings}
and so we label them as $2A,3A,4A$ respectively. One might ask how
this feature can survive screening by gluons, but it turns out that
for long flux tubes the effects of screening are extremely weak
\cite{AAMT16_strings}. 
The flux tubes in the other representations listed 
in Table~\ref{table_sigksigf_cont} are in principle not stable; for
example $\sigma_{2S} > 2\sigma_f$ and $\sigma_{3M} > \sigma_{2A} + \sigma_f$
(albeit not by much in this case). Since the least ambiguous string tensions 
are those that are associated with stable flux tubes, we will focus here on the 
$2A,3A,4A$ string tensions. 

\begin{figure}[htb]
\begin	{center}
\leavevmode
\input	{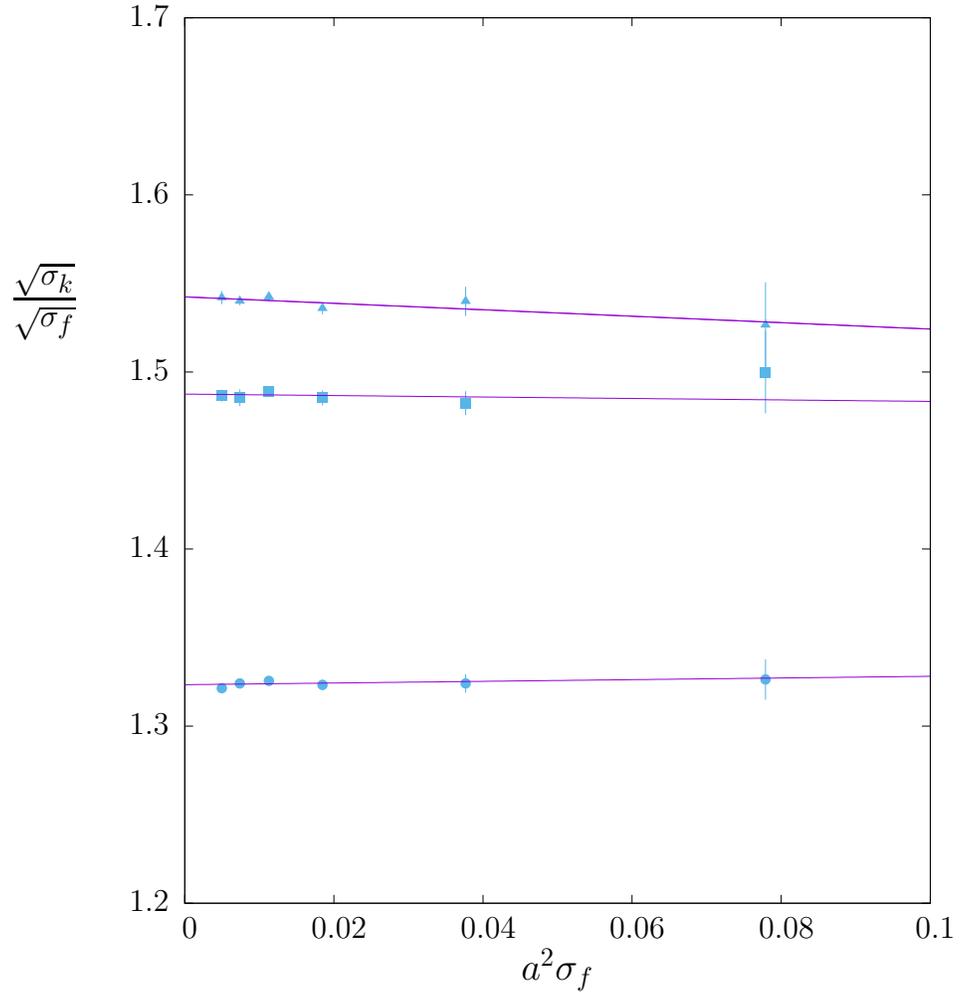}
\end	{center}
\caption{Lattice values of string tensions of $k=2$ ($\bullet$), $k=3$ ($\blacksquare$), and
  $k=4$ ($\blacktriangle$) flux tubes in $SU(8)$,}
\label{fig_kf_cont_su8}
\end{figure}

We obtain the continuum limit of $\surd\sigma_k/\surd\sigma_f$ from our
calculated values using the standard fit
\begin{equation}
\left.\frac{\surd\sigma_k}{\surd\sigma_f}\right|_{a}
=
\left.\frac{\surd\sigma_k}{\surd\sigma_f}\right|_{a=0}
+ c_1 a^2\sigma_f\,,
\label{eqn_sigksigfcont}
\end{equation}
where we include only an $O(a^2)$ correction term since that suffices
to obtain an acceptable fit in all cases. In Fig.~\ref{fig_kf_cont_su8}
we show the fits for $k=2,3,4$ in the case of $SU(8)$. We see that
the $a$-dependence is so small as to be consistent with zero.
Our continuum extrapolations are listed in Table~\ref{table_sigksigf_cont}.

We begin by fitting $\sigma_{2A}/\sigma_f$ for $N\geq 4$. In addition
to the values in Table~\ref{table_sigksigf_cont} we also add the constraint 
$\lim_{N\to\infty}\sigma_{2A} = 2  \sigma_f$ from the theoretical
expectation that $\lim_{N\to\infty}\sigma_{k} = k  \sigma_f$. We attempt 
separate fits: one that is in powers of $1/N$, and one in powers of $1/N^2$.
These give the best fits
%
%
%
\begin{equation}
\frac{\sigma_{2A}}{\sigma_f}
\stackrel{N\geq4}{=}
2 - \frac{1.406(49)}{N} - \frac{4.68(21)}{N^2}    \qquad \chi^2/n_{df}=1.5\,,
\label{eqn_sigk2A_Ndep}
\end{equation}
\begin{equation}
\frac{\sigma_{2A}}{\sigma_f}
\stackrel{N\geq4}{=}
2 - \frac{17.52(26)}{N^2} - \frac{5.70(75)}{N^4}   \qquad \chi^2/n_{df}=27.6\,.
\label{eqn_sigk2A_NNdep}
\end{equation}
Since $n_{df}=3$, the first fit is entirely acceptable, but the second is not.
The second fit is not much improved if we restrict ourselves to $N\geq 6$.
It is clear that our calculations imply that the leading correction
to $\sigma_{2A}/\sigma_f$ is $\propto 1/N$ rather than  $\propto 1/N^2$.

We repeat the exercise for $\sigma_{3A}/\sigma_f$, this time adding the
constraint $\lim_{N\to\infty}\sigma_{3A} = 3  \sigma_f$. Fitting to $N\geq 6$
we obtain
\begin{equation}
\frac{\sigma_{3A}}{\sigma_f}
\stackrel{N\geq6}{=}
3 - \frac{4.24(17)}{N} - \frac{16.3(1.2)}{N^2}   \qquad \chi^2/n_{df}=0.13\,,
\label{eqn_sigk3A_Ndep}
\end{equation}
\begin{equation}
\frac{\sigma_{3A}}{\sigma_f}
\stackrel{N\geq6}{=}
3 - \frac{68.0(1.0)}{N^2} - \frac{963(43}{N^4}   \qquad \chi^2/n_{df}=38.6\,.
\label{eqn_sigk3A_NNdep}
\end{equation}
Again this clearly points to the leading correction being $\propto 1/N$
and not $\propto 1/N^2$.

Finally we look at  $\sigma_{4A}/\sigma_f$ for $N\geq 8$, where we obtain 
\begin{equation}
\frac{\sigma_{4A}}{\sigma_f}
\stackrel{N\geq8}{=}
4 - \frac{9.08(46)}{N} - \frac{31.1(3.9)}{N^2}  \qquad \chi^2/n_{df}=0.14 
\label{eqn_sigk4A_Ndep}
\end{equation}
while the best fit with $N\to N^2$ has an unacceptable $\chi^2/n_{df}=12.1$. 

In the above we have restricted our fits to $N\geq 2k$ since the center symmetry allows
a  $k$-string to mix with a $(N-k)$-string, and the latter will have a lower
string tension if $N < 2k$, and so it will then provide the lightest flux
tube in the $k$-sector. So, for example, if we extrapolate eqn(\ref{eqn_sigk2A_Ndep})
to $SU(3)$, we expect to find ${\sigma_{2A}}\stackrel{su3}{=}{\sigma_f}$. 
Interestingly enough, substituting $N=3$ in  eqn(\ref{eqn_sigk2A_Ndep}) does
indeed give a value very close to unity. We can extend the argument to $SU(2)$,
where $k=2$ can  mix with the $k=0$ vacuum, and indeed the $O(1/N)$ fit in
eqn(\ref{eqn_sigk2A_Ndep}) gives us ${\sigma_{2A}}\stackrel{su2}{\sim} 0 $.
Similarly we find ${\sigma_{3A}}\stackrel{su4}{\sim}{\sigma_f}$ and
${\sigma_{3A}}\stackrel{su5}{\sim}{\sigma_{k=2}}$ (where we estimate
the value of $\sigma_{k=2}$ in $SU(5)$ using eqn(\ref{eqn_sigk2A_Ndep})).
Of course this rough agreement is only indicative: as we go to smaller
$N$ we must expect that the omitted higher order terms in $1/N$ will
become significant. Indeed a constructive way to look at this
is to use the constraints from $N < 2k$ to fix, with no extra work, the next 
higher order terms in the expansion in powers of $N$, which our fitting for 
$N\geq 2k$ is not sensitive to.

Another interesting feature becomes apparent when we look at the coefficient 
$c_1(k)$ of the leading $1/N$ correction in the above equations. We 
observe that 
\begin{equation}
c_1(k=4):c_1(k=3):c_1(k=2) \approx 6:3:1 
= \frac{4(4-1)}{2}:\frac{3(3-1)}{2}:\frac{2(2-1)}{2}
\label{eqn_c1k}
\end{equation}
i.e. $c_1(k) \propto k(k-1)/2$, which is just the number of pairwise interactions
amongst the $k$ fundamental strings which are bound into the $k$-string.  
A similar rough proportionality appears to hold when we look at the coefficient
$c_2(k)$ of the second, $1/N^2$, correction term. This appears to suggest that the
interactions binding the $k$ fundamental strings into the $k$-string are
predominantly pairwise.

\begin{figure}[htb]
\begin	{center}
\leavevmode
\input	{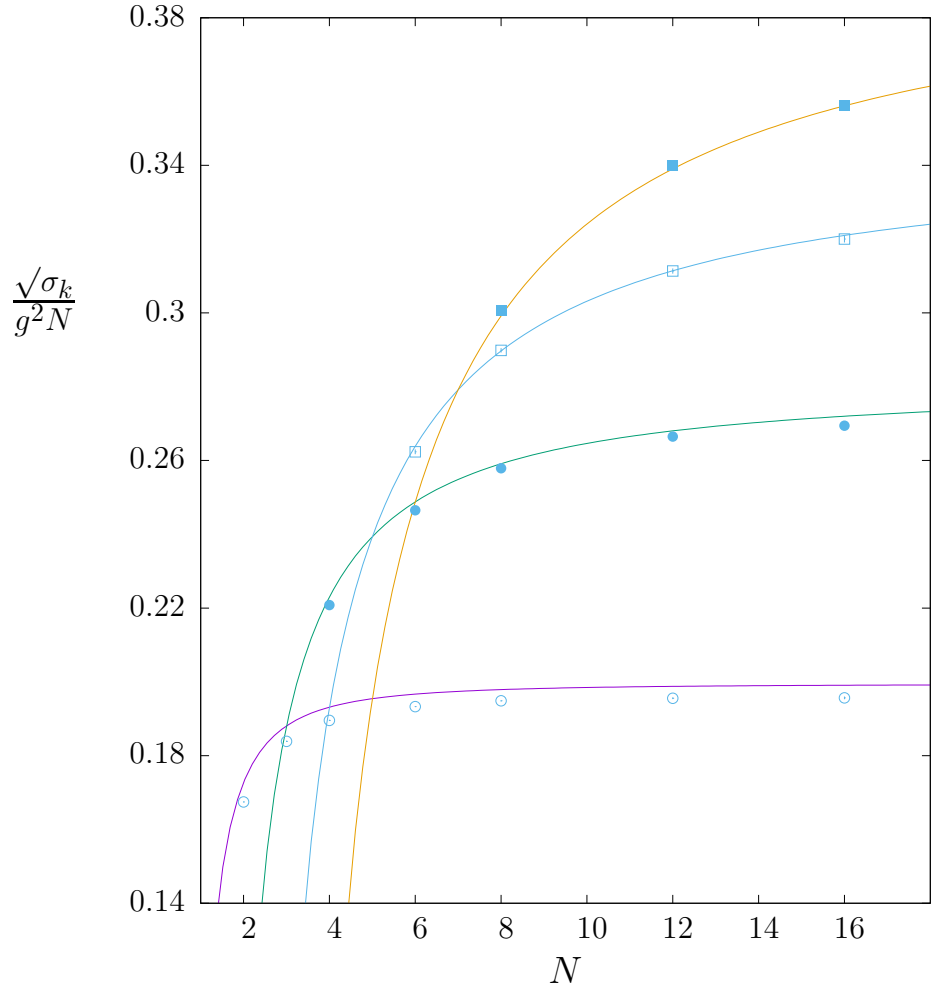}
\end	{center}
\caption{String tensions for various $SU(N)$: $k=1$, $\circ$; $k=2A$, $\bullet$;
 $k=3A$, $\Box$; $k=4$ $\blacksquare$. Lines are predictions from \cite{Nair}.}
\label{fig_ksig_Nair}
\end{figure}

It is also interesting to compare our results to the Karabali-Kim-Nair (KKN) analysis
\cite{Nair}
which provides a prediction not only for the $k$ and $N$ dependence of $\sigma_k$,
but also predicts its absolute magnitude:
\begin{equation}
\frac{\surd\sigma_k}{g^2N} \stackrel{KKN}{=}\sqrt{\frac{k(N-k)}{N-1}} 
\times \sqrt{\frac{1-\frac{1}{N^2}}{8\pi}}.
\label{eqn_KKN}
\end{equation}
Although we know from earlier studies
\cite{BBMT06_string} 
that eqn(\ref{eqn_KKN}) does not fit the calculated values of $\sigma_f/g^2N$ perfectly, 
it does come within $\sim 2\%$ for $N\geq 3$, and within $3\%$ even if we include $SU(2)$. 
The prefactor in eqn(\ref{eqn_KKN}) is simply the quadratic Casimir of the represention 
which, for reasons obvious from the above discussion, we take to be the totally 
antisymmetric one. From the values of $\surd\sigma_f/g^2N$ in Table~\ref{table_sigf_cont} 
and $\surd\sigma_k/\surd\sigma_f$ in Table~\ref{table_sigksigf_cont} we form the 
ratio  $\surd\sigma_k/g^2N$
which we plot in Fig.\ref{fig_ksig_Nair}. We observe very close agreement between the 
string tensions predicted by eqn(\ref{eqn_KKN}) and our calculated values. Indeed
the agreement is clearly better for $\sigma_{k\geq 2}$ than for $\sigma_f$. Of course,
since $\lim_{N\to\infty}\sigma_k = k\sigma_f$, any disagreement
will be the same for $\sigma_k$ and for $\sigma_f$ in the large $N$ limit.
Nonetheless it is clear that the simple formula in eqn(\ref{eqn_KKN})
provides a remarkably good approximation to all our $k$-string tensions for all
values of $N$.

\section{Glueball spectra}
\label{section_glueballs} 

We calculate glueball masses as described in Section~\ref{subsection_glue_calc}.
A particular concern is the finite volume states which might appear in
our spectrum of excited glueball states, as discussed in
Section~\ref{subsection_sys_errors}. The lightest
such state is composed of a ground state flux loop and its conjugate, so it will
have a mass $m_T\sim 2\sigma_f l$, and where it might appear is in the $0^{++}$ and 
$2^{++}$ spectra. Since our glueball operators are single trace while these finite
volume states are double trace, large-$N$ counting tells us that such states will 
decouple from  our correlators at larger $N$. So our strategy is to have a
large enough volume $l^2$ at small $N$ so that $m_T$ is heavier than the states
we calculate, and then to relax the volume to a smaller value at larger $N$
where we expect such states to become invisible. A summary of the `standard'
volumes we have chosen to use, expressed in the relevant physical units, is
presented in Table~\ref{table_sizeN}.
We check these choices with some explicit comparisons between smaller
and larger volumes. For example in $SU(8)$ at $\beta=306.25$ we compare the 
spectrum on our `standard' $44^2 48$ volume to the one on a larger $60^2 48$ volume.
The resulting glueball masses are given in Table~\ref{table_msu8_Vcomp}, where
we include all the excitations for which we will attempt to obtain continuum
limits. We see that the spectrum on the $l=44a$ volume matches that on the $l=60a$
volume within say $2\sigma$, suggesting that for this moderately large value of
$N$ the choice of $l\surd\sigma_f \sim 3.8$ is sufficient to exclude finite volume
corrections (of whatever source) at the level of our statistical errors.
In  Table~\ref{table_msu16_Vcomp} we perform
a similar comparison in $SU(16)$ at $\beta=800$ where our `standard' volume
is $l=22a$, corresponding to to $l\surd\sigma_f \sim 3.1$. Again we see
that the glueball spectra agree, within say $2\sigma$, indicating that
our reduction in $l\surd\sigma_f$ with increasing $N$ (based on the idea of
the large-$N$ suppression of finite volume corrections) is indeed appropriate.
The specific lattices and $\beta$-values used are listed in  
Tables~\ref{table_param_su2}-\ref{table_param_su16}, along with the
values of the string tension and the mass gap. For $N\in[2,6]$ there
are two sets of volumes listed: the larger are used for glueball calculations
while the smaller are used for string tension calculations. The reason for
doing this is that the energy of the flux loop grows with $l$, so we can perform
a (statistically) more precise calculation with smaller $l$. This is only
possible, of course, because we have a very precise theoretical control of
finite volume corrections in this case.

Having  obtained our `infinite' volume glueball masses in this way, for various
lattice spacings, we extrapolate the results to the continuum limit. Mostly
we do so for the dimensionless ratio $am/a\surd\sigma_f = m/\surd\sigma_f $ since
in general $a\surd\sigma_f$ is our most accurately calculated physical quantity
on the lattice. We perform the continuum extrapolation in the standard way
\begin{equation}
\left.\frac{M}{\surd\sigma_f}\right|_{a}
=
\left.\frac{M}{\surd\sigma_f}\right|_{a=0}
+ c_1 a^2\sigma_f + c_2 (a^2\sigma_f)^2+ ... \,,
\label{eqn_MKcont}
\end{equation}
and we find in nearly all cases that the first $O(a^2)$ correction
suffices for a good fit. The results of these continuum extrapolations
are listed in Tables~\ref{table_mksu2_cont}-\ref{table_mksu16_cont},
together with the $\chi^2$ and number of degrees of freedom $n_{dof}$ 
of the fit. In Table~\ref{table_mksu8_cont} we also show the
values one finds for the coefficient $c_1$ in eqn(\ref{eqn_MKcont});
they are very similar for other values of $N$. We see that the
lattice corrections are very modest even at our coarsest lattice
spacing, where $a^2\sigma_f \sim 0.1$.
Note that at the coarser values of $a$, the masses 
of  higher excited states can be too large for them to be identified, 
and then the value of $n_{dof}$ is smaller than for the ground states.

In addition to these continuum fits we also calculate in one or two
cases the continuum limit of the ratio $m/g^2N$ just as we did for
the string tension in eqn(\ref{eqn_sigf_cont}). This might seem an
attractive way to calculate continuum physics because the error on 
the denominator $g^2_I N$ comes from the average  plaquette and
is negligible. In practice, however, this advantage is more than 
outweighed by the fact that the leading correction is $O(a)$ rather than
$O(a^2)$, so we generally focus on the extrapolations using eqn(\ref{eqn_MKcont}). 
We will also look at some of the masses in units of the mass 
gap, which is of interest for the reasons discussed in
\cite{AARLMT_SONSUN}.

Finally we extrapolate our continuum mass ratios to $N=\infty$ using
\begin{equation}
\left.\frac{M}{\surd\sigma_f}\right|_N
=
\left.\frac{M}{\surd\sigma_f}\right|_\infty
+ \frac{c_1}{N^2} + \frac{c_2}{N^4} + ...
\label{eqn_MKN}
\end{equation}
with the results listed in Tables~\ref{table_mksuN_J0}-\ref{table_mksuN_J1}.
We note that in general fits with just the leading $O(1/N^2)$ 
correction are acceptable, although for our lightest and most accurately 
calculated masses, it helps to include a further   $O(1/N^4)$ correction
if one wishes to include the $SU(2)$ values in the fits.

\subsection{$|J|=0, 4, ...$ glueballs}
\label{subsection_J0glueballs}

Since the argument for parity doubling breaks down for $J=0$ glueballs,
the $0^{++}$, $0^{--}$, $0^{-+}$, and $0^{+-}$ glueball masses are, in
principle, all unrelated. The lightest glueball state is the  $0^{++}$ ground
state and, as we have seen in Fig.\ref{fig_meff_JPC_su6}, its mass and that 
of the lowest few $0^{++}$ excitations can be calculated very accurately. The
$0^{--}$ is the next lightest ground state state and the lowest few $0^{--}$ masses
can also be calculated accurately. The lightest $0^{-+}$ and $0^{+-}$ are quite
heavy and the mass estimates for them are correspondingly less reliable.

\begin{figure}[htb]
\begin	{center}
\leavevmode
\input	{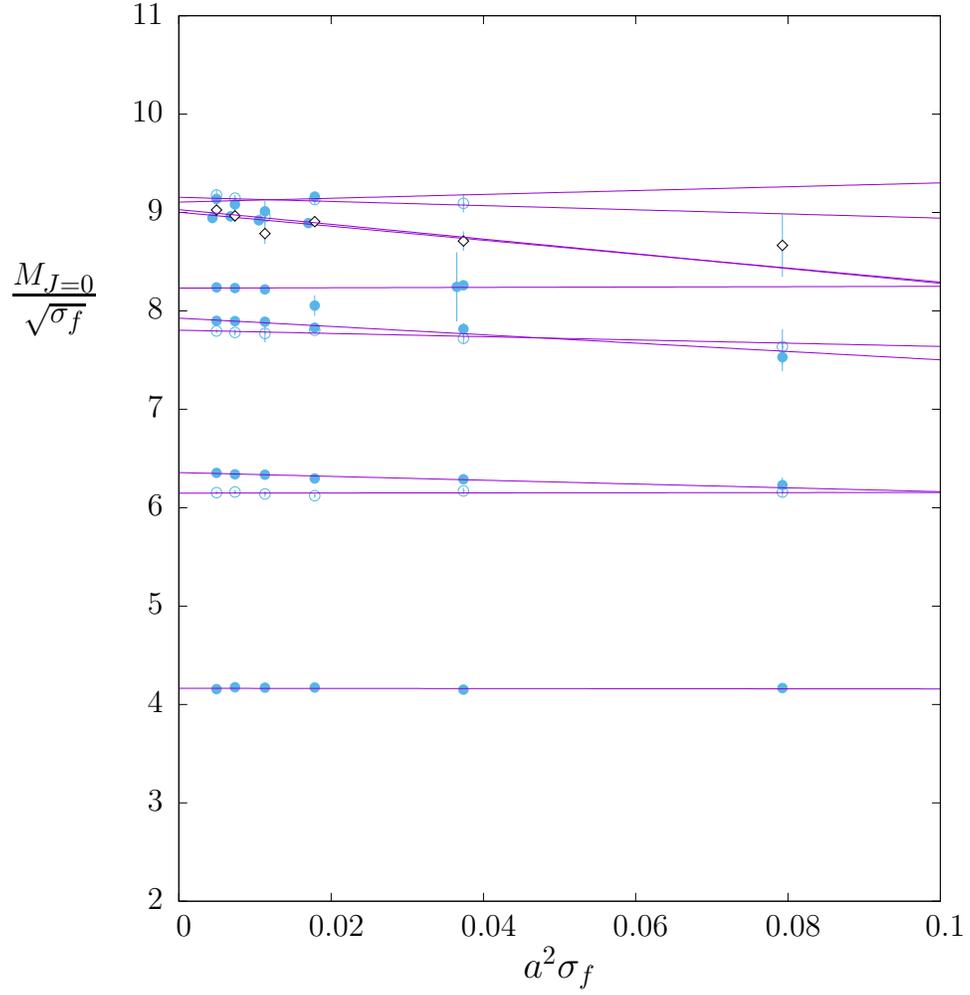}
\end	{center}
\caption{Lightest six $0^{++}$ ($\bullet$), lightest three $0^{--}$ ($\circ$),
and ground state $0^{-+}$ ($\Diamond$) glueballs with
continuum extrapolations. For $SU(6)$.}
\label{fig_mJ0_cont_su6}
\end{figure}

In Fig.\ref{fig_mJ0_cont_su6} we plot the lattice values of
$m/\surd\sigma_f$, calculated in $SU(6)$, against the value 
of $a^2\sigma_f$ for the lightest 
six $0^{++}$ states, the lightest three $0^{--}$ states, and for the $0^{-+}$ 
ground state. We also show the continuum extrapolations, and note that in all 
cases a leading $O(a^2)$ correction suffices to give an acceptable fit. 

We observe some striking regularities in Fig.\ref{fig_mJ0_cont_su6}. Firstly,
each of the three $0^{--}$ states is nearly (but not exactly) degenerate
with an excited $0^{++}$ state: to be specific, the first, second and fifth 
excited  $0^{++}$ states.
Secondly, the ground state $0^{-+}$ is degenerate, within errors, with the
fourth excited $0^{++}$. In fact earlier analyses
\cite{HMMT_spin}
have revealed that the lightest `$0^{-+}$' state is in fact a $4^{-+}$ state
so, by parity doubling, there should be a degenerate  $4^{++}$ state which
will appear in what we label as the set of `$0^{++}$' states. In other words,
we believe that what we have called the fourth $0^{++}$ excitation is in fact 
a $4^{++}$ state. Moreover this $4^{++}$ state is nearly degenerate
with the fifth  $0^{++}$ excitation (and also with the second excited $0^{--}$).
We also note that the ground state $0^{+-}$ appears to be nearly degenerate
with the first excited $0^{-+}$, and that in those cases where we have
calculations (i.e. for $N=4,12,16$) the first excited $0^{+-}$ appears to 
be nearly degenerate with the second excited $0^{-+}$. (With the caveat that
for these very massive states the errors are large.)

As we can see in Fig.\ref{fig_mJ0_cont_su6} the gaps between the states that
are `nearly degenerate' are much smaller than the typical gaps between other
states. That is to say, these regularities appear to be real rather than statistical
coincidences.

\begin{figure}[htb]
\begin	{center}
\leavevmode
\input	{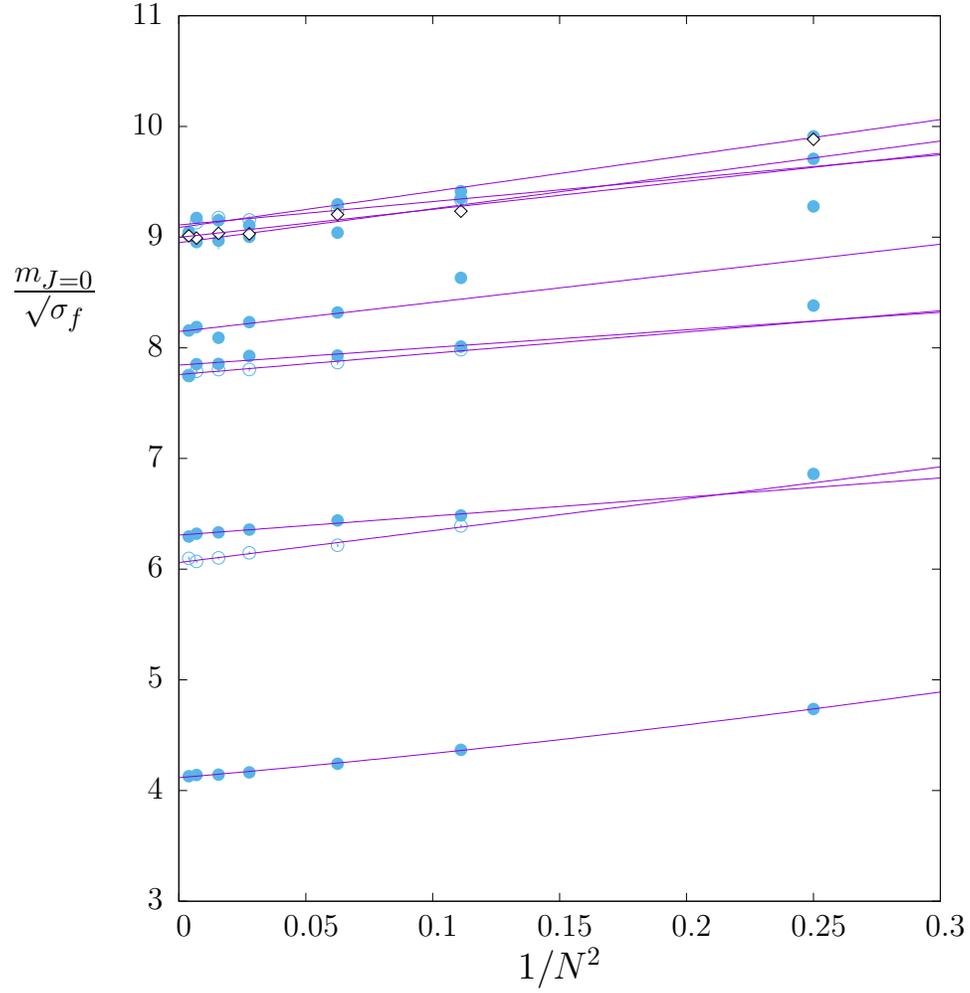}
\end	{center}
\caption{Lightest six $0^{++}$ ($\bullet$), lightest three $0^{--}$ ($\circ$),
and ground state $0^{-+}$ ($\Diamond$) glueballs with
large-$N$ extrapolations. For $SU(N)$ with $N\in[2,16]$.} 
\label{fig_m0K_N}
\end{figure}

While we have chosen to use $SU(6)$ in Fig.~\ref{fig_mJ0_cont_su6} to display these features
they are in fact common to all $SU(N)$ as we see from the continuum
limits listed in Tables~\ref{table_mksu2_cont}-\ref{table_mksu16_cont} and the large-$N$ 
extrapolations in Table~\ref{table_mksuN_J0}, together with Fig.\ref{fig_m0K_N} 
where we show the continuum values of $m/\surd\sigma_f$ for these states for all 
our values of $N$.

\subsection{$|J|=2, 6, ...$ glueballs}
\label{subsection_J2glueballs}

For $J=2$ glueballs we have parity doubling so we expect to have
$m_{2^{++}} = m_{2^{-+}}$ and $m_{2^{--}} = m_{2^{+-}}$ up to lattice
spacing and finite volume corrections.
We list the $J=2$ glueball masses in units of the string tension, and 
after an extrapolation to the continuum limit, in
Tables~\ref{table_mksu2_cont}-\ref{table_mksu16_cont} and we see 
that the parity doubling expectations are broadly satisfied, albeit
with an apparent exception in the case of the third state in the $2^{\pm +}$
sectors in both $SU(3)$ and $SU(8)$, as well as the fourth excited
state in $SU(3)$. (For $SU(3)$ these may be finite volume effects,
while in the case of $SU(8)$, where finite volume corrections should 
be large-$N$ suppressed, it may just be an unlikely statistical fluctuation.)

As an example we plot in Fig.\ref{fig_mJ2_cont_su12} our values in $SU(12)$ of
the lightest few $2^{++}$  and $2^{--}$ glueball masses as a function of 
$a^2\sigma_f$ together with the corresponding linear continuum extrapolations.
Just as for $J=0$ we observe a striking  pattern of near-degeneracies:
the first, second and fourth   $2^{++}$ excited states appear to be
nearly degenerate with the lightest three  $2^{--}$ states.
We also see from Tables~\ref{table_mksu2_cont}-\ref{table_mksu16_cont}
that this near-degeneracy holds for all $N$ except in $SU(3)$ and
$SU(8)$ for those excited states where the parity doubling is poor,
and, in any case, in those cases the near-degeneracy holds between the $2^{-+}$
and $2^{\pm -}$. 

The large-$N$ extrapolations of our $J=2$ spectra are listed in 
Table~\ref{table_mksuN_J2}. In the  $2^{++}$ sector
the second and fourth excited states have such poor fits
that we do not attempt to provide any estimate based on
a large-$N$ extrapolation. Instead we provide an estimate
based on the average of the $SU(12)$ and $SU(16)$ values
(with errors enhanced to encompass the two values.) Since
the fits to the supposedly degenerate $2^{-+}$ parity partners are mostly 
acceptable, it is these that we plot in  Fig.\ref{fig_m2K_N} against 
$1/N^2$, together with the lightest $2^{--}$ states. We clearly see the
near-degeneracies discussed above. We also see that the large-$N$ limits
of the second and third excited $2^{-+}$ states are very similar.
All this is very similar to what we observed in the $J=0$ sector,
displayed in  Fig.\ref{fig_m0K_N}. 

One might ask if there is a $J=6$ state located in the $J=2$ sector,
just like the $J=4$ state in the $J=0$ sector. The answer given in
\cite{HMMT_spin}
for $SU(2)$, suggests that $m_{gs}^{J=6} \sim 1.5 m_{gs}^{J=2}$ which
means that it will be heavier than the states we consider here.

\subsection{$|J|=1, 3, ...$ glueballs}
\label{subsection_J1glueballs}

For $J=1$ (and indeed any odd $J$) we expect to have exact parity doubling, not
only in the continuum limit but even on a square lattice in a finite volume
as long as the latter respects $\pi/2$ rotational invariance. Our calculated
masses are consistent with parity doubling, as we see for example in
Tables~\ref{table_msu4_opcomp},~\ref{table_msu6_opcomp} and
Tables-\ref{table_msu8_Vcomp},-\ref{table_msu16_Vcomp}.

We list in Tables~\ref{table_mksu2_cont}-\ref{table_mksu16_cont} the values
that we obtain for the $J=1$ glueball masses expressed in units of the string
tension, after an extrapolation to the continuum limit. For simplicity we refer
to the $|J|=odd$ states as $J=1$ both here and in the Tables, but will shortly
return to the question of their true spin.) A striking feature of the $J=1$
spectrum is that for $N\geq 3$ the lightest two $1^{++}$ states are nearly
degenerate, within our errors, and that the next two states are also nearly 
degenerate. (By parity doubling the same is true for the $1^{-+}$ states.)
The gap between these two pairs of states is much larger than any
gap within each pair, making it plausible that the near-degeneracy is not just
a statistical accident but has a dynamical origin. Moreover the fact that all
these states are very massive, with correspondingly large statistical errors,
creates some technical problems for the variational procedure, where the ordering
of the states may differ in different (jack-knife) data bins so that the observed
small splittings between the states in each pair may be enhanced, or even
entirely driven, by the statistical fluctuations. This may be the reason for
the fact that many of the continuum fits are quite poor, and that we occasionally
see discrepancies between the $P=\pm$ $J=1$ masses.

Another feature of the $J=1$ spectrum, that differs from the $J=0,2$ spectra, is that 
the lightest state is the $C=-$ ground state. Moreover the first excited $C=-$ state
appears to be roughly degenerate with the pair of nearly degenerate $C=+$ ground states 
and, albeit now within larger errors, the third excited $C=-$ state is nearly degenerate 
with the next pair of  nearly degenerate $C=+$ states. All this is very reminiscent of
what we saw for $J=0,2$, but with $C=-$ and $C=+$ interchanged.

\begin{figure}[htb]
\begin	{center}
\leavevmode
\input	{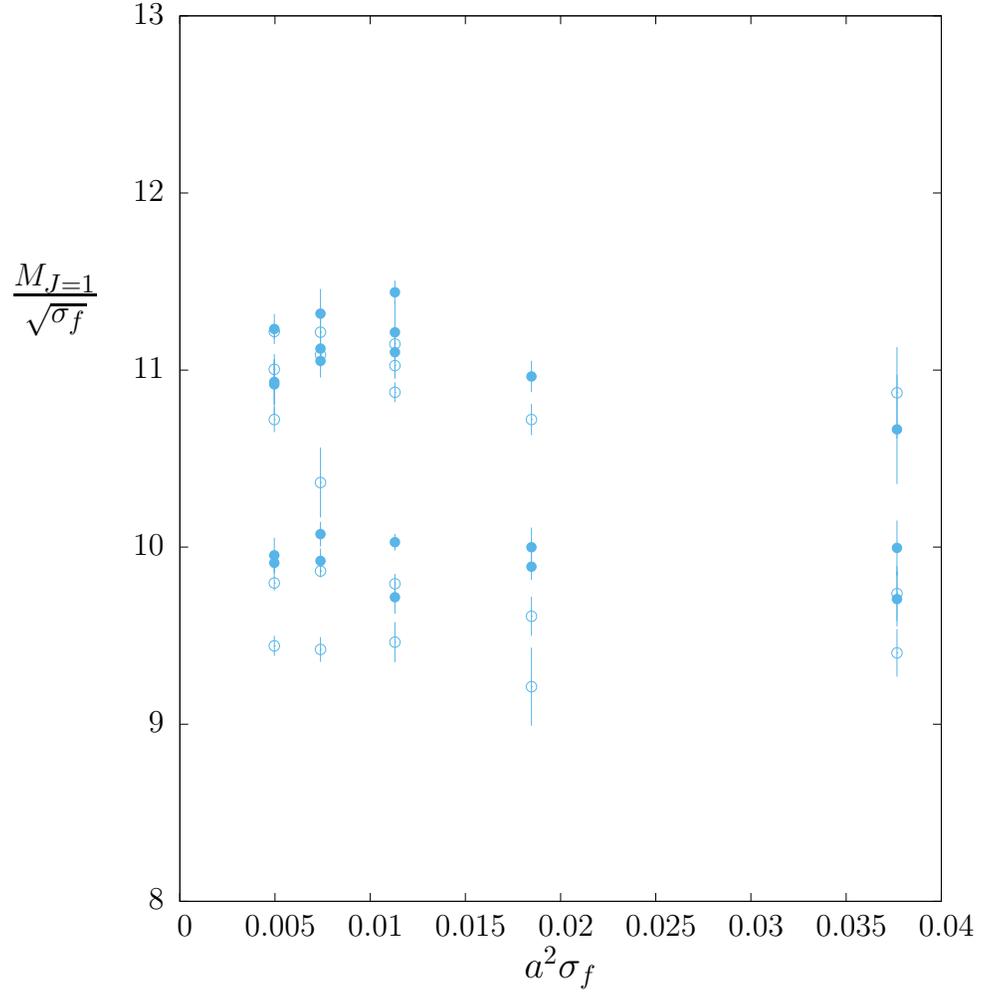}
\end	{center}
\caption{Lightest four $1^{\pm +}$ ($\bullet$) and $1^{\pm-}$ ($\circ$)
glueballs versus $a^2\sigma_f$ in $SU(8)$.}
\label{fig_mJ1Pav_cont_su8}
\end{figure}

To illustrate these features of the $J=1^{\pm +}$ and $J=1^{\pm -}$ spectra, we plot 
in Fig.~\ref{fig_mJ1Pav_cont_su8} the masses of the $J=1$ states against $a^2\sigma_f$
for our $SU(8)$ calculation. We have averaged the $P=\pm$ masses since they should
be degenerate, and this should help to suppress fluctuations. The fact that the lightest
state is $C=-$ is unambiguous. We also see good evidence that the lightest two
$C=+$ states are (nearly) degenerate, and that the first excited $C=-$ state
is (nearly) degenerate with them. The next two $C=+$ states have similar masses to each
other and may be degenerate although the errors on these massive states are so
large that this is something of a conjecture. Equally, the next two $C=-$ states have
similar masses to each other and also, interestingly, to the second pair of
$C=+$ states.

\begin{figure}[htb]
\begin	{center}
\leavevmode
\input	{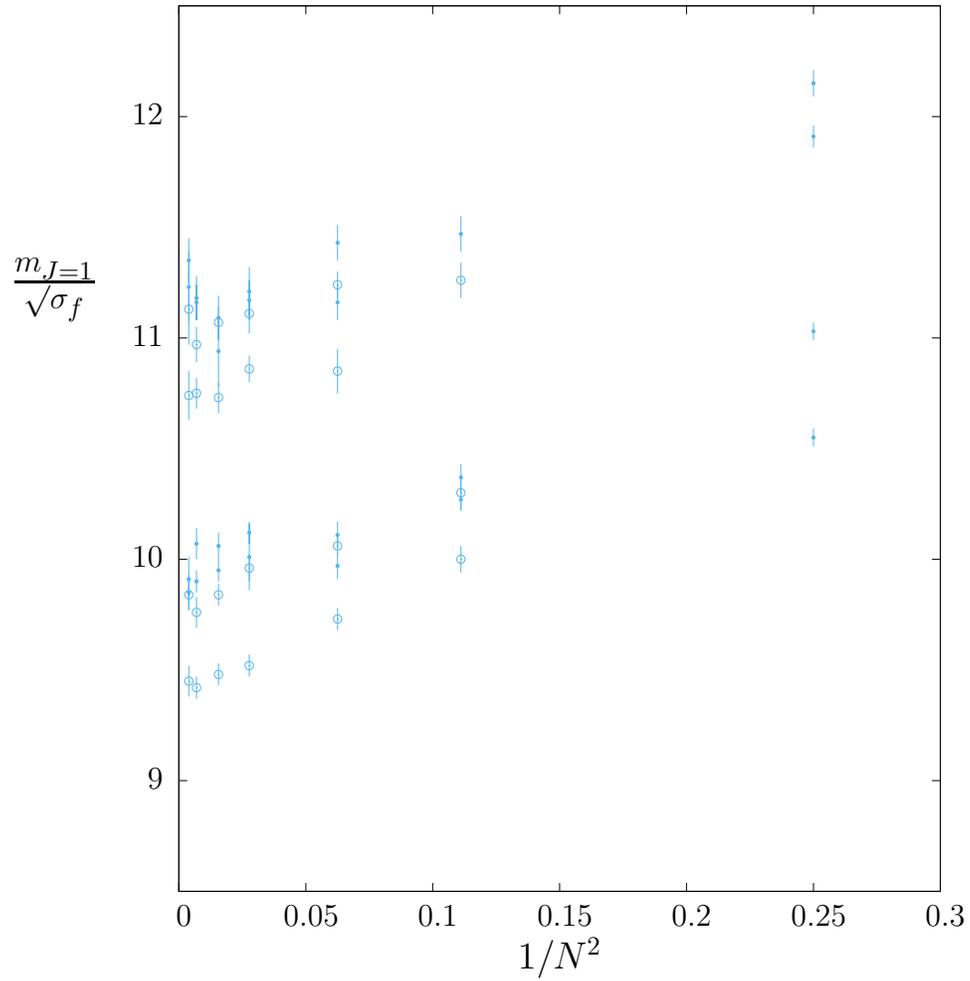}
\end	{center}
\caption{Lightest few $1^{\pm +}$ ($\bullet$) and $1^{\pm -}$ ($\circ$) glueballs versus $1/N^2$.}
\label{fig_m1K_N}
\end{figure}

These features are characteristic of all our $SU(N\geq 3)$ spectra and hence also of the
$N\to\infty$ extrapolations listed in Table~\ref{table_mksuN_J1}. These
extrapolations have reasonably small errors, and the (very near) degeneracy
of the lightest two $C=+$ states is convincing as is the approximate degeneracy of
the next two excited  $C=+$ states. It is also clear that the $C=-$ ground state
is the lightest $J=1$ state and that it is not degenerate with any other state.
However the first excited $C=-$ state appears to be (nearly) degenerate with the
lightest pair of $C=+$ states. The second $C=-$ excitation does not appear to
be very close to any other states, but the third $C=-$ excitation appears to
be nearly degenerate with the second pair of  $C=+$ states. One can, as for $SU(8)$,
reduce the fluctuations a little by averaging the $P=\pm$ values. We show the
result of doing so in  Fig.~\ref{fig_m1K_N} where we plot these averaged values
for all our $SU(N)$ continuum limits against $1/N^2$. Apart from $SU(2)$ where
there are no  $C=-$ states, and where the  lightest four $C=+$ states definitely
do not form nearly degenerate pairs, all the other $SU(N)$ spectra display
the features emphasised above.

Given these striking regularities, it is interesting to ask whether there is any
evidence that some of these states might not be $|J|=3,5,...$ rather than $J=1$.
Now the analysis for $SU(2)$ in 
\cite{HMMT_spin}
and the extension to $SU(3)$ and $SU(5)$ in Section 6.4 of the thesis referred to there
does in fact provide evidence that the lightest two states in the `$1^{++}$' sector
are $J=3$ and not $J=1$, while in the `$1^{--}$' sector the ground state is $J=1$ but
the first excited state is $J=3$. If we take these assignments as correct then
this will alter our conclusions as follows:
the lightest $J=1^{P=\pm}$ state has $C=-$ and is lighter than the lightest $J=3$ state
and is much lighter than the lightest $C=+$ $J=1$ state. The lightest $J=3^{P=\pm}$ states
consist of three nearly degenerate states: two with $C=+$ and one with $C=-$. For the
higher excited states we have no evidence concerning their spin, other than that it is odd.

%
%
%
%
\section{Discussion}
\label{section_conclusion} 

A striking feature of the fundamental string tension
expressed in units of the coupling, is how small are the lattice corrections,
as we can see in Fig.~\ref{fig_k1g_cont}. Even at the coarsest lattice spacings,
where $a\surd\sigma_f \sim 0.3$, the  correction is less than $10\%$.
Indeed a simple $O(ag^2_IN)$ correction term is all that 
is needed despite the precision of our lattice calculations of $a^2\sigma_f$ and the
substantial range of $a$ being fitted. Moreover the same is true of the mass gap, 
as we see in Table~\ref{table_sigf_cont}.
The lattice corrections are even smaller when we consider the masses of the lightest
glueballs expressed in units of the string tension, $m_G/\surd\sigma_f$, as we see
in Figs.~\ref{fig_mJ0_cont_su6},~\ref{fig_mJ2_cont_su12} and \ref{fig_mJ1Pav_cont_su8}. 
The same is true for the tensions of the stable higher representation flux tubes, 
as we see in Fig.~\ref{fig_kf_cont_su8} in the case of $SU(8)$.

This remarkably precocious scaling may have to do with the fact that the theory is 
super-renormalisable: the dimensionless running coupling decreases linearly as
the distance scale is reduced, $\tilde{g}^2(l) = lg^2$, so that corrections that
are higher powers of $g^2$ are necessarily accompanied by higher powers in $a$.
(In contrast to 4 dimensions where the leading $O(a^2)$ lattice correction is
a power series in $g^2$.) 

A striking feature of our (continuum extrapolated) values of 
$m_G/\surd\sigma_f$, $m_G/g^2N$ and $\surd\sigma_f/g^2N$, is how weak is their 
variation with $N$, over the whole range of $N\geq 2$, as we see in
Figs.~\ref{fig_m0K_N},~\ref{fig_m2K_N} and \ref{fig_m1K_N}. A possible explanation
for this has been given in our earlier paper
\cite{AARLMT_SONSUN}
where we discuss the constraints on the $N$-dependence that arise when we
consider $SO(N)$ as well as $SU(N)$ gauge theories.

In any case all this suggests that there may be some underlying simplicity in
the dynamics of $D=2+1$ $SU(N)$ gauge theories, and this motivates a close
examination of our results for unexpected regularities. We have indeed 
found evidence for a number of these (some already remarked upon in 
earlier less precise calculations)
which we shall now summarise, beginning with the string tensions.

In addition to the stable fundamental flux tube, it is known that
the ground states of $k$-strings (i.e. flux tubes which carry the flux of 
local static sources consisting of $k$ fundamental charges) are also
stable, and that these flux tubes are in the totally anti-symmetric representation
(when not very short). We confirm all this with our more accurate calculations.
Furthermore, our large range of $N$ allows us to make an unambiguous statement that the
leading corrections to the $N=\infty$ limit of $\sqrt{\sigma_k/\sigma_f}$ are
$O(1/N)$ rather than $O(1/N^2)$.
Moreover the $O(1/N)$ binding energy is consistent with being produced
by a pairwise interaction between the $k$ fundamental strings that are bound
into a $k$-string. And we confirm previous observations that to a very good 
approximation the string tensions
are proportional to the quadratic Casimir of the representation. 
It is intriguing that the absolute value of $\surd\sigma_k/g^2N$
is very close to that predicted in 
\cite{Nair} 
as we see in Fig.\ref{fig_ksig_Nair}.

Turning now to our results for glueballs, here the striking feature is a number
of unexpected near-degeneracies amongst glueballs with different $J^{PC}$
quantum numbers. (This is in addition to the expected parity doubling for 
$J\neq 0$ which we also observe.) 

For $J=0$ we saw that the first, second and fifth excited 
$0^{++}$ glueballs are nearly degenerate with the lightest three $0^{--}$
glueball states. The fourth excited `$0^{++}$' state is presumably
the $4^{++}$ ground state given its near degeneracy with the `$0^{-+}$'
ground state which is believed to be, in reality, the $4^{-+}$ ground state
\cite{HMMT_spin}.
The only $0^{++}$ states that are not nearly degenerate with another $J=0$ 
state (amongst the lightest
6 states) are the ground state and the third excited state. We note that
the mass of the latter is very nearly twice the mass of the  former.
(Although this may well be accidental.)

In the $J=2$ glueball sector we observe a nearly identical pattern of degeneracies.
The first, second and fourth excited $2^{++}$ glueballs are nearly degenerate with 
the lightest three $2^{--}$ glueball states. Only the ground and third excited
of these lightest $2^{++}$ states are not nearly degenerate with some other 
$J=2$ state..

In the $J=1$ sector the lightest glueball is a $1^{--}$. There are two very nearly 
degenerate `$1^{++}$' ground states which are nearly degenerate with the
first excited `$1^{--}$' state. But there is in fact good evidence 
\cite{HMMT_spin}
that these three nearly degenerate states are $J=3$ rather than $J=1$.
It also appears that the next two pairs of excited states in the $1^{++}$ and
$1^{--}$ sectors are nearly degenerate and nearly degenerate with each other,
although their large masses mean that this observation is more
speculative.

We finish by recalling that in the simplest closed flux tube model of glueballs
\cite{NIJP}, 
the $C=+$ and $C=-$ glueball states in two spatial dimensions are degenerate.
We also note that while
we have a number of near-degeneracies, they do not appear to be exact.
That is to say, even if they point to some kind of relatively simple dynamics,
this is likely to be the property of a field theory from which the $SU(N)$
gauge theory is -- at the very least -- a small perturbation. Helping to identify
such a  neighbouring field theory which one may hope to be analytically
tractable, is a  motivation for the present study.

\section*{Acknowledgements}

AA has been partially supported by an internal program of the University of Cyprus 
under the name of BARYONS. In addition, AA acknowledges the hospitality of the 
Cyprus Institute where part of this work was carried out. 
MT acknowledges partial support under STFC grant ST/L000474/1.
The numerical computations were carried out on the computing cluster 
in Oxford Theoretical Physics.

\clearpage

\begin{appendix}
\setcounter{table}{0}
\renewcommand{\thetable}{A\arabic{table}}
%
%
%
%
\section{Tables: some lattice data and operators}
\label{section_appendix_results}

\vspace*{1.5cm}

\begin{table}[ht]
\vspace{-0.5cm}
\begin{center}
\begin{tabular}{cccccc}
 $\parbox{0.65cm}{\rotatebox{0}{\includegraphics[height=0.65cm]{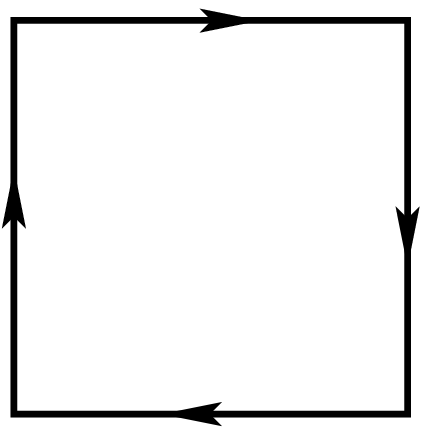}}}$ & $\parbox{1.3cm}{\rotatebox{0}{\includegraphics[height=1.3cm]{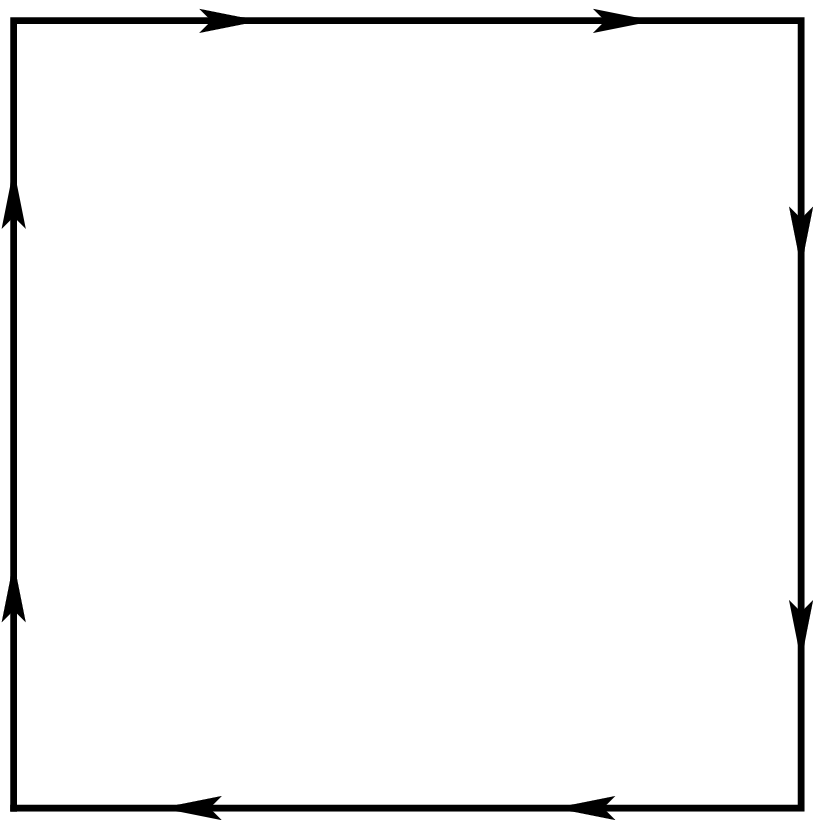}}}$ & $\parbox{1.3cm}{\rotatebox{0}{\includegraphics[height=1.3cm]{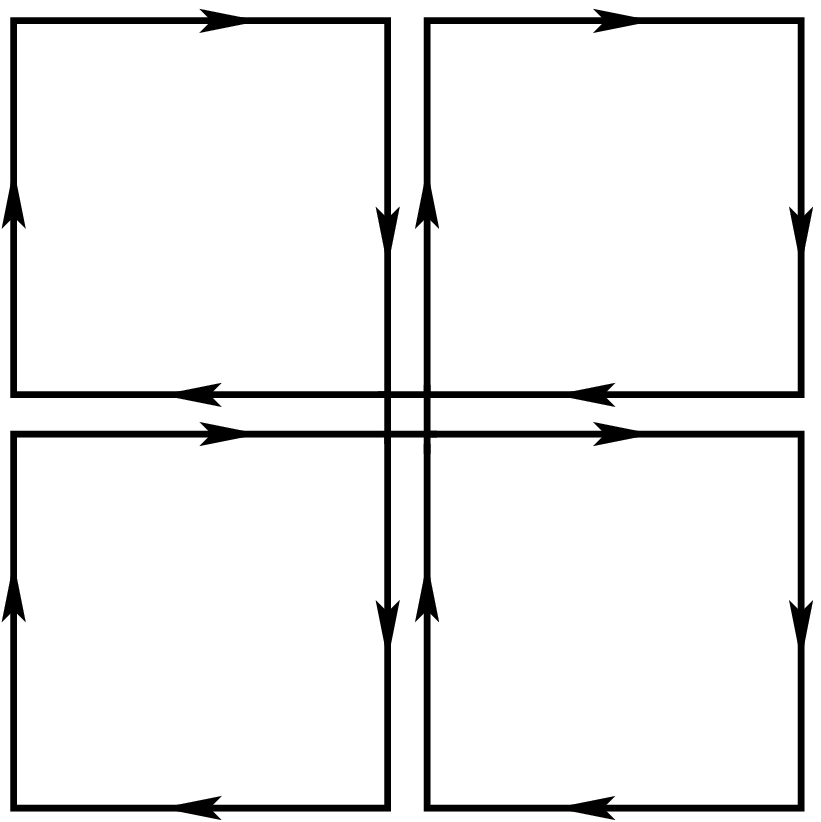}}}$ & $\parbox{1.3cm}{\rotatebox{0}{\includegraphics[height=1.3cm]{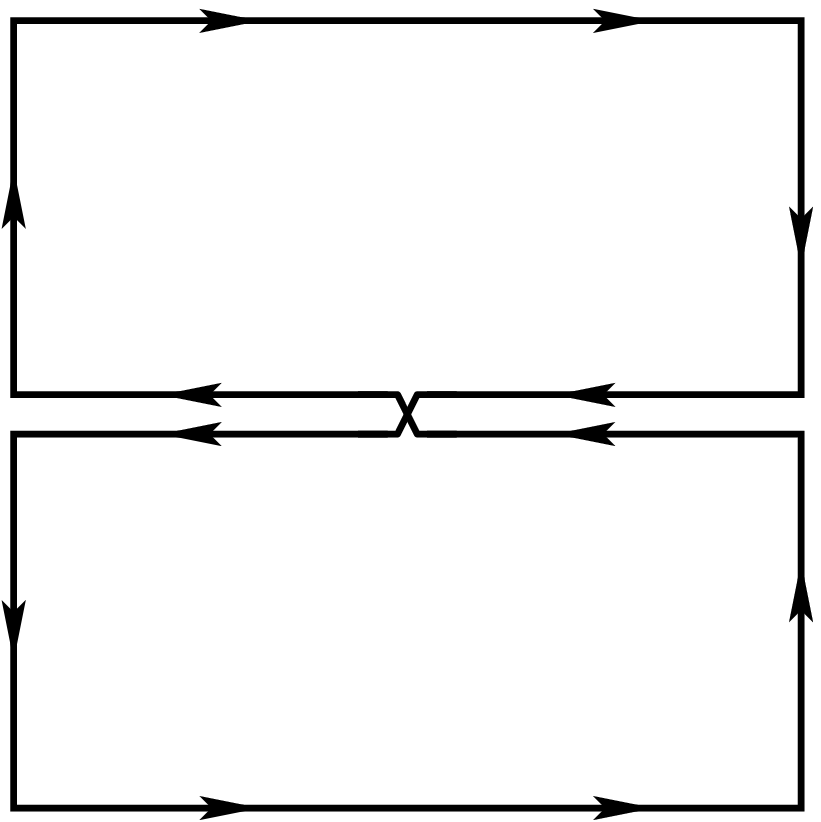}}}$ & $\parbox{1.3cm}{\rotatebox{0}{\includegraphics[height=0.65cm]{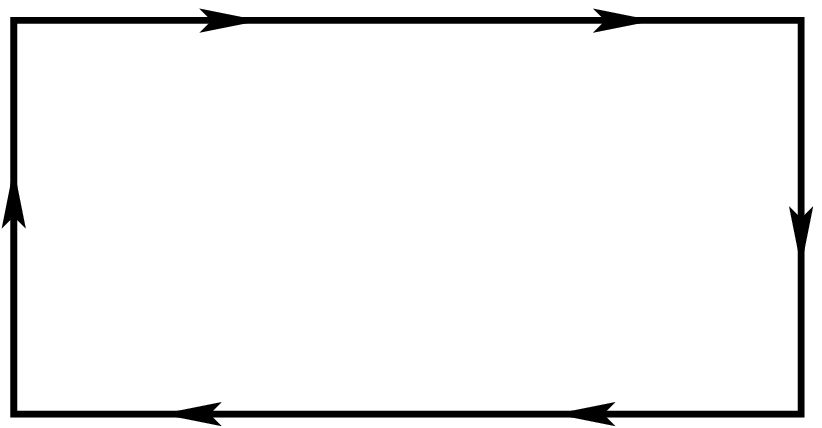}}}$  & $\parbox{1.95cm}{\rotatebox{0}{\includegraphics[height=0.65cm]{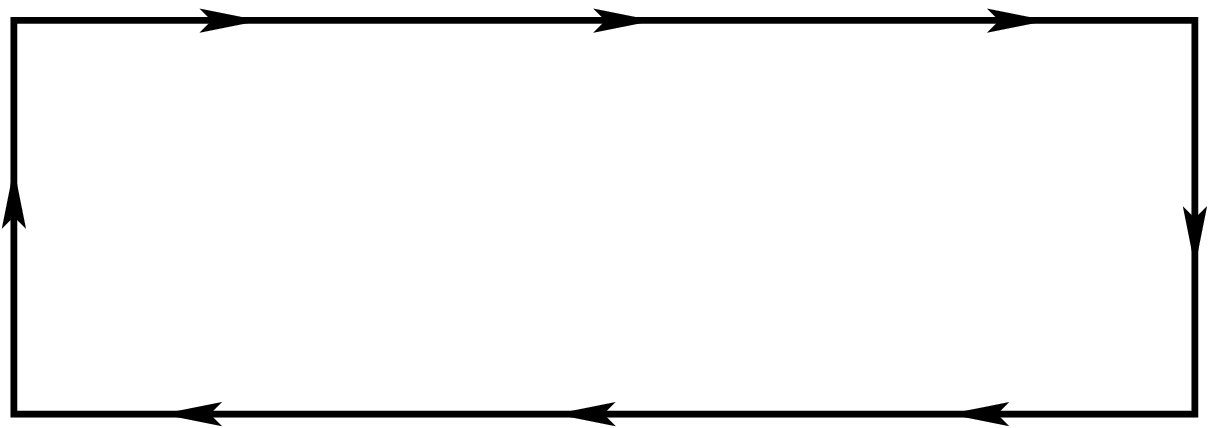}}}$ \\ \\ 
$\parbox{1.3cm}{\rotatebox{0}{\includegraphics[height=0.65cm]{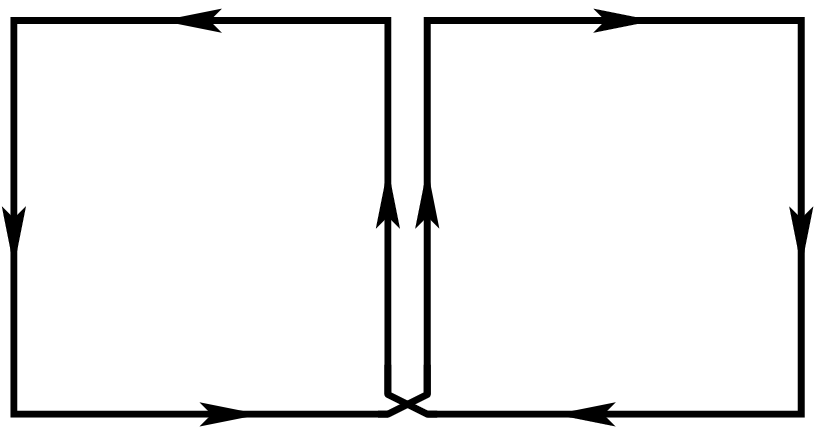}}}$ & $\parbox{1.3cm}{\rotatebox{0}{\includegraphics[height=1.3cm]{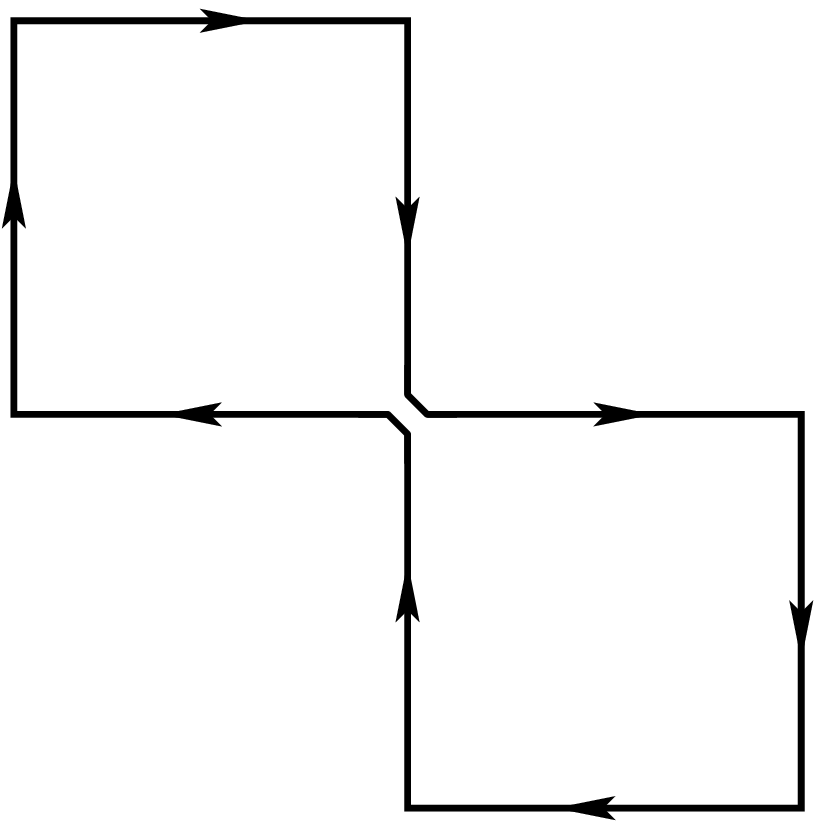}}}$ & $\parbox{1.3cm}{\rotatebox{0}{\includegraphics[height=1.3cm]{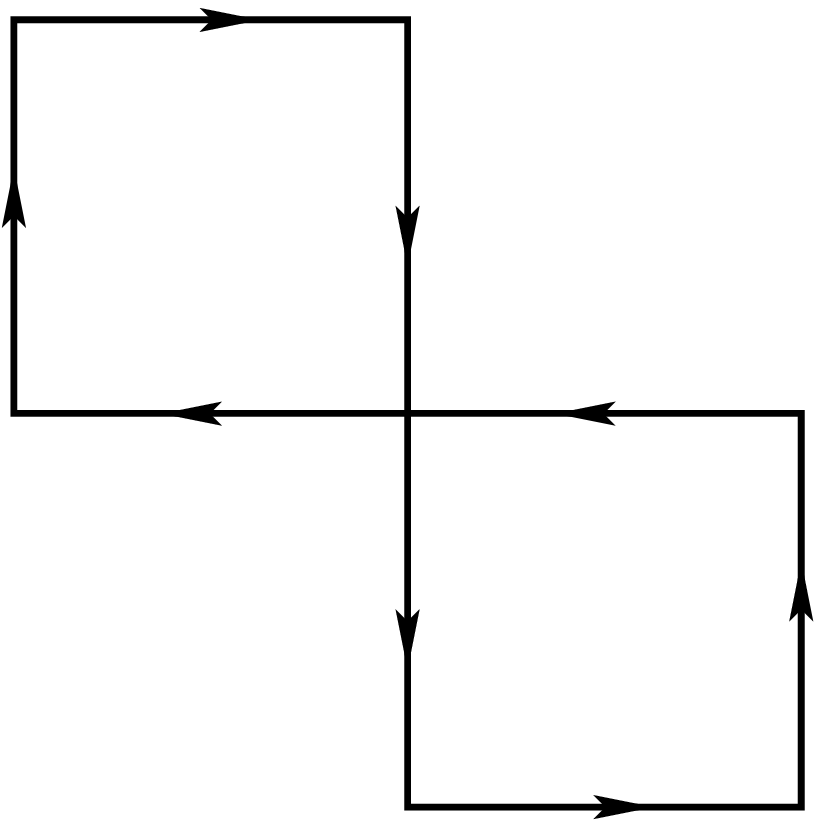}}}$  & $\parbox{1.95cm}{\rotatebox{0}{\includegraphics[height=1.95cm]{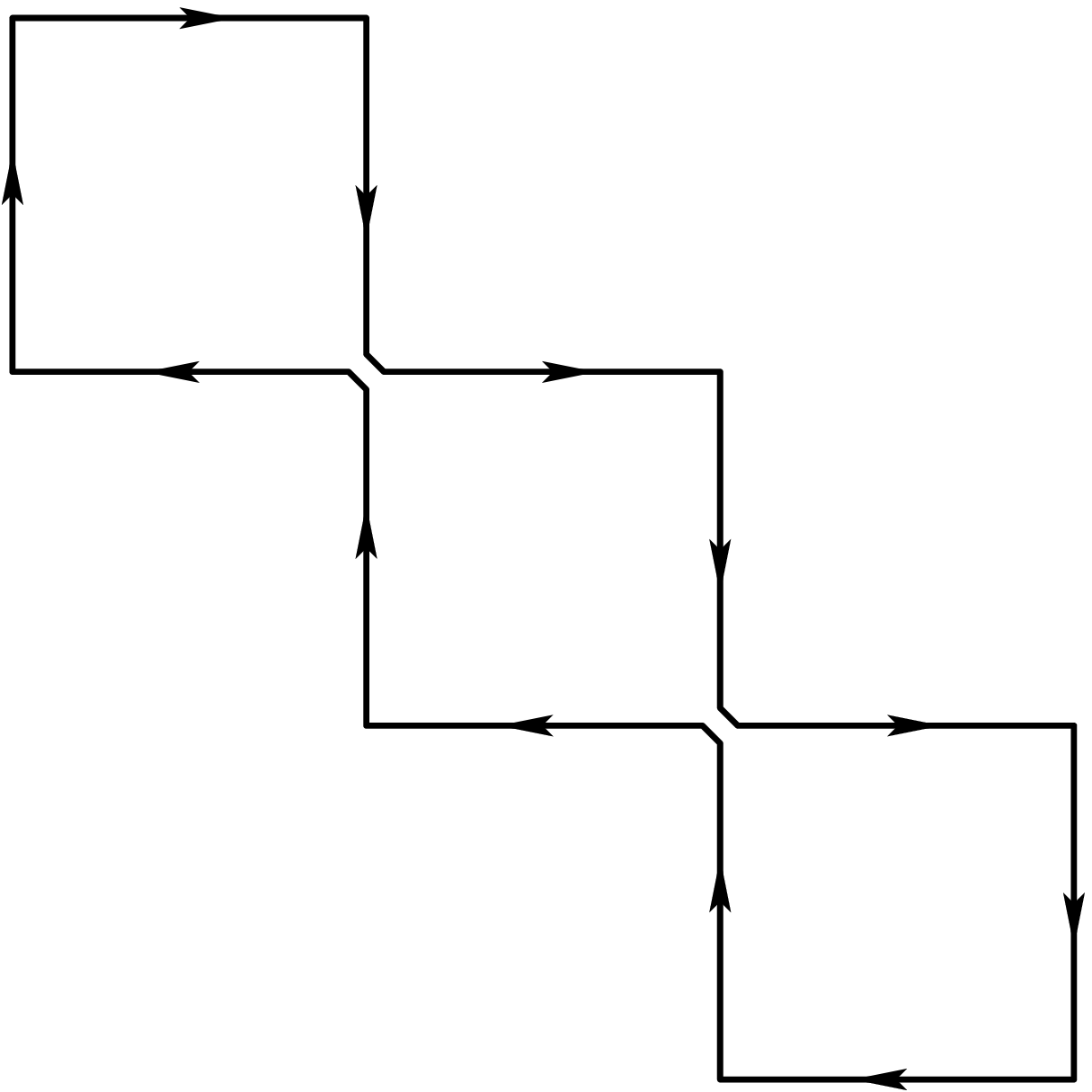}}}$  & $\parbox{1.95cm}{\rotatebox{0}{\includegraphics[height=1.95cm]{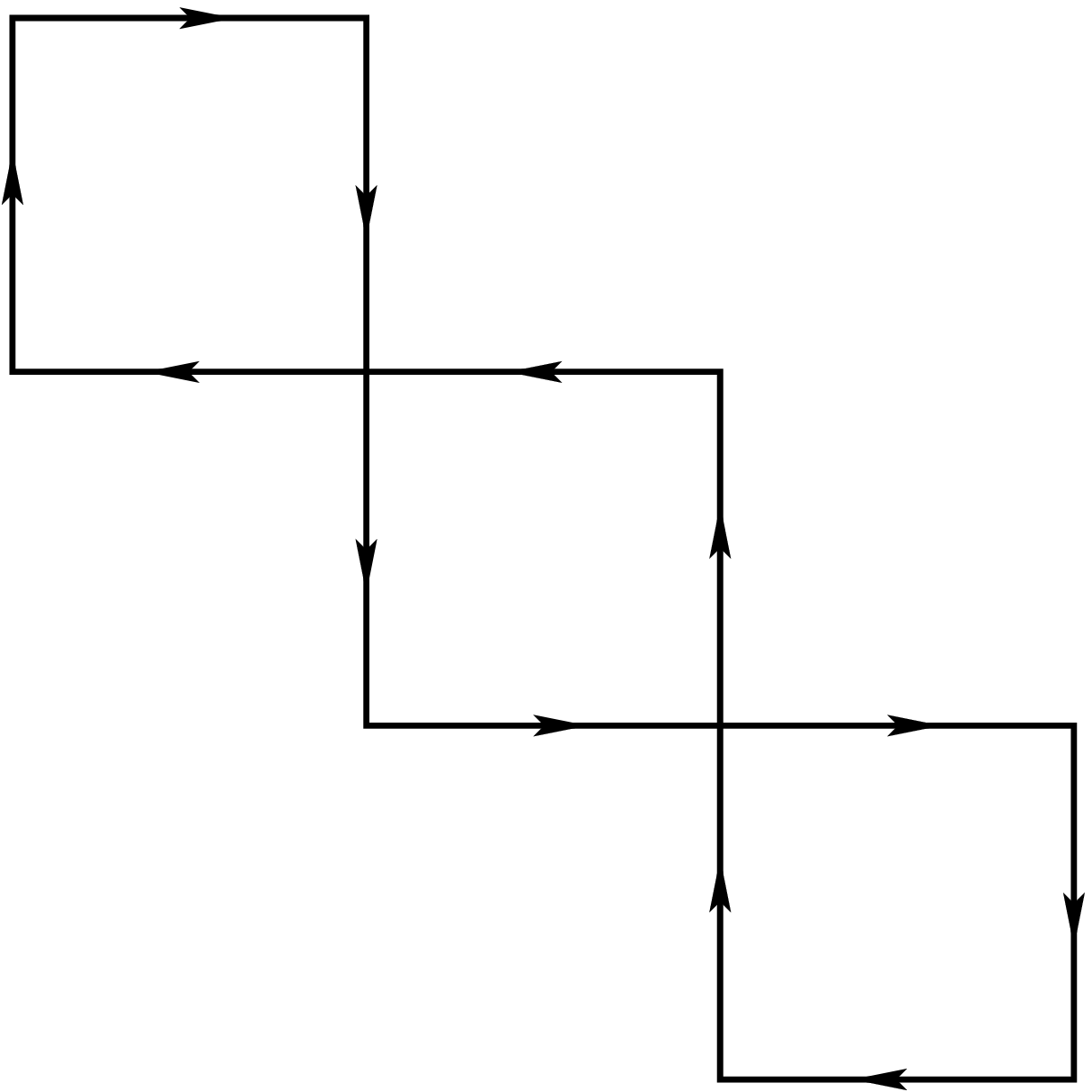}}}$  & $\parbox{1.95cm}{\rotatebox{0}{\includegraphics[height=1.95cm]{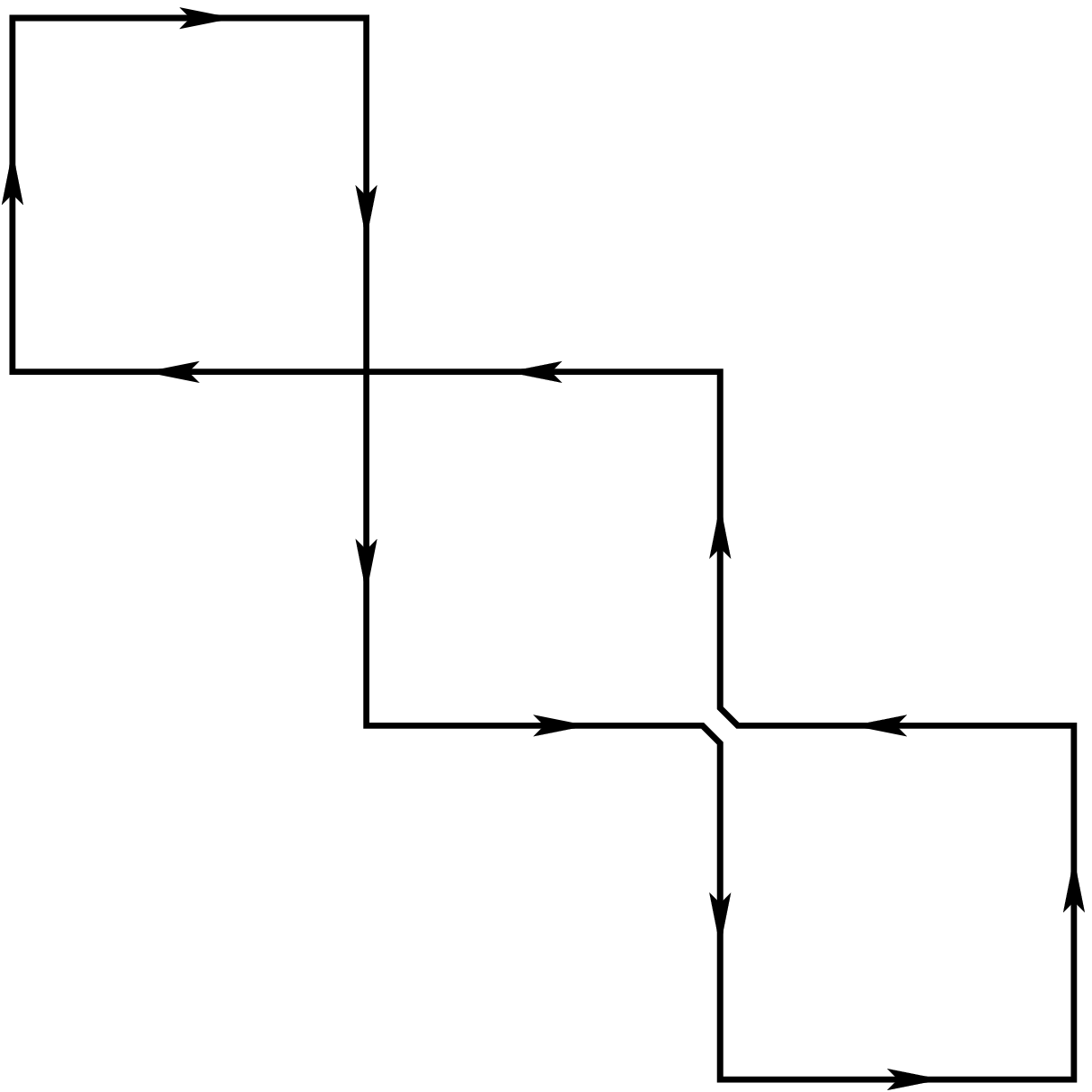}}}$ \\ \\ 
$\parbox{1.95cm}{\rotatebox{0}{\includegraphics[height=1.3cm]{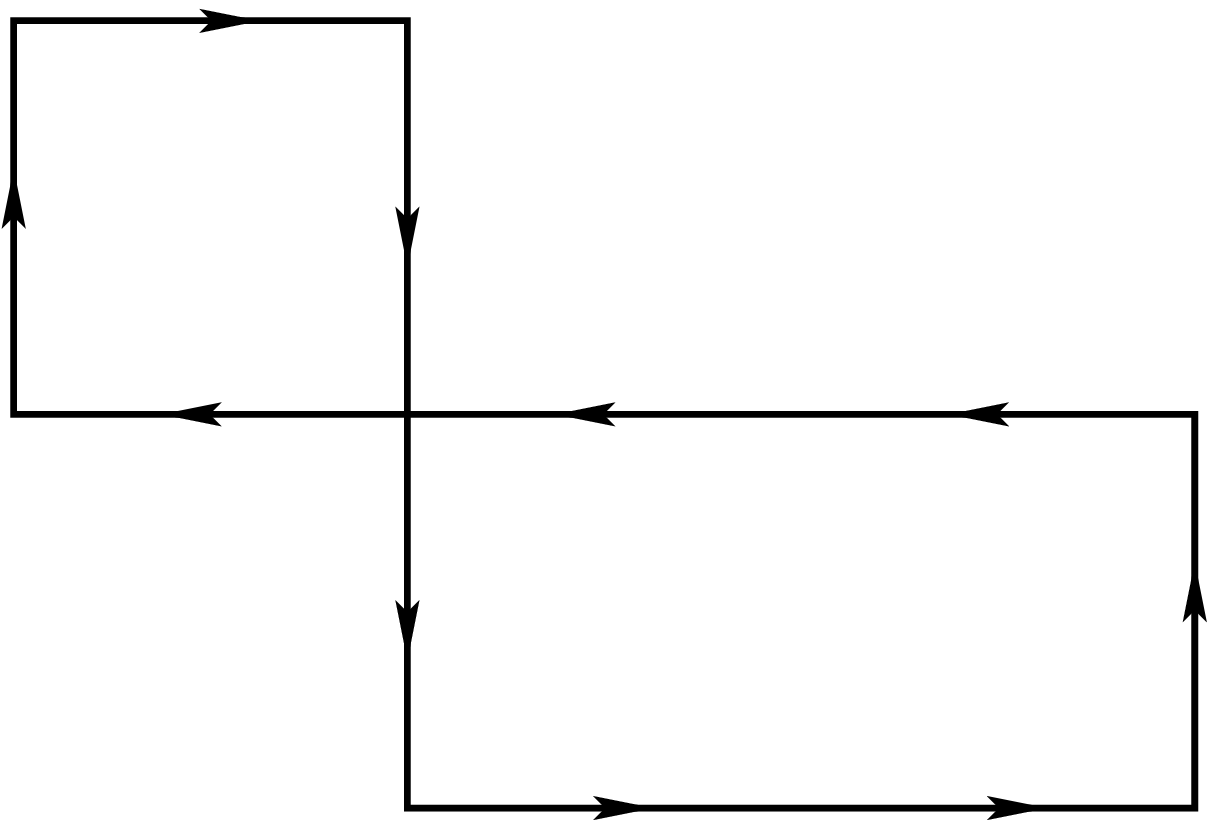}}}$ & $\parbox{1.95cm}{\rotatebox{0}{\includegraphics[height=1.3cm]{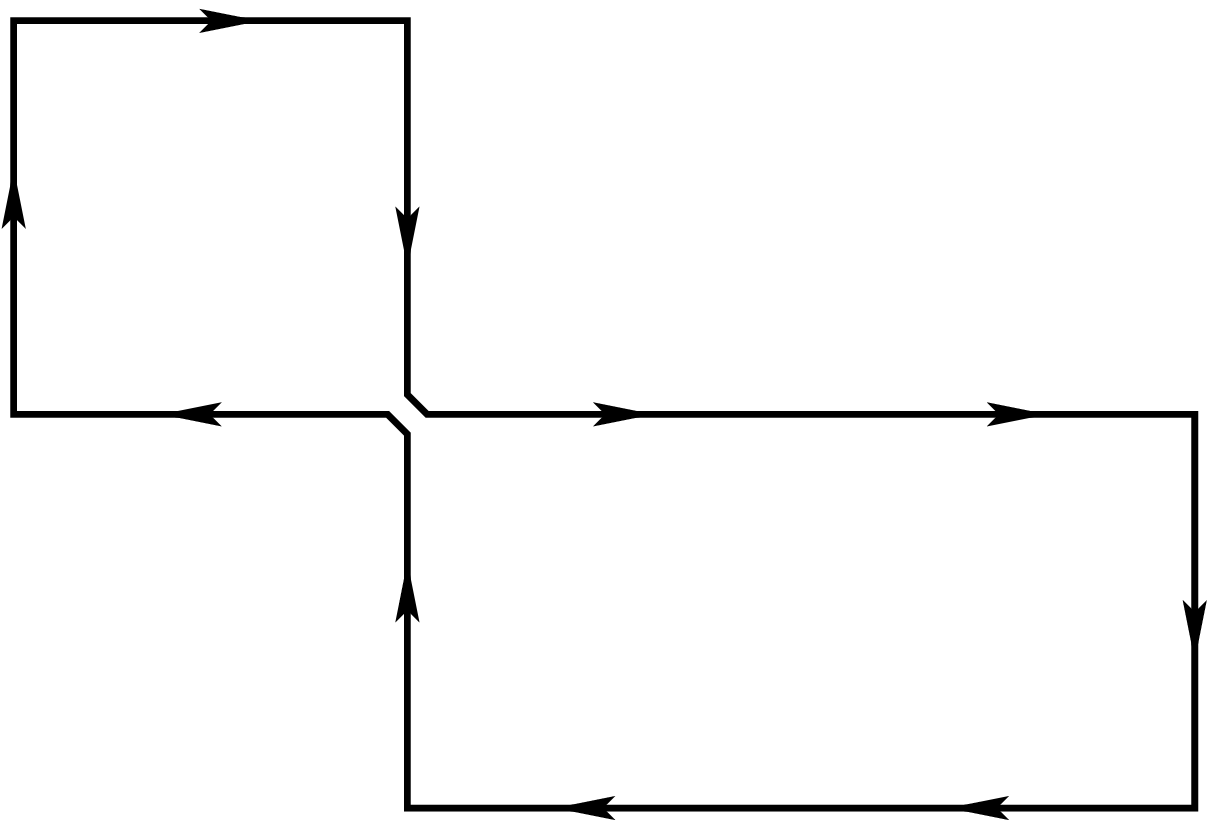}}}$  & $\parbox{1.3cm}{\rotatebox{0}{\includegraphics[height=1.3cm]{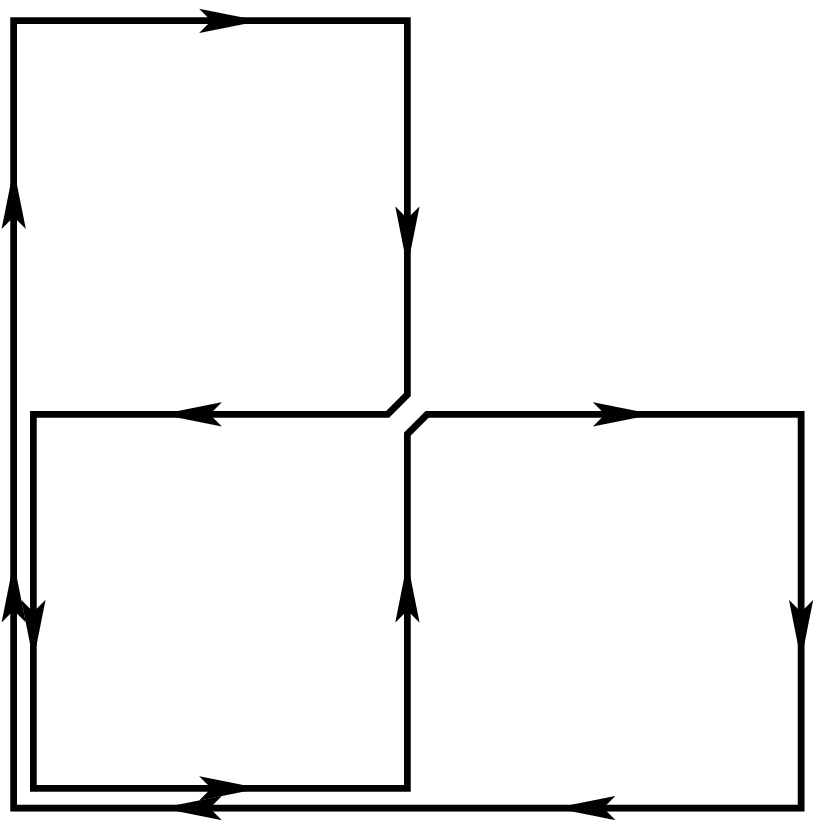}}}$  & $\parbox{1.3cm}{\rotatebox{0}{\includegraphics[height=1.3cm]{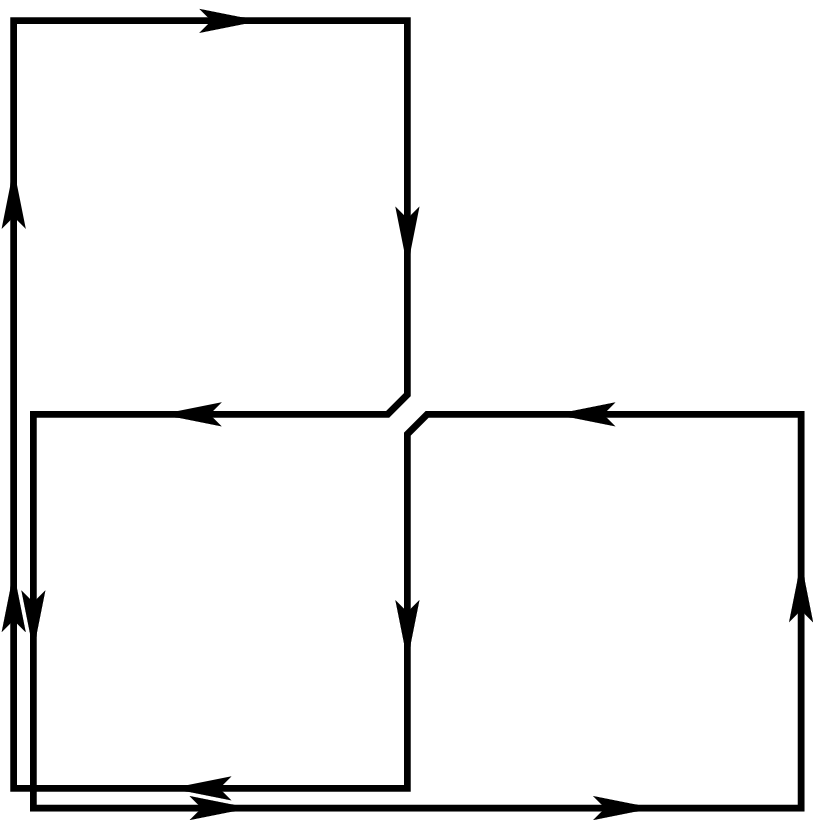}}}$  & $\parbox{1.95cm}{\rotatebox{0}{\includegraphics[height=0.65cm]{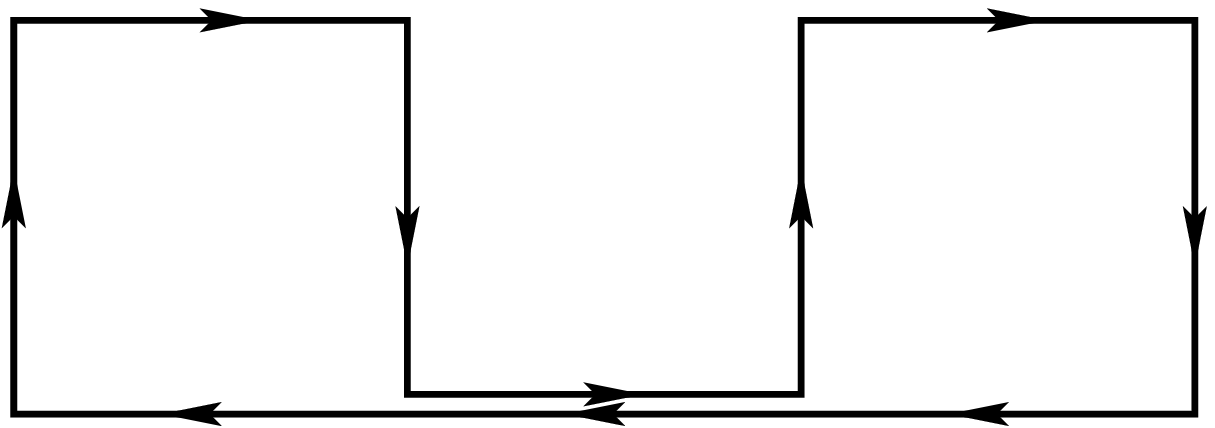}}}$ & $\parbox{1.95cm}{\rotatebox{0}{\includegraphics[height=0.65cm]{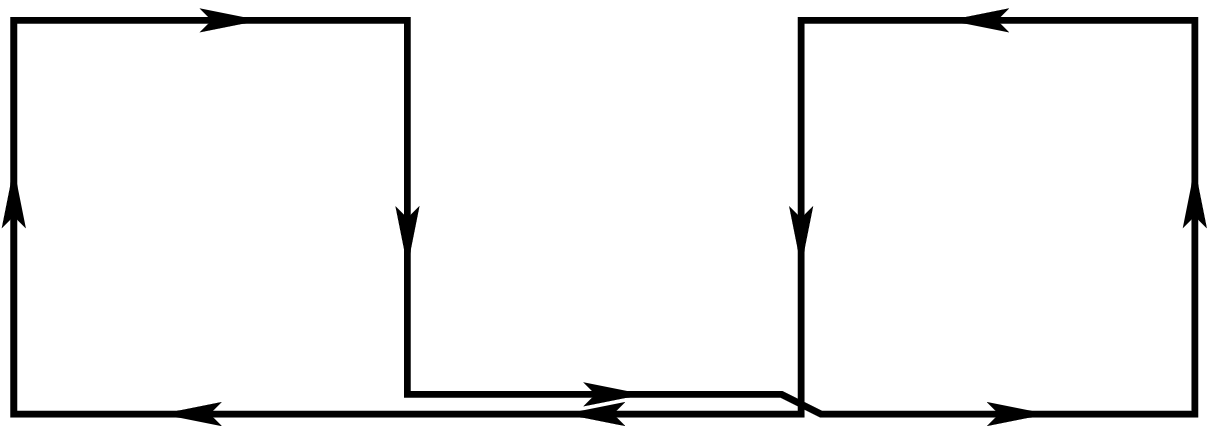}}}$ \\ \\ 
$\parbox{1.95cm}{\rotatebox{0}{\includegraphics[height=1.95cm]{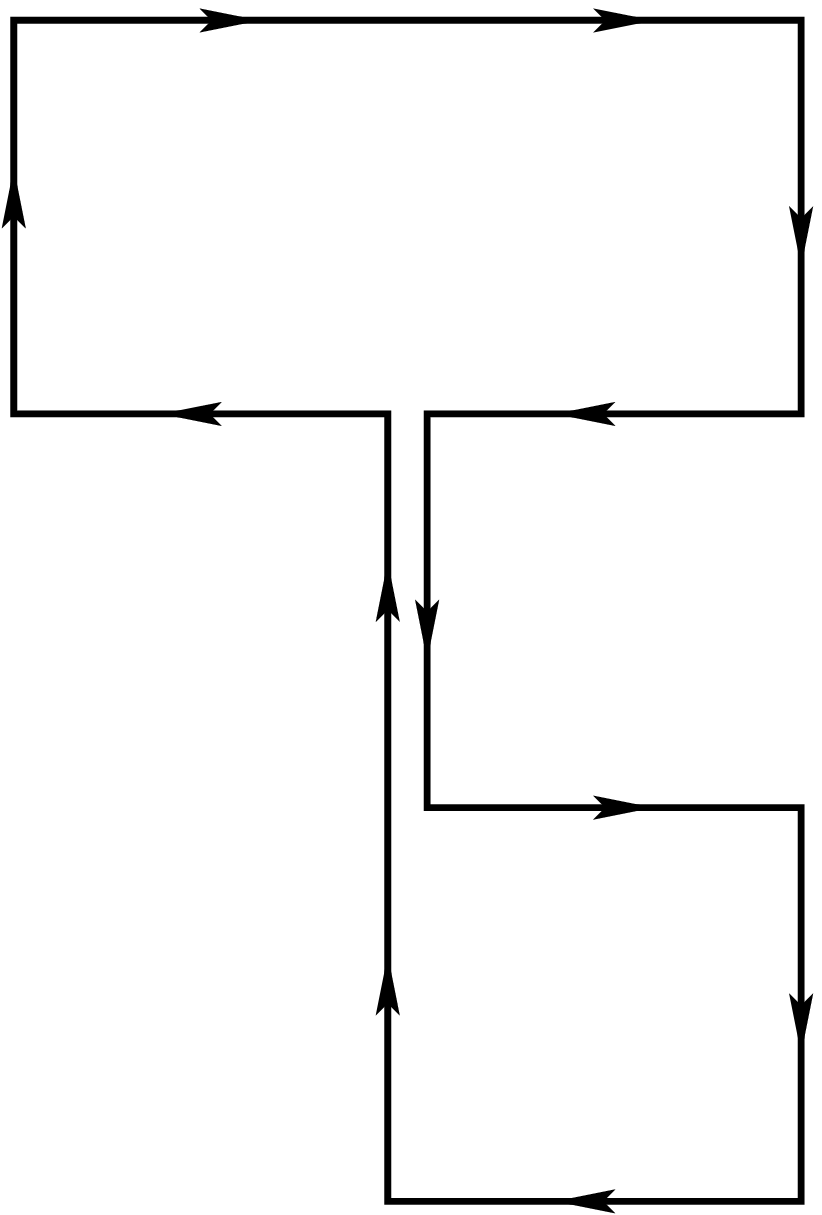}}}$ & $\parbox{1.95cm}{\rotatebox{0}{\includegraphics[height=1.95cm]{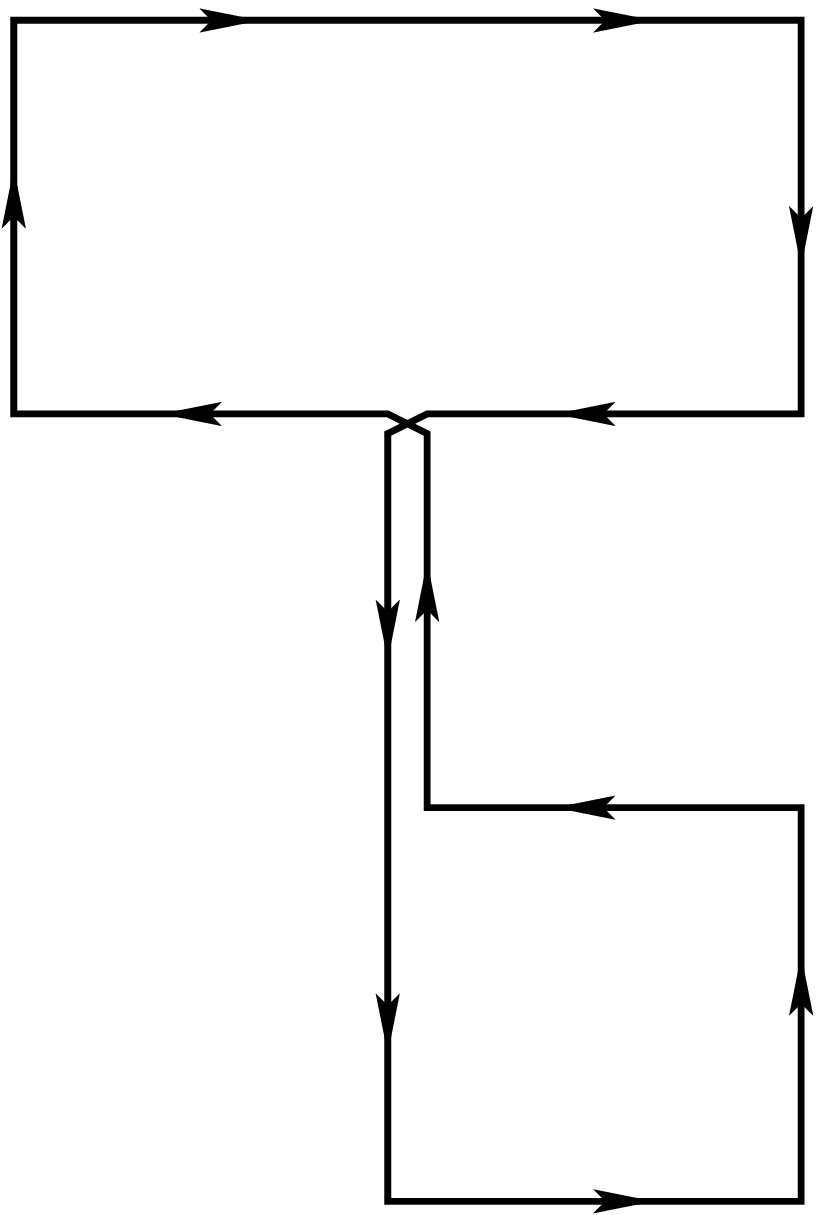}}}$ & $\parbox{1.95cm}{\rotatebox{0}{\includegraphics[height=1.95cm]{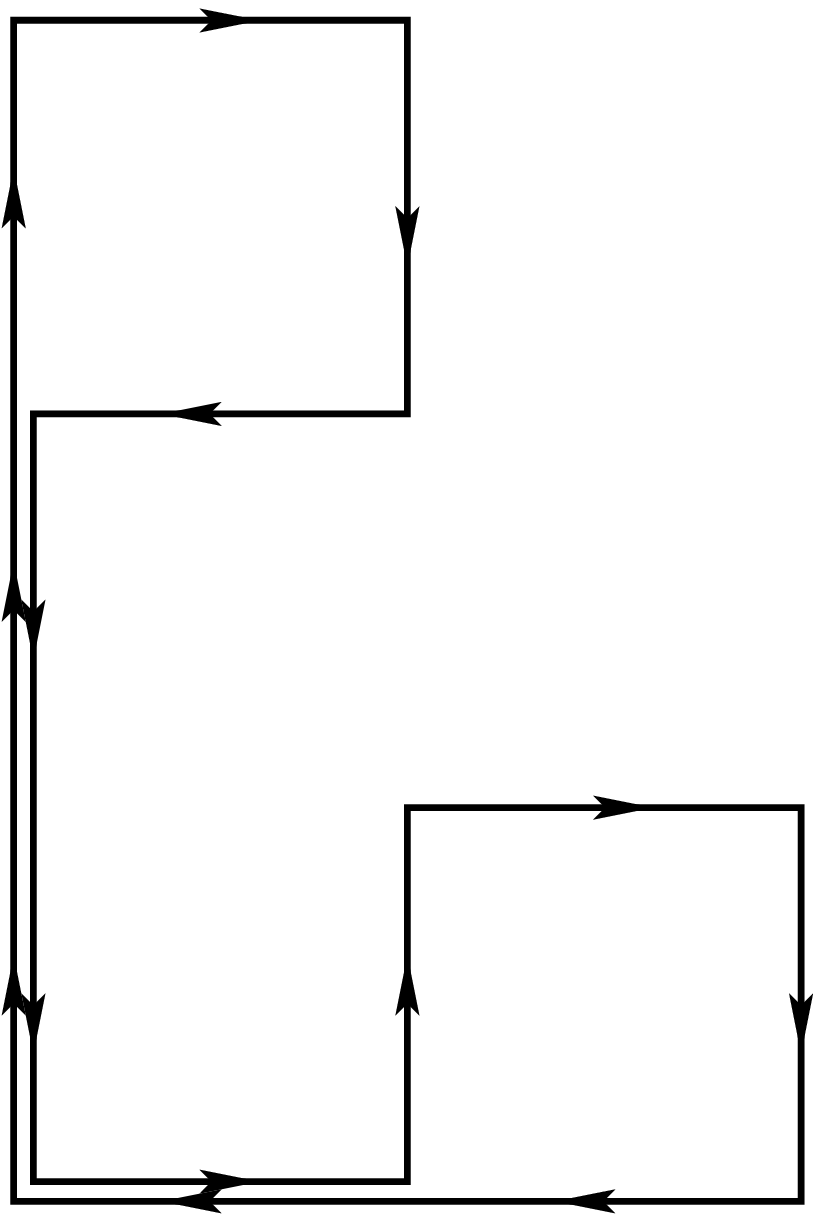}}}$  & $\parbox{1.95cm}{\rotatebox{0}{\includegraphics[height=1.95cm]{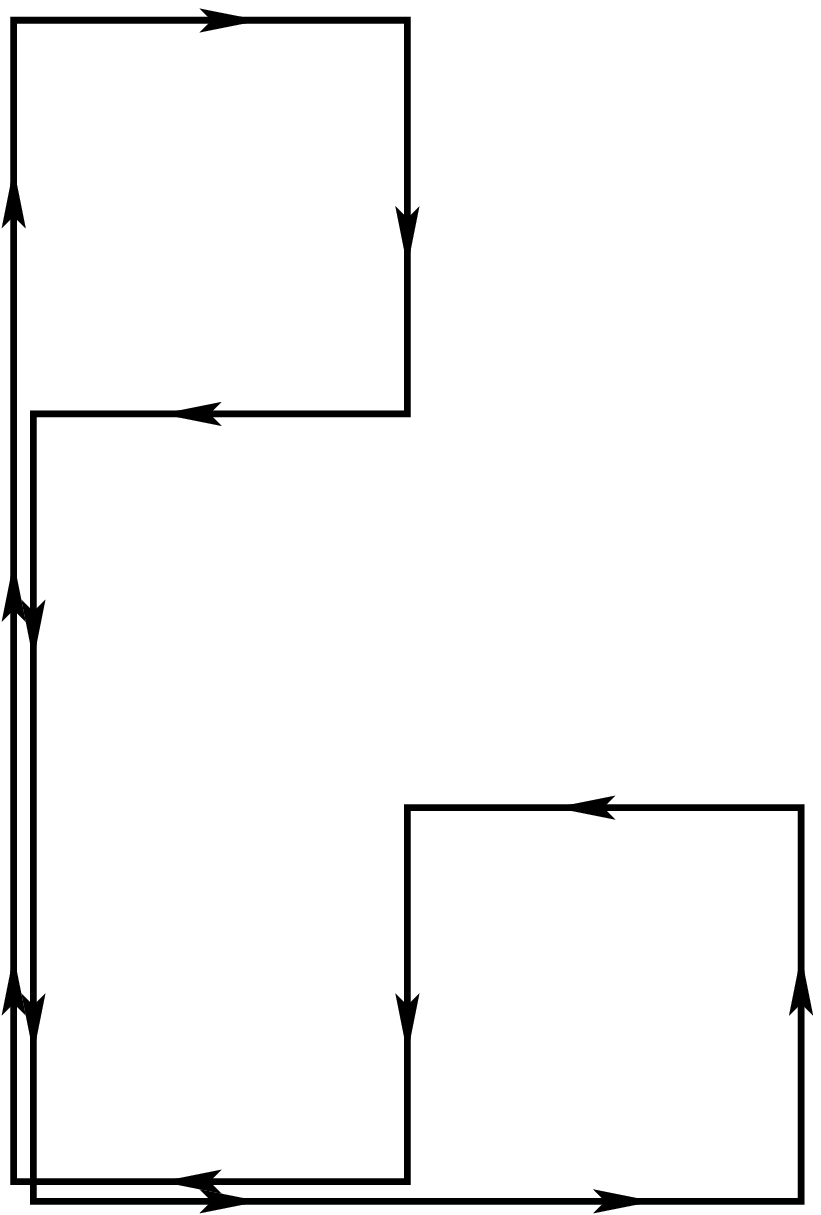}}}$  & $\parbox{1.95cm}{\rotatebox{0}{\includegraphics[height=1.95cm]{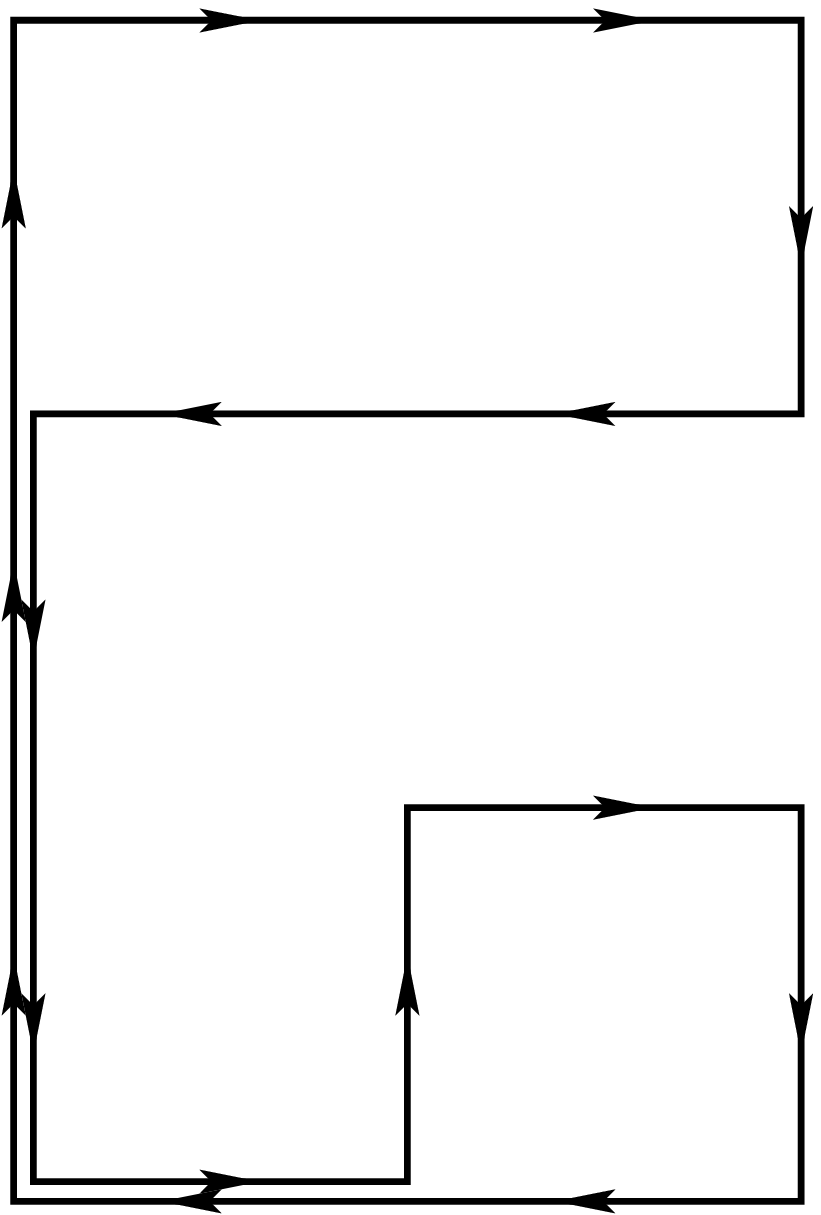}}}$  & $\parbox{1.95cm}{\rotatebox{0}{\includegraphics[height=1.95cm]{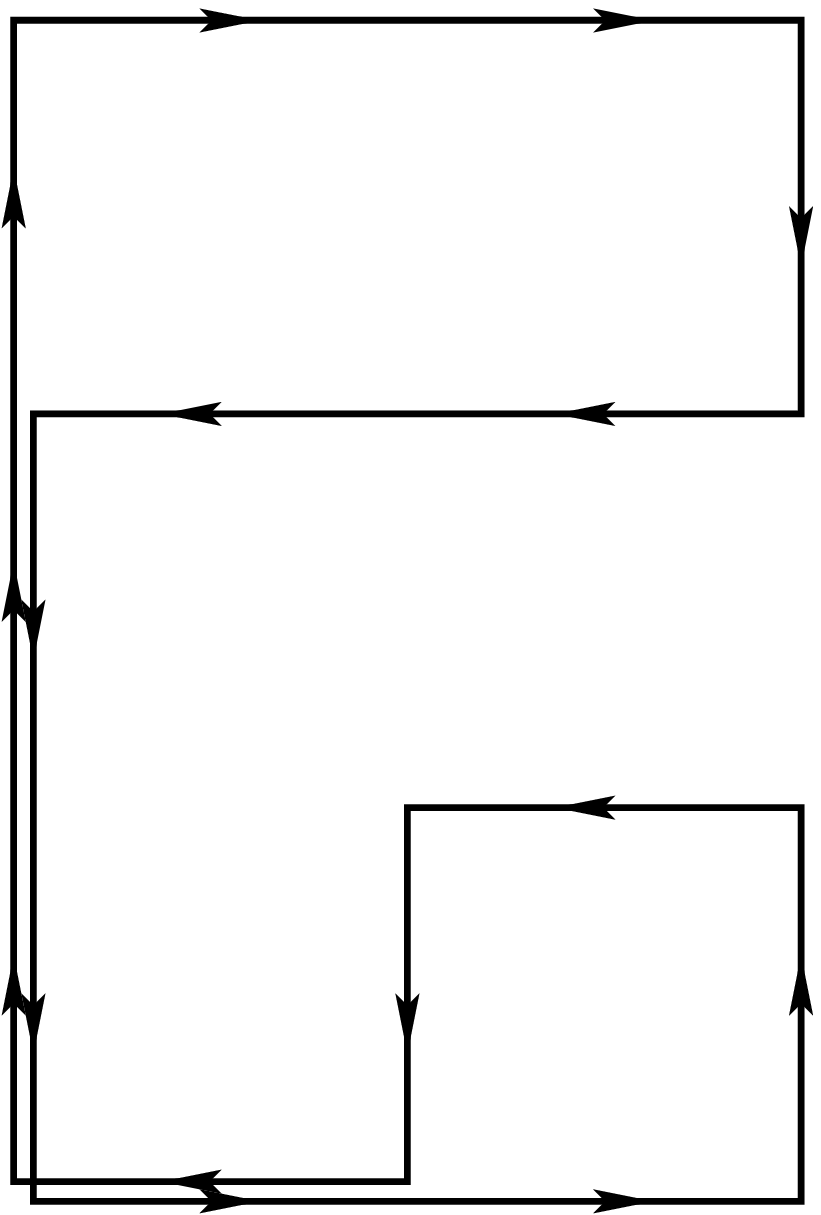}}}$ \\ \\ 
$\parbox{1.3cm}{\rotatebox{0}{\includegraphics[height=1.3cm]{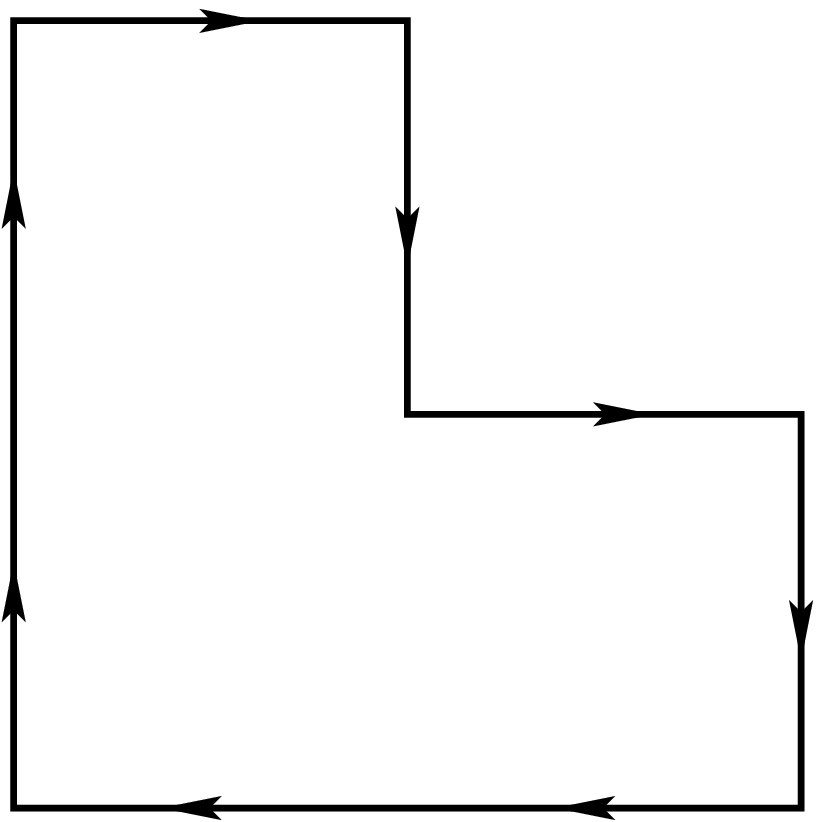}}}$ & $\parbox{1.95cm}{\rotatebox{90}{\includegraphics[height=1.3cm]{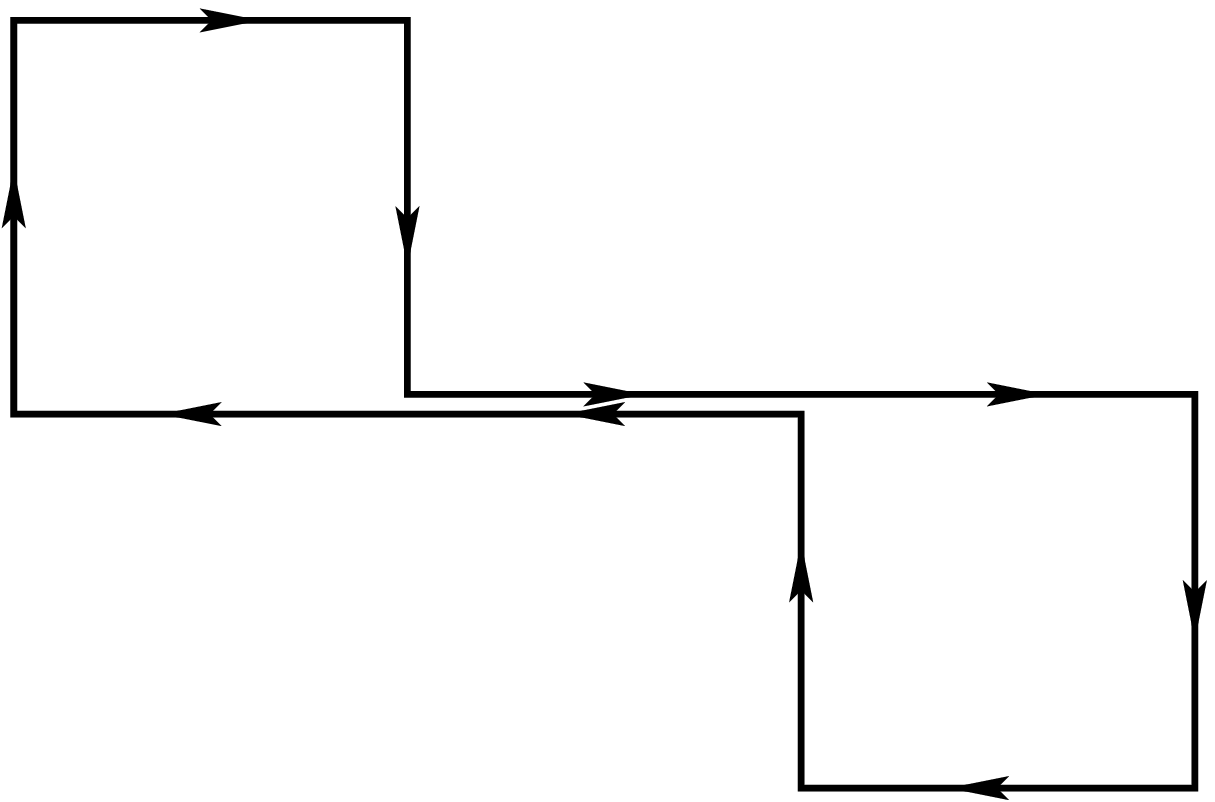}}}$ & $\parbox{1.95cm}{\rotatebox{90}{\includegraphics[height=1.3cm]{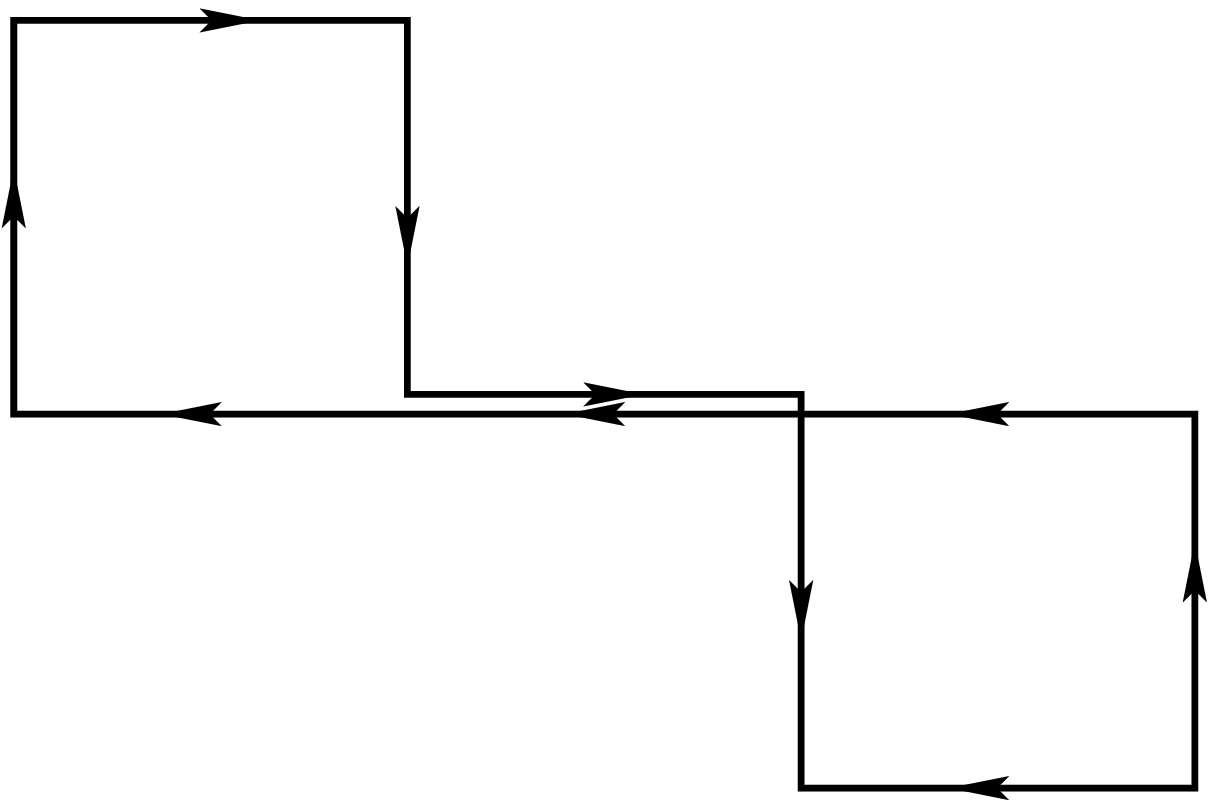}}}$  & $\parbox{1.95cm}{\rotatebox{90}{\includegraphics[height=1.3cm]{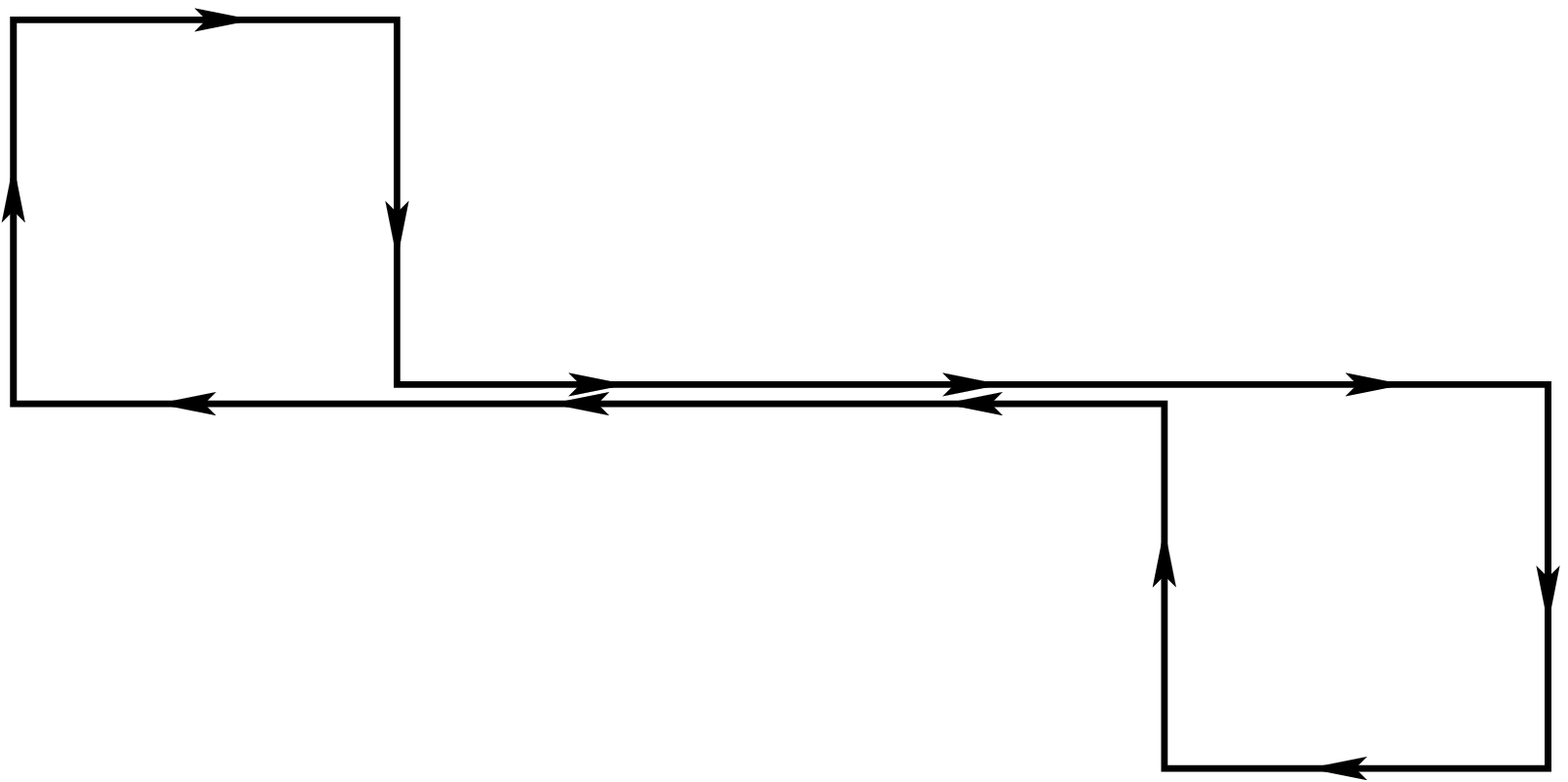}}}$  & $\parbox{1.95cm}{\rotatebox{90}{\includegraphics[height=1.3cm]{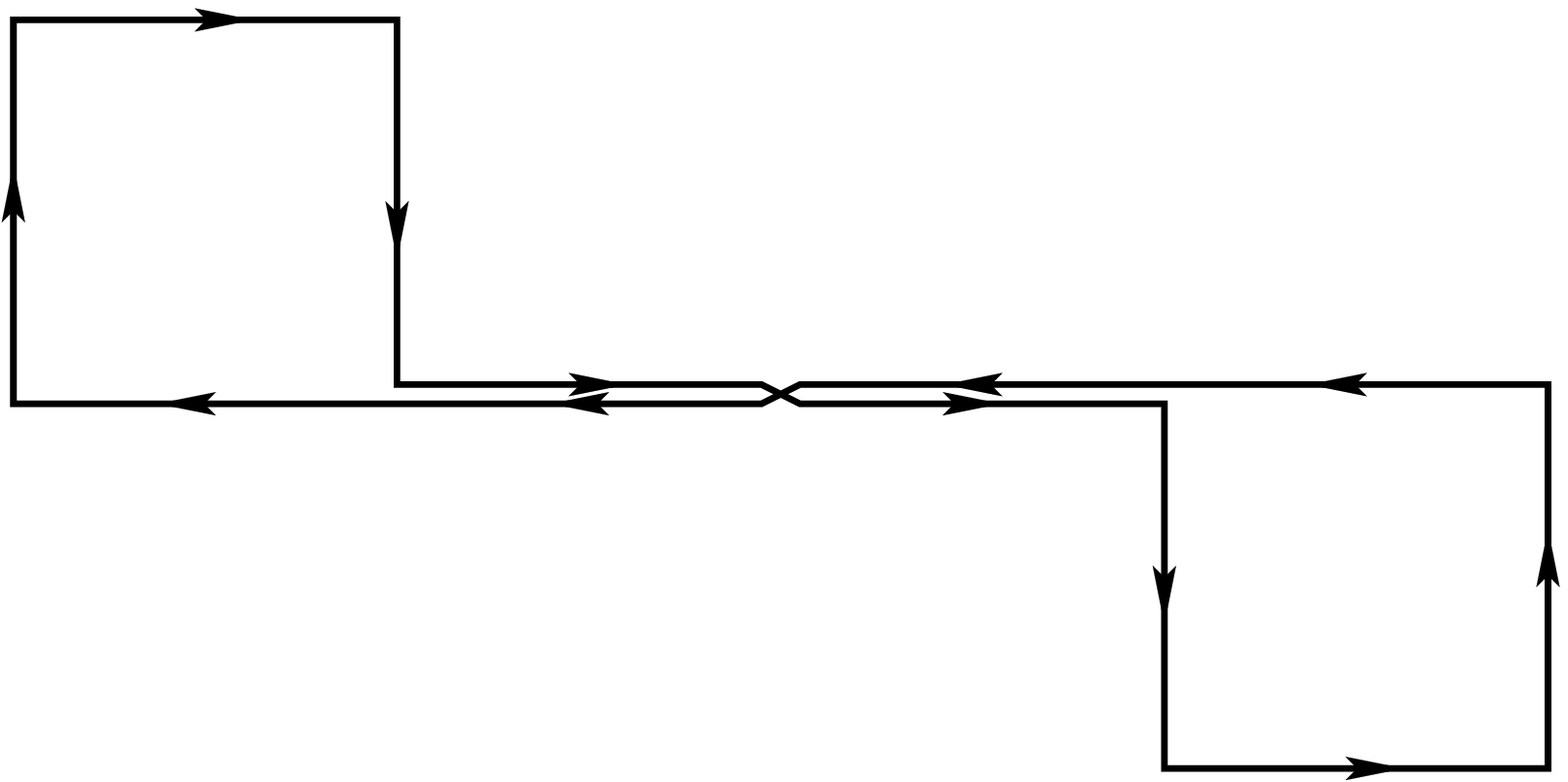}}}$  & $\parbox{1.95cm}{\rotatebox{90}{\includegraphics[height=1.95cm]{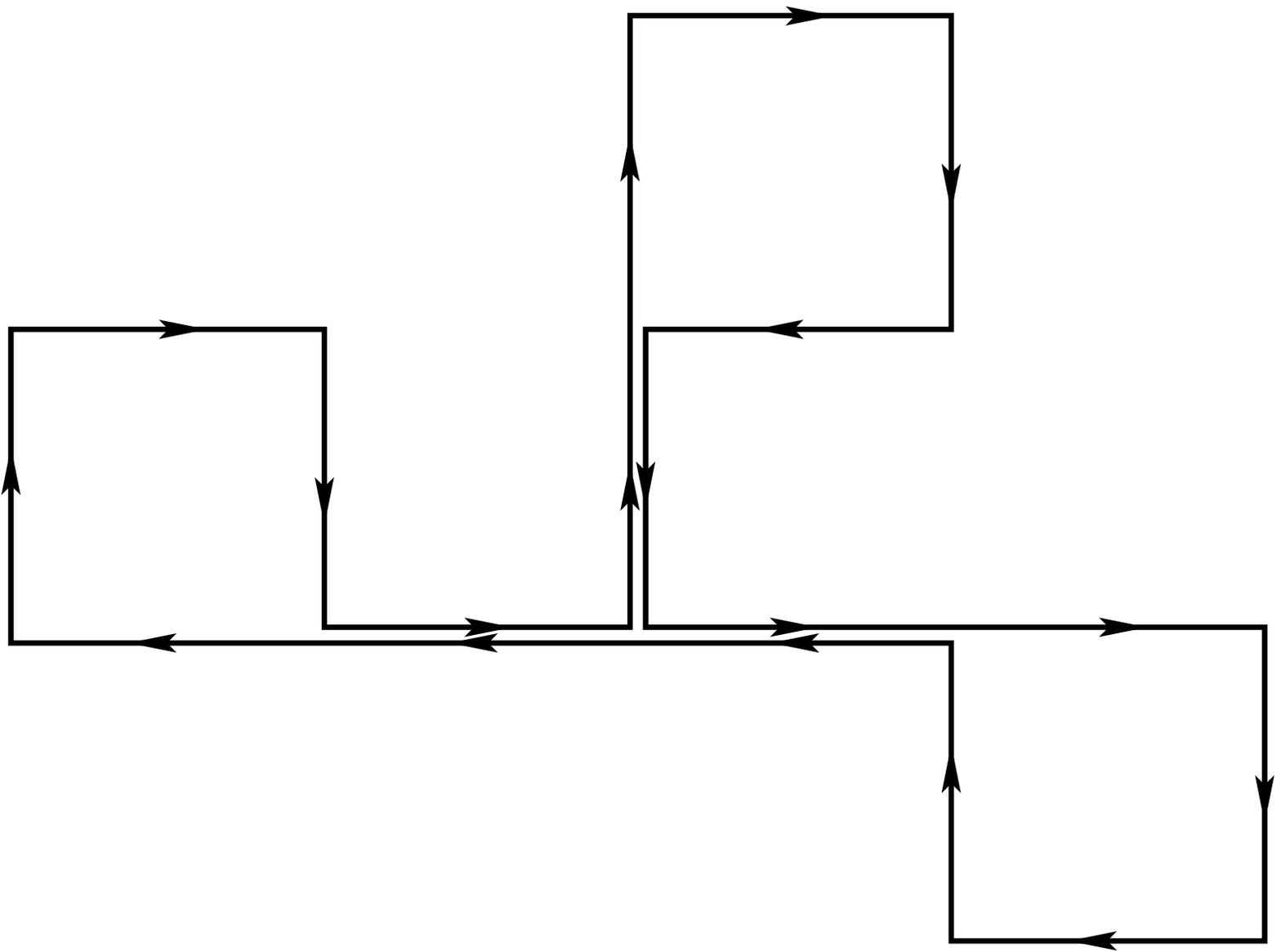}}}$ \\ \\ 
$\parbox{1.95cm}{\rotatebox{0}{\includegraphics[height=1.3cm]{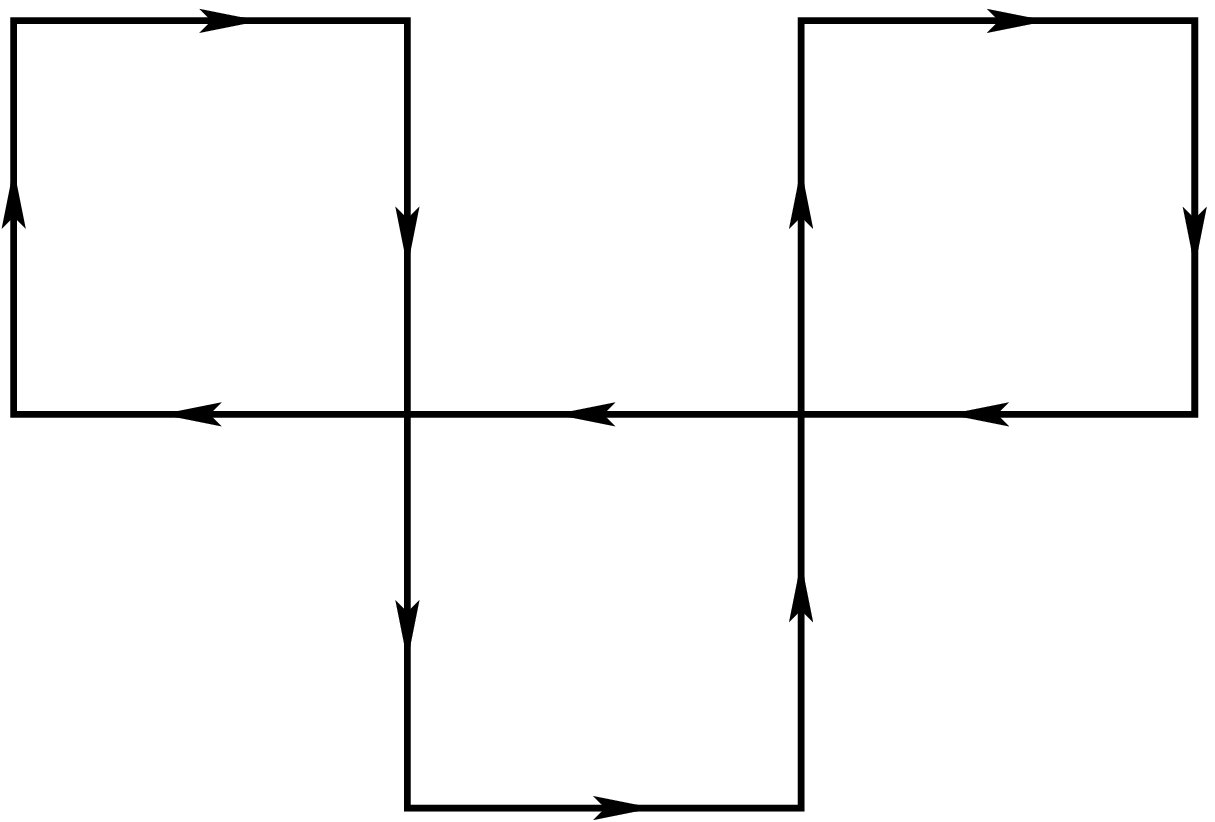}}}$ & $\parbox{1.95cm}{\rotatebox{0}{\includegraphics[height=1.3cm]{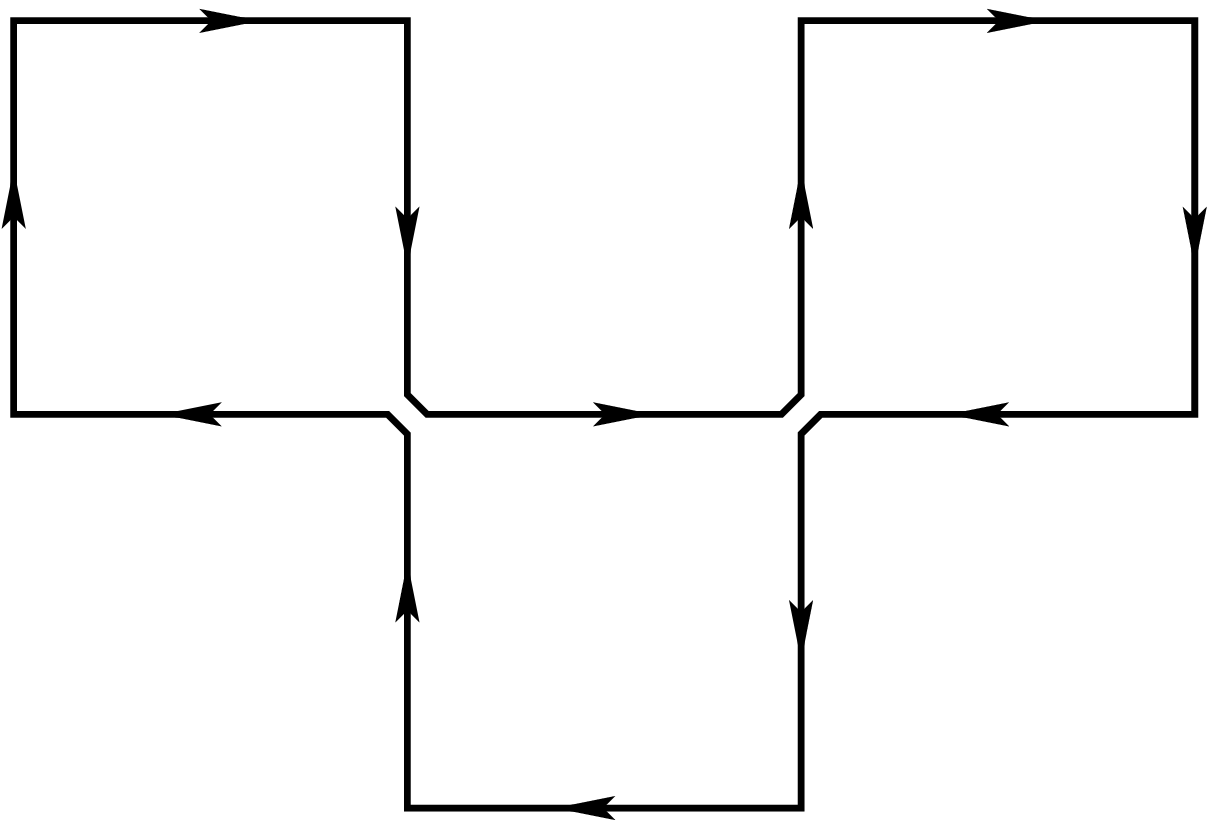}}}$ & $\parbox{1.95cm}{\rotatebox{0}{\includegraphics[height=1.3cm]{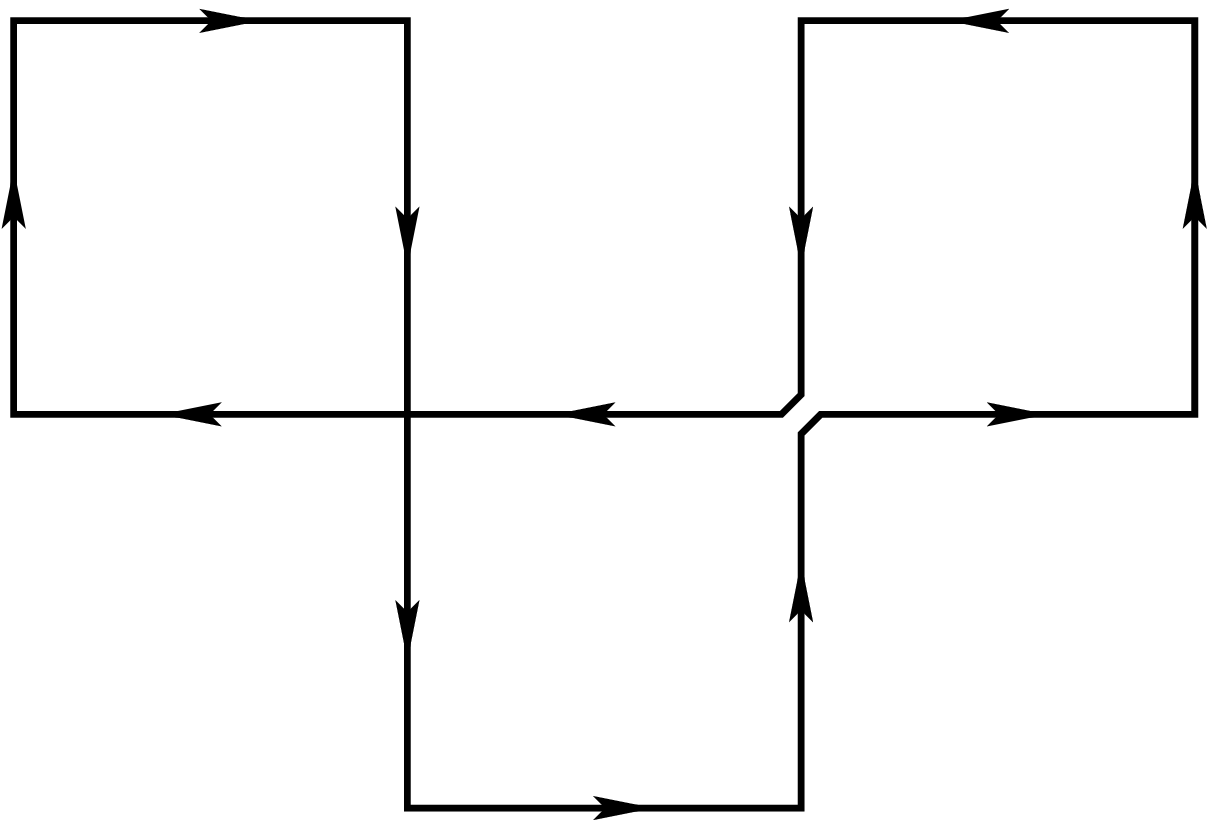}}}$  & $\parbox{1.95cm}{\rotatebox{90}{\includegraphics[height=0.65cm]{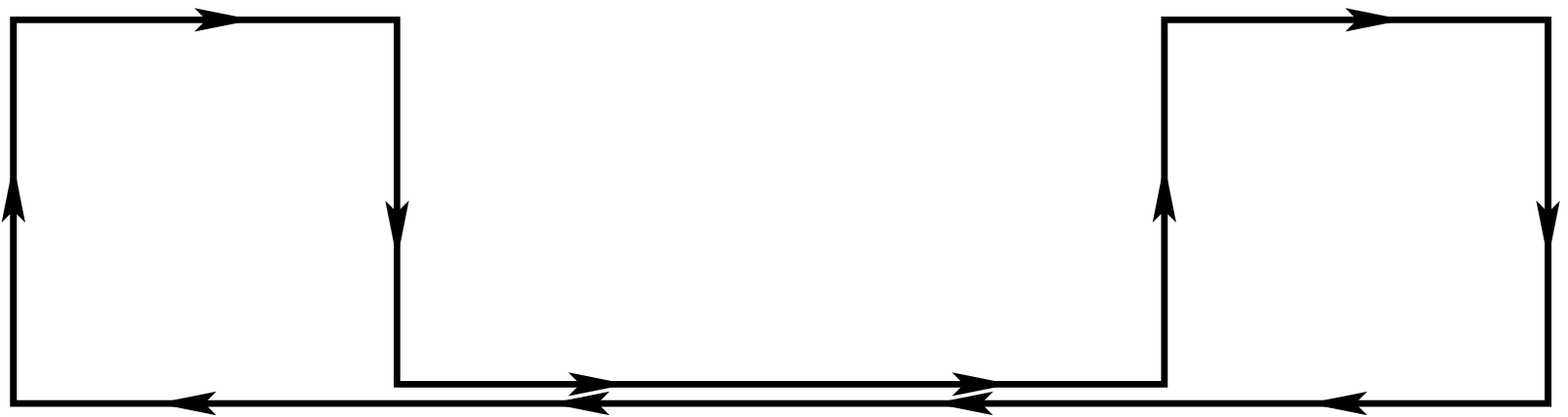}}}$ & $\parbox{1.95cm}{\rotatebox{90}{\includegraphics[height=0.65cm]{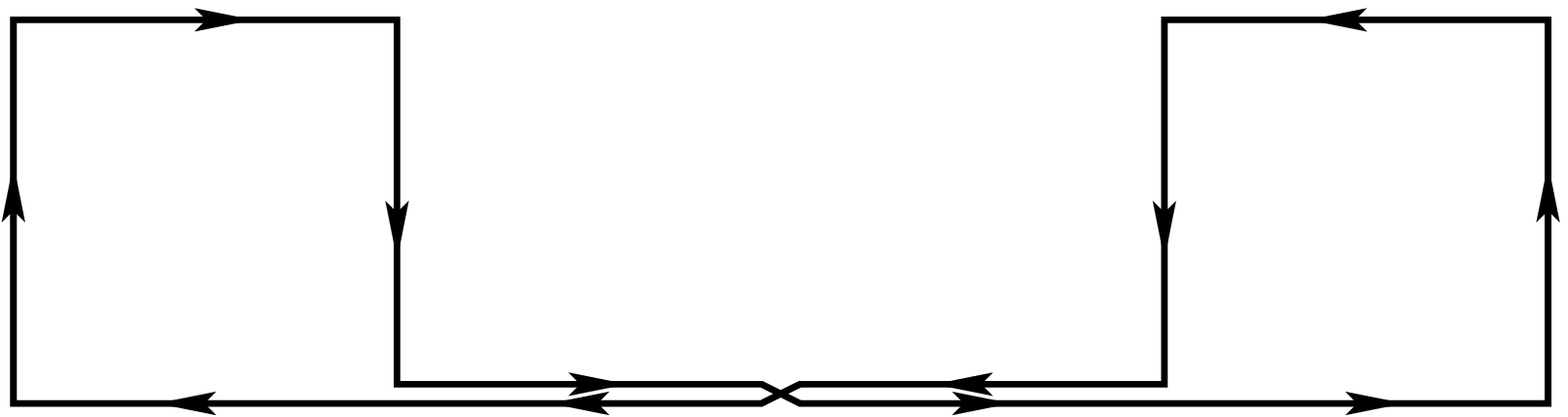}}}$ & $\parbox{1.95cm}{\rotatebox{90}{\includegraphics[height=1.95cm]{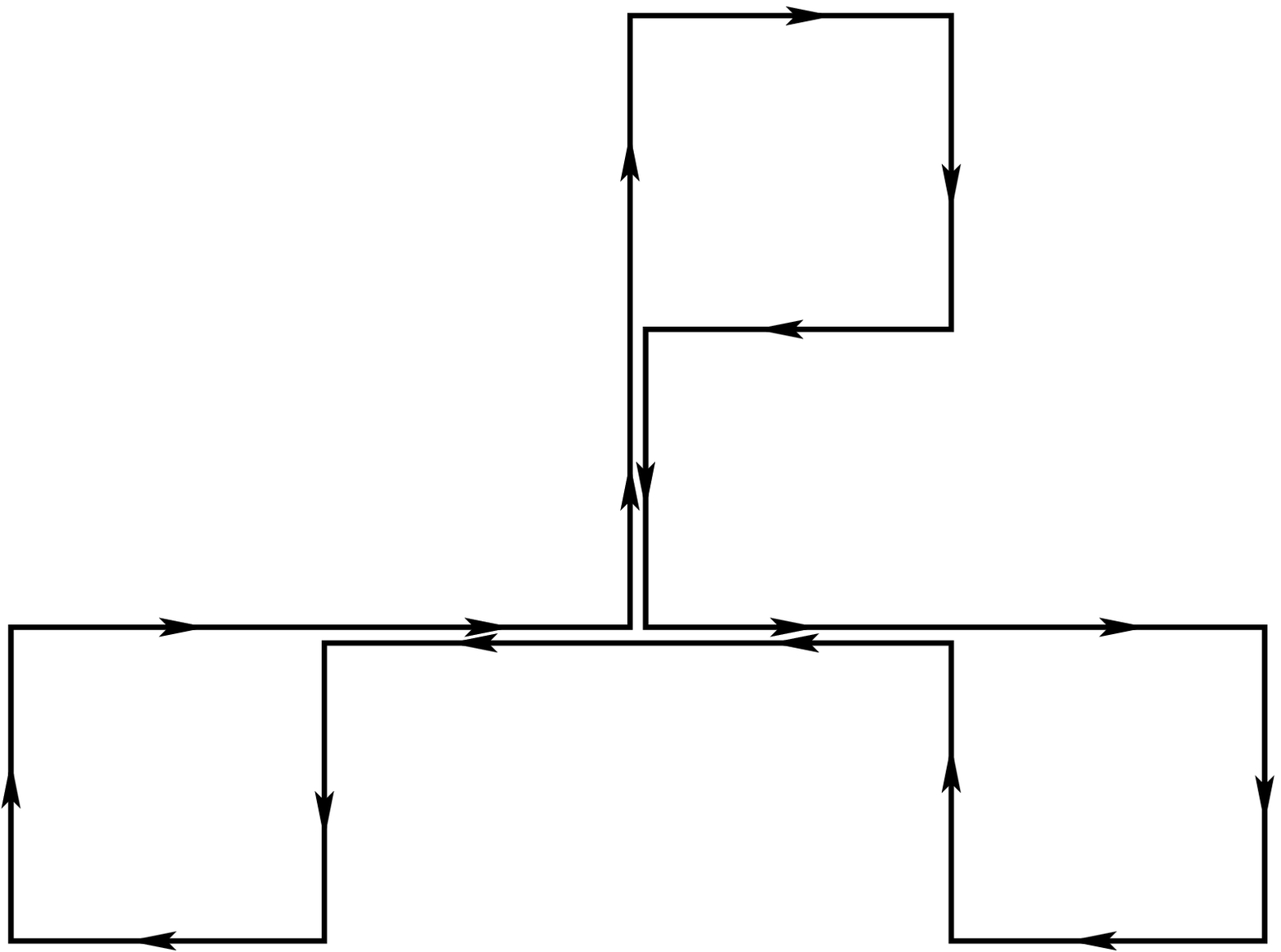}}}$ \\ \\ 
\end{tabular}
\end{center}
\caption{Examples of closed contractible paths used for the construction of glueball operators.}
\label{table_glueloops}
\end{table}

\begin{table}[h]
\begin{center}
\begin{tabular}{|cll|cll|}\hline
\multicolumn{6}{|c|}{ $aM$ : $SU(4)$ $\beta = 51.0$ } \\ \hline
$J^{PC}$  & I  & II  & $J^{PC}$  &  I   &  II  \\ \hline
$0^{++}$  & 0.5471(20)  & 0.5441(20)  & $0^{--}$  & 0.8049(23)  & 0.7996(34)  \\
        & 0.8197(29) & 0.8247(27)  &   & 0.995(12)  & 1.0129(30)  \\
        & 1.009(10)  & 1.023(4)  &   & 1.188(6)   & 1.1869(40)  \\
        & 1.027(32)  & 1.075(6)  &   & 1.262(6)   & 1.260(6)  \\
        & 1.141(16)  & 1.139(17) &   & 1.302(8)   & 1.299(8)  \\
        & 1.198(6)   & 1.182(4)  &   & 1.322(19)  & 1.337(6)  \\
        & 1.229(18)  & 1.235(8)  &   &   &   \\
        & 1.253(6)   & 1.259(7)  &   &   &   \\ \hline
$0^{-+}$  & 1.128(14)  & 1.167(15)  & $0^{+-}$  & 1.252(22) & 1.310(6)  \\ 
        & 1.317(8)  & 1.317(6)  &   & 1.487(11)  & 1.478(11)  \\
        & 1.394(37) & 1.455(8)  &   & 1.596(11)  & 1.575(11)  \\ \hline
$2^{++}$  & 0.9048(30)  & 0.9047(27)  & $2^{-+}$  & 0.9054(44)  & 0.9105(39)  \\
        & 1.092(5)  & 1.082(16)  &   & 1.071(12) & 1.091(12)  \\
        & 1.241(9)  & 1.242(7)  &   & 1.248(6)  & 1.251(7)  \\
        & 1.253(23) & 1.281(6)  &   & 1.210(22) & 1.279(7)  \\
        & 1.274(22) & 1.318(24) &   & 1.399(9)  & 1.388(9)  \\ \hline
$2^{--}$  & 1.067(9)  & 1.077(6) & $2^{+-}$  & 1.052(14)  & 1.067(12)  \\ 
        & 1.260(7)  & 1.233(6)  &   & 1.255(9)  & 1.256(6)  \\
        & 1.421(8)  & 1.364(34) &   & 1.414(12) & 1.343(32)  \\
        & 1.48(3)   & 1.42(3)  &   & 1.44(5)   &  1.38(3) \\
        & 1.541(10) & 1.497(7) &   & 1.509(8)  &  1.545(11) \\ \hline
$1^{++}$  & 1.292(7)  & 1.276(7)  & $1^{-+}$  & 1.290(8)  & 1.240(32)  \\ 
        & 1.300(8)  & 1.305(8)  &   & 1.297(7)  & 1.311(9)  \\
        & 1.471(8)  & 1.404(41) &   & 1.468(8)  & 1.447(10)  \\
        & 1.475(6)  & 1.458(10) &   & 1.490(8)  & 1.473(13)  \\
        & 1.461(25) & 1.426(34) &   & 1.497(8)  & 1.505(12)  \\ \hline
$1^{--}$  & 1.243(6)  & 1.247(7)  & $1^{+-}$  & 1.241(7)  & 1.242(5)  \\ 
        & 1.292(7)  & 1.280(8)  &   & 1.232(22) & 1.276(7)  \\
        & 1.338(26) & 1.40(1)  &   & 1.403(8)  & 1.421(8)  \\
        & 1.455(7)  & 1.411(9) &   & 1.455(7)  & 1.403(9)  \\
        & 1.463(10) & 1.446(9) &   & 1.462(7)  & 1.453(8)  \\ \hline
\end{tabular}
\caption{Glueball masses in SU(4) for two different operator bases, I and II, on a $40^248$ lattice at $\beta=51.0$.}
\label{table_msu4_opcomp}
\end{center}
\end{table}

\begin{table}[h]
\begin{center}
\begin{tabular}{|ccc|ccc|}\hline
\multicolumn{6}{|c|}{  $aM$ : $SU(6)$  $\beta\sim 206$ } \\ \hline
$J^{PC}$  & I  & II  & $J^{PC}$  &  I   &  II  \\ \hline
$0^{++}$  & 0.2922(7)   & 0.2948(10) &  $0^{--}$ & 0.4326(11) & 0.4331(10)   \\
     & 0.4467(9)   & 0.4466(17) &   & 0.5481(14) &  0.5493(14)  \\
     & 0.5554(19)  & 0.5502(21) &   & 0.6453(39) &  0.6496(19)  \\
     & 0.5793(14)  & 0.5795(26) &   & 0.6826(41) &  0.6861(18)  \\
     & 0.6289(38)  & 0.6331(27) &   & 0.7154(19) &  0.7077(45)  \\
     & 0.6426(28)  & 0.6337(33) &   & 0.7452(43) &  0.7111(64)  \\
     & 0.6433(61)  & 0.652(13)  &   & 0.7808(50) &  0.7631(51)  \\ \hline
$0^{-+}$  & 0.6346(34) & 0.6296(50) &  $0^{+-}$ & 0.7121(36) & 0.7202(43)   \\
         & 0.7112(36) & 0.7109(66) &     & 0.8100(47) & 0.7949(45)  \\
         & 0.8107(52) & 0.789(9)   &     & 0.8557(58) & 0.8611(52)  \\
         & 0.804(13)  & 0.812(14)  &     & 0.8821(79) & 0.8847(77)  \\ \hline
$2^{++}$  & 0.4914(16)  & 0.4932(16)  & $2^{-+}$  & 0.4926(14)  & 0.4933(19)  \\
    & 0.5970(16)  & 0.5992(16)  &   & 0.5964(18)  &  0.5980(32) \\
    & 0.6846(23)  & 0.6752(13)  &   & 0.6766(31)  &  0.6707(74) \\
    & 0.6882(82)  & 0.7029(48)  &   & 0.7013(34)  &  0.6963(44) \\
    & 0.7728(49)  & 0.7237(86)  &   & 0.7801(46)  &  0.7661(52) \\ \hline
$2^{--}$  & 0.5773(42)  & 0.5871(17)  & $2^{+-}$  & 0.5830(26)  & 0.5869(34)  \\ 
    & 0.6766(37)  & 0.6787(21)  &   & 0.6861(26)  & 0.6793(34)  \\
    & 0.7622(47)  & 0.7477(61)  &   & 0.7782(19)  & 0.7681(46)  \\
    & 0.8276(28)  & 0.7942(51)  &   & 0.7880(49)  & 0.7759(96)  \\
    & 0.8361(55)  & 0.781(12)   &   & 0.8438(63)  & 0.8573(26)  \\ \hline
$1^{++}$  & 0.7074(42)  & 0.7077(33)  & $1^{-+}$  & 0.7086(32)  & 0.6924(69)  \\ 
    & 0.7068(39)  & 0.7082(33)  &   & 0.7124(35)  & 0.7119(86)  \\
    & 0.7783(98)  & 0.7794(54)  &   & 0.7771(75)  & 0.7884(27)  \\
    & 0.7739(73)  & 0.7818(65)  &   & 0.7806(96)  & 0.776(12)  \\ \hline
$1^{--}$  & 0.6685(40)  & 0.6716(41)  & $1^{+-}$  & 0.6672(29)  & 0.6695(36)  \\ 
    & 0.6896(44)  & 0.6966(36)  &   & 0.6989(46)  & 0.6959(43)  \\
    & 0.7637(46)  & 0.7563(47)  &   & 0.7603(34)  & 0.7645(47)  \\
    & 0.7771(47)  & 0.7725(40)  &   & 0.7839(44)  & 0.7832(24)  \\ \hline
\end{tabular}
\caption{Glueball masses in SU(6) for two different operator bases: I on a $60^3$ lattice at $\beta=206.84$
  and II on a $54^260$ lattice at $\beta=206.0$.}
\label{table_msu6_opcomp}
\end{center}
\end{table}

\begin{table}[htb]
\centering
\begin{tabular}{|c|c|c|} \hline
$N$ & $l\surd\sigma_f\approx$ & $2\sigma l/m_{0^{++}} \approx$ \\ \hline
2  & 5.5  & 2.4  \\
3  & 5.2  & 2.4  \\
4  & 5.1  & 2.4  \\
6  & 4.2  & 2.0  \\
8  & 3.8  & 1.8  \\
12  & 3.6  & 1.7  \\
16  & 3.1  & 1.5  \\ \hline
\end{tabular}
\caption{Lattice sizes used in our $SU(N)$ calculations, expressed in physical units}
\label{table_sizeN}
\end{table}

\begin{table}[h]
\begin{center}
\begin{tabular}{|cll|cll|}\hline
\multicolumn{6}{|c|}{  $aM$ : $SU(8)$  $\beta = 306.25$ } \\ \hline
$J^{PC}$  & $60^2 48$  &  $44^2 48$   & $J^{PC}$  &   $60^2 48$    &  $44^2 48$   \\ \hline
 $0^{++}$  & 0.3569(13) & 0.3552(13) &  $0^{--}$ & 0.5248(14) & 0.5227(13)  \\
          & 0.5428(18) & 0.5436(16) &  & 0.6630(44) & 0.6669(32)  \\
          & 0.6747(20) & 0.6745(20) &  & 0.7848(31) & 0.785(3)  \\
          & 0.6953(42) & 0.697(5)   &  & 0.822(5) & 0.835(3)  \\
          & 0.7652(59) & 0.762(6)   &  & 0.860(7) & 0.864(7)  \\
          & 0.7801(54) & 0.7805(26) &  &  &   \\
          & 0.789(11)  & 0.8172(22) &  &  &   \\ \hline
 $0^{-+}$  & 0.7678(47) & 0.7705(43) &  $0^{+-}$ & 0.876(8) & 0.885(3)  \\
           & 0.882(9)  & 0.877(7)   &  & 0.974(5) &  0.942(9) \\ \hline
 $2^{++}$  & 0.5965(22) & 0.5972(46) &  $2^{-+}$ & 0.5973(22)  & 0.5934(32)  \\
          & 0.7276(23) & 0.7218(41) &  & 0.7177(51) & 0.7233(31)  \\
          & 0.8204(59) & 0.8318(69) &  & 0.8303(54) & 0.806(13)  \\
          & 0.8542(23) & 0.8506(76) &  & 0.8543(29) & 0.838(6)  \\
          & 0.9188(85) & 0.901(18)  &  & 0.928(10)  & 0.920(7)  \\ \hline
 $2^{--}$  & 0.7022(50) & 0.7055(43) &  $2^{+-}$ & 0.7110(41) & 0.7087(49)  \\
          & 0.8221(26) & 0.8253(30) &  & 0.8289(27) & 0.8282(23)  \\
          & 0.9281(39) & 0.9237(33) &  & 0.9097(83) & 0.903(14)  \\
          & 0.9761(91) & 0.939(27)  &  & 0.959(10)  & 0.942(26)  \\ \hline
 $1^{++}$  & 0.8516(65) & 0.8504(58)  &  $1^{-+}$ & 0.8540(62) & 0.8550(62)  \\
          & 0.8719(65) & 0.8680(57) &  & 0.8639(59) & 0.8646(65) \\
          & 0.934(9)  & 0.941(8) &  & 0.954(10)   & 0.959(8)  \\ 
          & 0.963(10) & 0.957(8) &  & 0.952(9)   & 0.952(9)  \\ \hline
 $1^{--}$  & 0.798(14) & 0.8131(54) &  $1^{+-}$ & 0.8185(21) & 0.800(12)  \\
          & 0.850(6) & 0.8414(56) &  & 0.840(6) &  0.8500(30) \\
          & 0.896(25) & 0.884(17)  &  & 0.878(20) & 0.898(21)  \\
          & 0.945(10) & 0.9521(32)  &  & 0.944(4) & 0.9539(34)  \\ \hline
\end{tabular}
\caption{Glueball masses in SU(8) at $\beta=306.25$ on two different volumes.}
\label{table_msu8_Vcomp}
\end{center}
\end{table}

\begin{table}[h]
\begin{center}
\begin{tabular}{|cll|cll|}\hline
\multicolumn{6}{|c|}{  $aM$ : $SU(16)$  $\beta = 800$ } \\ \hline
$J^{PC}$  & $26^2 30$  &  $22^2 30$   & $J^{PC}$  &   $26^2 30$    &  $22^2 30$   \\ \hline
 $0^{++}$  & 0.5640(29)  & 0.5609(22) &  $0^{--}$ & 0.8270(45)  & 0.8311(35)  \\
          & 0.8643(56)  & 0.8532(38) &       & 1.057(8)  &  1.027(16) \\
          & 1.056(7)  & 1.060(6)  &          & 1.227(9)  &  1.237(8)  \\
          & 1.118(7)  & 1.103(8)  &          & 1.310(11) &  1.291(10) \\
          & 1.195(10) & 1.204(10) &          & 1.346(14) &  1.375(13) \\
          & 1.235(11) & 1.236(7)  &          & 1.398(17) &  1.409(9)  \\ \hline
 $0^{-+}$  & 1.232(12) & 1.211(10) &  $0^{+-}$ & 1.290(64)  & 1.325(46)  \\
          & 1.367(15) & 1.367(10) &  & 1.48(9)   &  1.36(5) \\
          & 1.556(22) & 1.540(16) &  & 1.647(21) &  1.638(16) \\ \hline
 $2^{++}$  & 0.9380(58) & 0.9357(39)  &  $2^{-+}$ & 0.9404(66) & 0.9494(43)  \\
          & 1.157(8)  & 1.149(7) &  & 1.152(10) & 1.150(6)  \\
          & 1.297(10) & 1.294(8) &  & 1.313(11) & 1.297(9)  \\
          & 1.343(17) & 1.353(8) &  & 1.339(16) & 1.281(33)  \\
          & 1.444(23) & 1.439(11) &  & 1.385(71) & 1.479(13)  \\ \hline
 $2^{--}$  & 1.119(9) & 1.111(5) &  $2^{+-}$ & 1.127(7) & 1.122(6)  \\
          & 1.299(17) & 1.298(8) &  & 1.301(11) & 1.310(9)  \\
          & 1.435(15) & 1.464(11) &  & 1.35(8)  & 1.466(12)  \\
          & 1.557(18) & 1.517(12) &  & 1.501(16) & 1.522(12)  \\ \hline
 $1^{++}$  & 1.353(15) & 1.344(10) &  $1^{-+}$ & 1.331(13) & 1.323(10)  \\
          & 1.393(15) & 1.400(9) &  & 1.370(10) & 1.366(12)  \\
          & 1.499(17) & 1.494(11) &  & 1.41(7) & 1.514(11)  \\ 
          & 1.526(16) & 1.518(11) &  & 1.45(9) & 1.504(13)  \\ \hline
 $1^{--}$  & 1.271(12) & 1.282(9) &  $1^{+-}$ & 1.270(11)  & 1.287(8)  \\
          & 1.317(19) & 1.335(11) &  & 1.330(14) & 1.330(9)  \\
          & 1.490(17) & 1.443(12) &  & 1.488(11) & 1.475(10)  \\
          & 1.478(20) & 1.485(14) &  & 1.475(15) & 1.449(56)  \\ \hline
\end{tabular}
\caption{Glueball masses in SU(16) at $\beta=800$ on two different volumes.}
\label{table_msu16_Vcomp}
\end{center}
\end{table}

\begin{table}[htb]
\centering
\begin{tabular}{|cl|cll|cl|} \hline
\multicolumn{7}{|c|}{$SU(2)$} \\ \hline
$\beta$ & $\tfrac{1}{N}\text{Tr}\langle U_p\rangle$ & 
$L_s^2L_t$ & $a\surd\sigma_f$ & $am_G$  & $L_s^2L_t$ & $a\surd\sigma_f$ \\ \hline
30.0  &  0.96639434(4) & $120^290$ & 0.04579(7)  & 0.2167(5) & $84^2100$  & 0.04573(6)  \\ 
26.5  &  0.96191273(4) & $104^280$ & 0.05194(7)  & 0.2459(6) & $72^290$   & 0.05186(7)  \\ 
23.5  &  0.95699656(4) & $96^264$  & 0.05869(8)  & 0.2775(6) & $68^280$   & 0.05869(7)  \\ 
20.0  &  0.94937113(8) & $80^264$  & 0.06929(7)  & 0.3272(10) & $56^272$   & 0.06926(7)  \\ 
16.0  &  0.93649723(10) & $68^248$ & 0.08760(11) & 0.4112(12) & $44^254$   & 0.08743(11)  \\ 
12.0  &  0.91482126(21) & $50^240$ & 0.11847(14) & 0.5594(14) & $34^240$   & 0.11832(17)  \\ 
 9.0  &  0.88544949(27) & $36^232$ & 0.16075(16) & 0.7624(15) & $22^232$   & 0.16082(16)  \\ 
 7.0  &  0.85112757(60) & $28^224$ & 0.21217(45) & 1.0012(22) & $18^232$   & 0.21214(25)  \\ 
 6.0  &  0.82477909(68) & $25^220$ & 0.25098(50) & 1.1908(26) & $18^220$   & 0.25260(25)  \\ 
 5.0  &  0.7868676(11) & $20^220$  & 0.3134(11)  & 1.4689(60) & $14^220$   & 0.3113(8)   \\ 
 4.5  &  0.7608386(13) & $18^220$  & 0.3557(17)  & 1.665(6)  & $14^220$   & 0.3524(13)   \\  \hline
\end{tabular}
\caption{Parameters, string tension and mass gap  of the $SU(2)$ calculations.} 
\label{table_param_su2}
\end{table}

\begin{table}[htb]
\centering
\begin{tabular}{|c|cll|cl|} \hline
\multicolumn{6}{|c|}{$SU(3)$} \\ \hline
$\beta$ & lattice & $\tfrac{1}{N}\text{Tr}\langle U_p\rangle$ & 
 $a\surd\sigma_f$ & lattice & $am_G$\\ \hline
49.450 & $52\times 74^2$  & 0.9452122(3) & 0.069694(69)& $74^3$ & 0.3046(11) \\
40.400 & $44\times 62^2$  & 0.9326799(3) & 0.086106(92)& $62^3$ & 0.3759(6) \\
33.130 & $34\times 48^2$  & 0.917513(2) & 0.10611(10)& $48^3$  & 0.4625(15) \\
26.788 & $27\times 40^2$  & 0.897308(2) & 0.13326(15)& $38^240$ & 0.5799(18) \\
19.154 & $18\times 40^2$  & 0.854215(2) & 0.19292(22)& $26^240$ & 0.8337(33) \\
13.621 & $13\times 36^2$  & 0.789856(2) & 0.28954(38)& $18^236$ & 1.236(6) \\ \hline
\end{tabular}
\caption{Parameters, string tension and mass gap of the $SU(3)$ calculations}
\label{table_param_su3}
\end{table}

\begin{table}[htb]
\centering
\begin{tabular}{|cl|cll|cl|} \hline
\multicolumn{7}{|c|}{$SU(4)$} \\ \hline
$\beta$ & $\tfrac{1}{N}\text{Tr}\langle U_p\rangle$ & 
$L_s^2L_t$ & $a\surd\sigma_f$ & $am_G$ & $L_s^2L_t$ & $a\surd\sigma_f$ \\ \hline
86.0  & 0.94081019(5)  & $70^280$ & 0.07378(8)   & 0.3115(11) & $50^280$ & 0.073711(46) \\
74.0  & 0.93099416(7)  & $60^268$ & 0.08637(10)  & 0.3650(12) & $44^268$ & 0.08634(8)  \\
63.0  & 0.91861509(13) & $50^256$ & 0.10252(11)  & 0.4350(15) & $36^256$ & 0.10228(10) \\
51.0  & 0.89878825(22) & $40^248$ & 0.12838(17)  & 0.5441(20) & $30^248$ & 0.12814(15) \\
40.0  & 0.86961166(24) & $30^236$ & 0.16758(20)  & 0.7096(18) & $22^232$ & 0.16745(16) \\
28.0  & 0.8093380(7)   & $20^224$ & 0.25205(33)  & 1.059(4)  & $14^224$ & 0.25147(26) \\ \hline
\end{tabular}
\caption{Parameters, string tension and mass gap of the $SU(4)$ calculations.}
\label{table_param_su4}
\end{table}

\begin{table}[htb]
\centering
\begin{tabular}{|c|cll|cl|} \hline
\multicolumn{6}{|c|}{$SU(6)$} \\ \hline
$\beta$ & lattice & $\tfrac{1}{N}\text{Tr}\langle U_p\rangle$ & $a\surd\sigma_f$ &
lattice & $am_G$ \\ \hline 
206.84  & $52.56^2$ & 0.9425827(1) & 0.07031(4)  &   $60^3$ & 0.2922(7)  \\
206.0   & $54^260$ & 0.9423440(1) & 0.07060(7)  & $54^260$ & 0.2948(10)  \\
171.0   &  ---      & 0.9302656(2) & 0.085827(21) & $50^3$ & 0.3583(12)  \\
139.870 & $34.48^2$ & 0.9142825(5) & 0.10630(12) & $40^3$  & 0.4434(18)  \\
113.176 & $27.40^2$ & 0.893280(2)  & 0.13360(17) & $32^240$  & 0.5574(27)  \\
81.019  & $18.40^2$ & 0.848402(2)  & 0.19334(23) & $22^240$  & 0.8025(27)  \\
58.906  & $13.36^2$ & 0.786018(2)  & 0.2816(5)   & $16^236$  & 1.1731(66)  \\  \hline
\end{tabular}
\caption{Parameters, string tension and mass gap of the $SU(6)$ calculations.}
\label{table_param_su6}
\end{table}

\begin{table}[htb]
\centering
\begin{tabular}{|cclll|} \hline
\multicolumn{5}{|c|}{$SU(8)$} \\ \hline
$\beta$ & lattice & $\tfrac{1}{N}\text{Tr}\langle U_p\rangle$ & $a\surd\sigma_f$ & $am_G$ \\ \hline
370.0  & $54^260$ & 0.94220766(4) & 0.07041(8)   & 0.2926(9) \\
306.25 &  ---     & ---  & 0.085880(24) & --- \\
306.25 & $44^248$ & 0.92989047(5) & 0.08597(10) &  0.3552(13) \\
250.0  & $36^240$ & 0.9136369(2)  & 0.10624(11) &  0.4408(17) \\
200.0  & $28^232$ & 0.8911821(2)  & 0.13590(23) &  0.5634(24) \\
145.0  & $20^224$ & 0.8473983(4)  & 0.19408(35) &  0.8053(31) \\
106.0  & $14^216$ & 0.7857764(9)  & 0.27911(67) &  1.1620(76) \\ \hline
\end{tabular}
\caption{Parameters, string tension and mass gap of the $SU(8)$ calculations.}
\label{table_param_su8}
\end{table}

\begin{table}[htb]
\centering
\begin{tabular}{|cclll|} \hline
\multicolumn{5}{|c|}{$SU(12)$} \\ \hline
$\beta$ & $L_s^2L_t$ & $\tfrac{1}{N}\text{Tr}\langle U_p\rangle$ & $a\surd\sigma_f$ & $am_G$ \\ \hline
830.0 & $50^260$ & 0.94150228(3) & 0.07105(8) &  0.2940(9) \\
700.0 & $42^250$ & 0.93037943(4) & 0.08485(8) &  0.3508(9) \\
565.0 & $34^240$ & 0.9132337(1) & 0.10647(15) &  0.4418(17) \\
450.0 & $28^230$ & 0.8901571(2) & 0.13649(19) &  0.5636(22) \\
355.0 & $20^224$ & 0.8591249(3) & 0.17767(27) &  0.7286(30)  \\
260.0 & $14^216$ & 0.8031790(5) & 0.25405(50) &  1.045(5) \\ \hline
\end{tabular}
\caption{Parameters, string tension and mass gap of the $SU(12)$ calculations.}
\label{table_param_su12}
\end{table}

\begin{table}[htb]
\centering
\begin{tabular}{|cclll|} \hline
\multicolumn{5}{|c|}{$SU(16)$}  \\ \hline
$\beta$ & $L_s^2L_t$ & $\tfrac{1}{N}\text{Tr}\langle U_p\rangle$ & $a\surd\sigma_f$ & $am_G$ \\ \hline
1180.0 & $34^248$ & 0.92624511(4) & 0.08984(10) &  0.3716(10) \\
980.0  & $28^240$ & 0.91071306(7) & 0.10976(15) &  0.4514(18) \\
800.0  & $20^2,22^2,26^230$ & 0.8897983(2) & 0.13664(14) & 0.5609(22) \\
560.0  & $15^224$ & 0.8394872(4) & 0.20369(31) &  0.8356(39) \\
430.0  & $11^216$ & 0.7861389(5) & 0.27859(40) &  1.1461(84) \\ \hline
\end{tabular}
\caption{Parameters, string tension and mass gap of the $SU(16)$ calculations.}
\label{table_param_su16}
\end{table}

\begin{table}[htb]
\centering
\begin{tabular}{|c|cccc|ccc|} \hline
\multicolumn{8}{|c|}{$SU(4)$} \\ \hline 
$\beta$ & $l/a$ & $aE_{k=0}$  & $aE_{k=2A}$  & $aE_{k=2S}$ & $l/a$ & $aE_{k=0}$  & $aE_{k=2A}$ \\ \hline
28.0 & 14 & 1.794(33) & 1.1627(66) & 1.99(2)    & 20  & 2.73(12)  &  1.672(9) \\
40.0 & 22 & 1.271(7) & 0.8101(24) & 1.4244(74)  & 30  & 1.797(14) &  1.126(5) \\
51.0 & 30 & 1.046(6) & 0.6519(28) & 1.150(20)   & 40  & 1.418(12) &  0.8846(28) \\
63.0 & 36 & 0.7749(58) & 0.4941(14) & 0.853(10) & 50  & 1.086(15) &  0.6998(26) \\
74.0 & 44 & 0.6870(56) & 0.4346(10) & 0.7623(72) & 60  & 0.952(7)  &  0.5973(17) \\
86.0 & 50 & 0.5669(35) & 0.3578(7)  & 0.6306(37) & 70  & 0.800(13) &  0.5067(18) \\ \hline
\end{tabular}
\caption{Ground state energies of flux tubes in various representations of $SU(4)$.}
\label{table_Egs_su4}
\end{table}

\begin{table}[htb]
\centering
\begin{tabular}{|cc|cccccc|} \hline
\multicolumn{8}{|c|}{$SU(8)$} \\ \hline 
$\beta$ & $l/a$ & $aE_{k=0}$  & $aE_{k=2A}$  & $aE_{k=2S}$  & $aE_{k=3A}$  & $aE_{k=3M}$  & $aE_{k=4}$ \\ \hline
106.0  & 14 & 2.25(11)  & 1.881(32) & 2.47(9)   & 2.42(8)   & 3.31(54)  & 2.50(8) \\
145.0  & 20 & 1.537(13) & 1.294(9)  & 1.654(15) & 1.629(14) & 2.204(27) & 1.760(18) \\
200.0  & 28 & 1.034(8) & 0.8867(26) & 1.1299(68) & 1.1221(54) & 1.457(30) & 1.2011(35) \\
250.0  & 36 & 0.8260(39) & 0.6992(12) & 0.8883(56) & 0.8864(29) & 1.193(9) & 0.9520(24) \\
306.25 & 44 & 0.6606(25) & 0.5582(17) & 0.7137(41) & 0.7056(43) & 0.9545(57) & 0.7593(17) \\
370.0  & 54 & 0.5466(26) & 0.4576(13) & 0.5933(23) & 0.5818(23) & 0.7919(23) & 0.6267(27) \\ \hline
\end{tabular}
\caption{Ground state energies of flux tubes in various representations of $SU(8)$.}
\label{table_Egs_su8}
\end{table}

\begin{table}[htb]
\centering
\begin{tabular}{|cc|cccccc|} \hline
\multicolumn{8}{|c|}{$SU(12)$} \\ \hline 
$\beta$ & $l/a$ & $aE_{k=0}$  & $aE_{k=2A}$  & $aE_{k=2S}$  & $aE_{k=3A}$  & $aE_{k=3M}$  & $aE_{k=4}$ \\ \hline
260.0 & 14 & 1.78(3)   & 1.658(14)  & 1.904(25)  & 2.23(5)    & 2.67(8)    & 2.66(8) \\
355.0 & 20 & 1.258(11) & 1.1449(51) & 1.3283(77) & 1.5787(73) & 1.878(12)  & 1.76(8) \\
450.0 & 28 & 1.0505(65) & 0.9480(39) & 1.106(10) & 1.3027(72) & 1.5891(58) & 1.518(33) \\
565.0 & 34 & 0.7721(43) & 0.7007(19) & 0.8177(48) & 0.9663(35) & 1.1667(43) & 1.1403(68) \\
700.0 & 42 & 0.6068(24) & 0.5512(15) & 0.6481(26) & 0.7562(40) & 0.9216(54) & 0.894(10) \\
830.0 & 50 & 0.5047(16) & 0.4569(16) & 0.5483(23) & 0.6261(25) & 0.7567(23) & 0.7536(44) \\ \hline
\end{tabular}
\caption{Ground state energies of flux tubes in various representations of $SU(12)$.}
\label{table_Egs_su12}
\end{table}

\begin{table}[htb]
\centering
\begin{tabular}{|cc|cccccc|} \hline
\multicolumn{8}{|c|}{$SU(16)$} \\ \hline 
$\beta$ & $l/a$ & $aE_{k=0}$  & $aE_{k=2A}$  & $aE_{k=2S}$  & $aE_{k=3A}$  & $aE_{k=3M}$  & $aE_{k=4}$ \\ \hline
430.0  & 11 & 1.652(24) & 1.553(14)  & 1.769(21)  & 2.253(43) & 2.346(47) & 2.796(84) \\
560.0  & 15 & 1.198(11) & 1.1552(73) & 1.2900(89) & 1.641(15) & 1.767(13) & 2.022(25) \\
800.0  & 22 & 0.7938(50) & 0.7537(28) & 0.8506(31) & 1.0775(50) & 1.2250(40) & 1.320(20) \\
980.0  & 28 & 0.6579(39) & 0.6196(15) & 0.7003(26) & 0.8846(30) & 1.0131(31) & 1.0836(74) \\
1180.0 & 34 & 0.5355(27) & 0.5025(23) & 0.5659(22) & 0.7162(45) & 0.8043(51) & 0.9010(47) \\ \hline
\end{tabular}
\caption{Ground state energies of flux tubes in various representations of $SU(16)$.}
\label{table_Egs_su16}
\end{table}



\clearpage

\setcounter{table}{0}
\renewcommand{\thetable}{B\arabic{table}}
%
%
%
%
\section{Tables: continuum and large-N limits}
\label{section_appendix_continuum}

\begin{table}[htb]
\centering
\begin{tabular}{|c|cc|ccc|} \hline
group & $\surd\sigma_f/g^2N$ & $\chi^2/n_{dof}$ & $m_G/g^2N$ & $c_1$ & $\chi^2/n_{dof}$ \\ \hline
 $SU(2)^a$ & 0.16745(11) & 4.7/9  & 0.7930(11) & -0.036(2) & 16.5/9 \\
 $SU(2)^b$ & 0.16780(15) & 7.4/6  & 0.7930(11) & -0.035(2) & 7.0/9 \\
 $SU(3)$   & 0.18389(17) & 1.0/4  & 0.8066(20) & -0.042(3) & 1.5/4 \\
 $SU(4)^a$ & 0.18957(16) & 2.8/4  & 0.8057(22) & -0.038(4) & 2.8/4 \\
 $SU(4)^b$ & 0.18968(21) & 2.0/4  & 0.8048(27) & -0.038(4) & 2.8/4 \\
 $SU(6)$   & 0.19329(11) & 3.3/5  & 0.8060(22) & -0.035(4) & 4.0/5 \\
 $SU(8)$   & 0.19486(16) & 11.4/4  & 0.8072(28) & -0.033(5) & 1.7/4 \\
$SU(12)$   & 0.19557(24) &  3.6/4  & 0.8120(26) & -0.040(4) & 1.5/4 \\
$SU(16)$   & 0.19549(27) &  1.4/3  & 0.8093(35) & -0.037(6) & 2.0/3 \\ \hline
$SU(\infty)$ & 0.19636(12) & 5.0/4  & 0.8102(12) & & 6.6/7 \\ \hline
\end{tabular}
\caption{Continuum limit of $\surd\sigma_f/g^2N$ and the mass gap, $m_G/g^2N$, 
  with the total $\chi^2$ of the fit to $n_{dof}$ degrees of freedom,
  and with the coefficient $c_1$ of the $ag^2_IN$ correction term.
Also the extrapolation to $N=\infty$. Superscripts $a,b$ indicate
medium and large lattice volumes respectively.}
\label{table_sigf_cont}
\end{table}

\begin{table}[htb]
\centering
\begin{tabular}{|c|cccccc|} \hline
 \multicolumn{7}{|c|}{$\surd\sigma_k/\surd\sigma_f$} \\ \hline
 group & $k=0$ & $k=2A$ & $k=2S$ & $k=3A$ & $k=3M$ & $k=4$ \\ \hline
 SU(2) & 1.497(10) & -- & -- & --  & --  & --   \\
 SU(4$)^a$ & 1.4617(37) & 1.1649(11) & 1.5377(38) & --  & --  & --  \\
 SU(4)$^b$ & 1.4600(60) & 1.1619(21) & --  & --  & --  & --  \\
 SU(6) & 1.4297(42) & 1.2753(22) & 1.4961(41) & 1.3573(33) & 1.6861(67) &  --  \\
 SU(8) & 1.4389(28) & 1.3234(18) & 1.4971(30) & 1.4875(28) & 1.7306(37) & 1.5424(25) \\
SU(12) & 1.4311(24) & 1.3626(19) & 1.4921(47) & 1.5920(30) & 1.7428(43) & 1.7385(52) \\
SU(16) & 1.4178(34) & 1.3771(22) & 1.4590(26) & 1.6356(37) & 1.7605(37) & 1.8210(54)  \\ \hline
\end{tabular}
\caption{Continuum limit of various string tension ratios. For $SU(4)$  superscripts $a,b$ 
  denote medium and large size lattices respectively. The $SU(6)$ values are from
\cite{AAMT16_strings}.}
\label{table_sigksigf_cont}
\end{table}

\begin{table}[h]
\begin{center}
\begin{tabular}{|clc|clc|}\hline
\multicolumn{6}{|c|}{ SU(2) continuum limit } \\ \hline
$J^{PC}$ & $M/\sqrt{\sigma}$ &  $\chi^2/n_{df}$ & $J^{PC}$ & $M/\sqrt{\sigma}$ & $\chi^2/n_{df}$   \\ \hline
$0^{++}$  & 4.7367(55) &  6.6/9 & $0^{-+}$ & 9.884(25)  & 12.5/9  \\
         & 6.861(14)  &  8.3/5 &          & 11.079(30) & 16.4/9 \\
         & 8.382(14)  &  8.1/9 &          & 11.50(5)   & 19.8/7 \\
         & 9.278(16)  &  7.7/9 &          & 12.39(7)   & 12.4/7 \\
         & 9.708(21)  &  7.2/8 &  &   & \\
         & 9.910(32)  &  7.0/5 &  &   & \\
         & 10.05(7)   &  1.1/4 &  &   & \\ \hline
$2^{++}$ & 7.762(10) &  12.6/9  & $2^{-+}$ & 7.795(12)  & 11.5/9 \\
        & 9.107(20) &  11.1/9  &      & 9.123(24) &   18.7/9 \\
        & 10.138(35) &  5.7/9  &      & 10.220(33) &  7.0/8 \\
        & 10.624(23) &  18.3/7 &      & 10.617(30) &  8.4/6 \\
        & 10.823(40) &  6.2/5  &      & 10.978(32) &  10.6/7 \\ \hline
$1^{++}$ & 10.553(31) &  18.9/9 & $1^{-+}$ & 10.538(28)  & 22.7/9 \\
        & 11.049(37) &  11.9/9 &       & 11.020(32) &   19.4/9 \\
        & 11.994(45) &  28.5/7 &       & 11.840(51) &   17.9/7 \\
        & 12.273(50) &  19.7/7 &       & 12.037(65) &   2.2/7 \\ \hline
\end{tabular}
\caption{Glueball spectrum: SU(2) continuum limit.}
\label{table_mksu2_cont}
\end{center}
\end{table}

\begin{table}[h]
\begin{center}
\begin{tabular}{|clc|clc|}\hline
\multicolumn{6}{|c|}{ SU(3) continuum limit } \\ \hline
$J^{PC}$ & $M/\sqrt{\sigma}$ & $\chi^2/n_{df}$ & $J^{PC}$ & $M/\sqrt{\sigma}$ & $\chi^2/n_{df}$   \\ \hline
$0^{++}$  & 4.3683(73) &  2.0/4 &  $0^{--}$  & 6.391(14) &  2.1/4\\
 & 6.486(13) &  5.7/4 &      & 7.983(26) &  3.8/3 \\
 & 8.012(27) &  5.3/4 &      & 9.338(38) &  0.2/3 \\
 & 8.632(29) &  3.6/4 &      & 10.145(52) &  2.4/3 \\
 & 9.353(34) &  1.3/3 &      & &   \\
 & 9.414(37) &  3.2/3 &      & &  \\ \hline
$0^{-+}$ &  9.234(43) &  2.0/4  & $0^{+-}$ & 10.55(6) &  1.1/4 \\
        &  10.47(8) &  1.1/2 &          & 11.55(12) &  1.4/3 \\
        &  11.44(10) &  15.5/2 &        & &  \\ \hline
$2^{++}$ &  7.241(17)  & 1.5/4 & $2^{-+}$ & 7.261(16) &  6.8/4 \\
        &  8.730(22) &  2.5/4 &         & 8.760(23) &  16.6/4 \\
        &  9.589(46) &  0.9/3 &         & 9.992(33) &  1.6/3 \\
        &  10.082(54) &  4.6/3 &        & 10.05(9) &  7.6/3 \\ 
        &  9.99(19)  &  9.6/3 &         & 10.86(9) &  6.9/3 \\ \hline
$2^{--}$ &  8.599(30) &  2.8/4 & $2^{+-}$ & 8.665(24) &  8.3/4 \\
        &  9.940(44) &  3.5/3 &         & 9.871(63) &  1.1/3 \\
        &  11.285(54)  & 7.2/3 &        & 11.17(9) &  2.5/2 \\
        &  12.09(12)  & 5.7/3 &        & &  \\ 
        & 12.09(12)  & 5.7/3 &        & &  \\ \hline
$1^{++}$ & 10.221(50) &  11.0/4 & $1^{-+}$ & 10.316(46) &  2.8/4 \\
        & 10.443(44) &  2.3/4 &          & 10.294(55) &  15.3/4 \\
        & 11.415(87) &  2.6/3 &          & 11.50(5) &  5.1/3 \\ \hline
$1^{--}$ & 10.039(44) &  7.0/4 & $1^{+-}$ & 9.964(58) &  4.8/4 \\
        & 10.341(60) &  2.1/4 &        & 10.260(86) &  3.7/3 \\
        & 11.25(8) &  3.7/2  &         & 11.27(8) &  5.0/2 \\ \hline
\end{tabular}
\caption{Glueball spectrum: SU(3) continuum limit.}
\label{table_mksu3_cont}
\end{center}
\end{table}

\begin{table}[h]
\begin{center}
\begin{tabular}{|clc|clc|}\hline
\multicolumn{6}{|c|}{ SU(4) continuum limit } \\ \hline
$J^{PC}$ & $M/\sqrt{\sigma}$ & $\chi^2/n_{df}$ & $J^{PC}$ & $M/\sqrt{\sigma}$ & $\chi^2/n_{df}$   \\ \hline
$0^{++}$  & 4.242(9) &   3.3/4 &  $0^{--}$ & 6.216(15) &   6.3/4 \\
         & 6.440(13) &   8.6/4 &         & 7.867(22) &   9.0/4 \\
         & 7.930(24) &   3.1/3 &         & 9.248(30) &   5.8/4 \\
         & 8.319(32) &   2.6/3 &         & 9.795(52) &   0.3/3 \\
         & 9.040(51) &   4.3/3 &         & 10.44(5) &   20.5/3 \\
         & 9.297(31) &   4.8/3 &         & 10.36(8) &   10.5/3 \\
         & 9.560(74) &   5.3/3 &         & &   \\ 
         & 9.988(88) &   0.5/2 &         & &   \\ \hline
$0^{-+}$ & 9.204(30) &   4.6/4  & $0^{+-}$ & 10.29(7) &   5.4/4 \\
        & 10.327(60) &   5.8/4 &         & 11.39(10) &   7.6/2 \\ 
        & 11.47(10) &   0.2/3 &          & 12.45(11) &   12.3/3 \\ \hline
$2^{++}$ & 7.091(17) &   2.7/4 & $2^{-+}$ & 7.096(20) &   2.9/4 \\
         & 8.597(27) &   7.3/4 &         & 8.531(32) &   13.7/4  \\
         & 9.785(32) &   3.5/4 &         & 9.736(45) &   3.6/4 \\
         & 9.981(55) &   5.4/3 &         & 10.117(65) &   6.7/3 \\
         & 10.85(7) &   8.9/3  &         & 11.01(6) &   1.6/3 \\ \hline
$2^{--}$ & 8.368(40) &   2.1/4 & $2^{+-}$ & 8.475(30) &   5.7/4 \\
         & 9.737(24) &   3.8/4  &        & 9.823(27) &   4.1/4 \\
         & 10.889(47) &   6.3/4 &        & 10.923(45) &   10.7/4 \\
         & 11.62(9) &   12.9/3 &         & 11.39(8) &   9.0/3 \\ 
         & 12.06(5) &   2.6/3  &         & 12.01(11) &   0.8/2 \\ \hline
$1^{++}$ & 9.952(62) &   5.0/3 & $1^{-+}$ & 9.997(51) &   6.4/4 \\
         & 10.09(6) &   0.7/3 &         & 10.14(6) &   8.0/4 \\
         & 11.20(7) &   18.5/3 &        & 11.12(10) &   9.2/4 \\
         & 11.27(8) &   0.2/3 &         & 11.58(6) &   15.7/3 \\ 
         & 11.59(8) &   2.3/3 &         & 11.49(9) &   3.9/3 \\ \hline
$1^{--}$ & 9.797(39) &   2.4/4 & $1^{+-}$ & 9.668(41) &   3.3/4 \\
         & 10.071(37) &   9.1/3 &        & 10.043(42) &   9.1/4 \\
         & 10.88(10) &   2.5/4 &         & 10.81(12) &   4.4/3 \\
         & 11.26(5) &   11.5/3 &         & 11.21(7) &   17.0/3 \\ 
         & 11.41(7) &   2.5/3 &          & 11.43(6) &   3.6/3 \\ \hline
\end{tabular}
\caption{Glueball spectrum: SU(4) continuum limit.}
\label{table_mksu4_cont}
\end{center}
\end{table}

\begin{table}[h]
\begin{center}
\begin{tabular}{|clc|clc|}\hline
\multicolumn{6}{|c|}{ SU(6) continuum limit } \\ \hline
$J^{PC}$ & $M/\sqrt{\sigma}$ & $\chi^2/n_{df}$ & $J^{PC}$ & $M/\sqrt{\sigma}$ & $\chi^2/n_{df}$   \\ \hline
$0^{++}$  & 4.164(8) &  2.3/4 &  $0^{--}$ & 6.148(13) &  2.7/4 \\
         & 6.357(13) &  1.2/4 &         & 7.804(21) &  1.0/4 \\
         & 7.927(18) &  1.5/4 &         & 9.155(39) &  5.0/3 \\
         & 8.232(32) &  3.3/3 &         & 9.777(74) &  4.6/2 \\
         & 9.003(56) &  2.4/3 &         & &  \\
         & 9.106(54) &  2.2/2 &         & &  \\
         & 9.24(13) &  7.2/2 &         & &  \\ \hline
$0^{-+}$  & 9.030(44) &  3.6/4 & $0^{+-}$ & 10.089(53) &  3.0/4 \\
         & 10.175(56) &  5.3/4 &         & 11.48(7) &  17.6/3 \\ \hline
$2^{++}$ & 6.983(19)  &  2.9/4 & $2^{-+}$ & 6.996(16) &  6.7/4  \\
         & 8.513(21) &  4.5/4 &         & 8.472(26) &  4.7/4 \\
         & 9.728(31) &  1.6/4 &         & 9.744(36) &  11.5/4 \\
         & 10.02(10) &  4.2/2 &         & 10.024(56) &  9.6/2 \\
         & 11.30(14) &  0.5/2 &         & 10.97(9) &  27.6/2 \\ \hline
$2^{--}$ & 8.318(34) &  7.8/4 & $2^{+-}$ & 8.275(34) &  2.3/4 \\
         & 9.626(46) &  10.3/4 &         & 9.700(57)  & 6.2/4 \\
         & 10.911(57) &  1.6/3 &         & 10.94(9) &  11.3/2 \\
         & 11.811(67) &  2.3/2 &         & 11.20(10) &  2.5/2 \\ \hline
$1^{++}$ & 10.163(45) &  9.0/4 & $1^{-+}$ & 9.863(57) &  3.0/4 \\
         & 10.084(44) &  9.0/4 &         & 10.154(46) &  3.1/4 \\
         & 11.194(89) &  3.4/4 &         & 11.15(10) &  6.2/4 \\
         & 11.21(11) &  9.1/2 &         & &  \\ \hline
$1^{--}$ & 9.568(43) &  1.8/4 & $1^{+-}$ & 9.469(43) &  1.8/4 \\
         & 9.872(57) &  6.6/4 &         & 10.040(50) &  6.9/4 \\
         & 10.874(57) &  1.4/3 &        & 10.838(61) &  8.3/3 \\
         & 11.11(9) &  1.5/2 &          & 11.10(9) &  1.1/2 \\ \hline
\end{tabular}
\caption{Glueball spectrum: SU(6) continuum limit.}
\label{table_mksu6_cont}
\end{center}
\end{table}

\begin{table}[h]
\begin{center}
\begin{tabular}{|cllc|cllc|}\hline
\multicolumn{8}{|c|}{ SU(8) continuum limit } \\ \hline
$J^{PC}$ & $M/\sqrt{\sigma}$ &  $c_1$ & $\chi^2/n_{df}$ & $J^{PC}$ & $M/\sqrt{\sigma}$ & $c_1$ & $\chi^2/n_{df}$   \\ \hline
$0^{++}$  & 4.144(10) & 0.2(6) &  1.5/4 & $0^{--}$ & 6.102(14) & 0.3(1.1) &  5.3/4 \\
         & 6.332(15) & -3.1(1.3) &  3.7/4 &         & 7.803(26) & -2.8(3.3) &  6.0/4 \\
         & 7.855(22) & -0.9(2.3) &  3.2/4 &         & 9.177(38) & -1.9(4.8) &  2.2/3 \\
         & 8.090(36) & -1.6(2.9) &  3.3/3 &         & 9.803(38) & -16.3(4.6) &  1.3/2 \\
         & 8.969(78) & -13.1(11.8) &  0.5/2 &         & 10.00(11) & -1.5(8.8) &  1.9/2 \\
         & 9.152(40) & -9.0(4.6)  &  0.2/2 &         & &  & \\
         & 9.58(6)   & -14.8(5.9) &  6.4/2 &         & &  & \\ \hline
$0^{-+}$ & 9.035(47) & -7.1(4.2) &  3.0/4 & $0^{+-}$ & 10.35(6)  & -14.9(5.3) &  11.2/4 \\
         & 10.14(7) & 2.3(7.7)  &  4.7/2 &          & 11.31(10) & -14.6(8.3) &  8.1/2 \\ \hline
$2^{++}$ & 6.952(18) & -0.5(1.7) &  6.7/4 & $2^{-+}$ & 6.942(21) & -1.2(1.7) &  4.8/4 \\
         & 8.412(43) & -1.7(3.7) &  4.2/4 &         & 8.486(38) & -3.9(3.3) &  4.8/4 \\
         & 9.392(73) & 1.4(5.3) &  1.5/3 &          & 9.760(37) & -5.8(3.9) &  0.3/3 \\
         & 9.811(60) & -3.2(6.0) &  7.9/3 &         & 9.982(66) & -12.2(6.0) &  3.9/3 \\
         & 10.724(70) & -0.7(6.3) &  4.0/3 &         & 10.98(9) & -11.2(9.6) &  5.0/3 \\ \hline
$2^{--}$ & 8.273(22) & -2.6(2.8) &  5.3/4 & $2^{+-}$ & 8.273(34) & -0.8(2.7) &  2.1/4 \\
        & 9.667(35) & -7.4(4.0) &  0.5/3 &         & 9.637(42) & -1.3(4.5) & 1.7/3 \\
        & 10.758(62) & -6.1(7.0) & 4.1/3 &        & 10.940(54) & -14.7(7.1) &  4.3/3 \\
        & 11.29(11) & -9.4(12.7) &  1.0/2 &         & 11.06(14) & -11.4(22.4) &  2.3/2 \\ \hline
$1^{++}$ & 9.941(60) & -6.6(5.3) &  2.2/4 & $1^{-+}$ & 9.959(50) & -7.5(4.5) &  7.8/4 \\
        & 10.074(70) & -5.1(5.1) &  5.8/4 &         & 10.054(61) & -0.2(5.6) &  3.0/4 \\
        & 11.12(9) & 1.3(7.6) &  11.3/3 &          & 11.05(12) & -11.2(9.4) &  1.5/3 \\
        & 10.94(16) & 30.9(15.0) &  11.6/3 &         & &  &  \\ \hline
$1^{--}$ & 9.449(50) & -1.2(5.0) &  2.8/4 & $1^{+-}$ & 9.505(47) & -4.2(4.7) &  4.0/4 \\
         & 9.822(52) & -5.5(4.1) &  1.3/4 &         & 9.859(43) & -2.8(5.3) &  5.1/4 \\
         & 10.71(7) & 7.4(6.6) &  9.1/3  &         & 10.76(8) & 1.0(7.4) & 3.8/3 \\
         & 11.06(7) & -3.7(9.2) & 3.4/2 &         & 11.07(6) & -4.1(6.9) & 6.1/2 \\ \hline
\end{tabular}
\caption{Glueball spectrum: SU(8) continuum limit, with coefficient $c_1$ of $a^2\sigma_f$ correction.}
\label{table_mksu8_cont}
\end{center}
\end{table}

\begin{table}[h]
\begin{center}
\begin{tabular}{|clc|clc|}\hline
\multicolumn{6}{|c|}{ SU(12) continuum limit } \\ \hline
$J^{PC}$ & $M/\sqrt{\sigma}$ & $\chi^2/n_{df}$ & $J^{PC}$ & $M/\sqrt{\sigma}$ & $\chi^2/n_{df}$   \\ \hline
$0^{++}$  & 4.140(9) &   2.5/4 &  $0^{--}$  & 6.070(15) &   3.3/4 \\
         & 6.320(14) &   7.7/4 &         & 7.787(25) &   1.6/4 \\
         & 7.852(20) &   5.9/4 &         & 9.132(40) &   7.5/3 \\
         & 8.188(20) &   4.2/4 &         & 9.656(40) &   0.4/3 \\
         & 8.955(44) &   4.9/3 &         & 9.987(71) &   0.7/2  \\
         & 9.173(40) &   2.5/2 &         & 10.263(61) &   2.3/2\\ \hline
$0^{-+}$ & 8.991(50) &   3.9/4 & $0^{+-}$ & 10.21(6) &   7.2/4 \\
         & 10.16(7) &   4.3/4 &         & 11.29(9) &   4.5/4 \\
         & 11.23(8) &   0.55/3 &        & 12.10(11) &   4.4/3 \\ \hline
$2^{++}$ & 6.938(18) &   3.4/4 & $2^{-+}$ & 6.947(16) &   0.3/4 \\
         & 8.384(50) &   0.6/4 &         & 8.506(25) &   6.2/4 \\
         & 9.505(45) &   1.5/4 &         & 9.611(54) &   6.5/4 \\
         & 9.862(60) &   2.2/3 &         & 9.802(60) &   2.8/3 \\
         & 10.727(76) &   2.4/2 &        & 10.826(70) &   5.8/2 \\ \hline
$2^{--}$ & 8.285(31) &   5.4/4 & $2^{+-}$ & 8.247(35) &   2.7/4 \\
         & 9.619(44) &   4.0/4 &         & 9.656(30) &   1.3/4 \\
         & 10.692(53) &   3.9/4 &        & 10.86(6) &   5.8/4 \\
         & 11.16(11) &   0.6/2 &         & 11.21(7) &   2.1/2 \\
         & 12.04(8) &   9.6/2 &          & 12.02(10) &   3.5/2 \\ \hline
$1^{++}$ & 9.876(50) &   5.4/4 & $1^{-+}$ & 9.924(52) &   1.0/4 \\
        & 10.08(8) &   17.1/4 &         & 10.07(7) &   3.0/4 \\
        & 11.08(8) &   15.2/4 &         & 11.24(6) &   19.4/4 \\
        & 11.21(10) &   20.8/2 &        & 11.16(11) &   0.7/2 \\ \hline
$1^{--}$ & 9.388(47) &   2.5/4 & $1^{+-}$ & 9.448(50) &   3.8/4 \\
         & 9.724(51) &   3.5/4 &         & 9.805(75) &   1.6/4 \\
         & 10.70(7) &   7.8/4 &         & 10.80(7) &   2.9/4 \\
         & 10.98(7) &   3.1/2 &         & 10.95(9) &   3.7/2 \\ \hline
\end{tabular}
\caption{Glueball spectrum: SU(12) continuum limit.}
\label{table_mksu12_cont}
\end{center}
\end{table}

\begin{table}[h]
\begin{center}
\begin{tabular}{|clc|clc|}\hline
\multicolumn{6}{|c|}{ SU(16) continuum limit } \\ \hline
$J^{PC}$ & $M/\sqrt{\sigma}$ & $\chi^2/n_{df}$ & $J^{PC}$ & $M/\sqrt{\sigma}$ & $\chi^2/n_{df}$   \\ \hline
$0^{++}$  & 4.129(11) &   1.9/3 &  $0^{--}$  & 6.098(19) &   0.4/3 \\
         & 6.295(18) &   0.7/3 &           & 7.751(32) &   6.6/3 \\
         & 7.746(45) &   2.9/3 &           & 9.022(66) &   3.8/3 \\
         & 8.157(41) &   2.3/2 &           & 9.716(50) &   5.6/2 \\
         & 9.037(61) &   0.2/2 &           & 9.935(86) &   3.9/2 \\
         & 9.047(54) &   2.1/2 &           & 10.40(10) &   1.1/2 \\ \hline
$0^{-+}$ & 9.012(54) &   1.2/3 & $0^{+-}$ & 10.07(8) &   0.5/3 \\
         & 10.28(10) &   5.6/3 &         & 11.44(14) &   1.0/3 \\
         & 11.63(9) &   1.6/3 &         & 12.63(17) &   0.1/2 \\ \hline
$2^{++}$ & 6.937(30) &   3.4/3 & $2^{-+}$ & 6.991(30) &   3.8/3 \\
         & 8.458(38) &   5.1/3 &         & 8.498(35) &   0.3/3 \\
         & 9.721(56) &   4.5/3 &         & 9.81(6)  &   10.8/3 \\
         & 9.923(52) &   3.6/2 &         & 9.99(6)  &   3.9/2 \\ 
         & 10.93(9) &   0.5/2 &         & 10.85(11) &   5.0/2 \\ \hline
$2^{--}$ & 8.164(54) &   6.6/3 & $2^{+-}$ & 8.305(37) &   2.4/3 \\
        & 9.614(46) &   1.8/3 &         & 9.610(73) &   2.6/3 \\
        & 10.71(8) &   0.1/2 &          & 10.92(11) &   6.1/2 \\
        & 11.19(13) &   0.6/2 &         & 11.53(10) &   7.4/2 \\ \hline
$1^{++}$ & 9.90(7)  &   1.0/3 & $1^{-+}$ & 9.80(9) &   0.4/3 \\
        & 9.88(13) &   24.5/3 &         & 9.94(8) &   3.2/3 \\
        & 11.19(17) &   0.5/2 &         & 11.43(9) &   0.5/2 \\
        & 11.23(14) &   8.1/2 &         & 11.23(17) &   9.2/2 \\ \hline
$1^{--}$ & 9.53(6) &   6.1/3 & $1^{+-}$ & 9.36(8) &   9.4/3 \\
         & 9.85(6) &   4.0/3 &         & 9.82(8) &   5.5/3 \\
         & 10.76(11) &   2.1/2 &       & 10.71(12) &   1.8/3 \\ 
         & 11.20(19) &   1.1/2 &       & 11.07(16) &   2.0/2 \\ \hline
\end{tabular}
\caption{Glueball spectrum: SU(16) continuum limit.}
\label{table_mksu16_cont}
\end{center}
\end{table}

\begin{table}[h]
\begin{center}
\begin{tabular}{|c|lcccc|}\hline
\multicolumn{6}{|c|}{ $SU(N\to\infty)$ : J=0 } \\ \hline
$J^{PC}$ & $M/\sqrt{\sigma}$ & $c_1$ & $c_2$ & $N\geq $ & $\chi^2/n_{df}$ \\ \hline
$0^{++}$  & 4.116(6)  & 2.00(15)  & 1.94(53) & 2 & 4.1/4 \\
         & 6.308(10)  & 1.72(20)  & & 3 & 5.8/4 \\
         & 7.844(14)  & 1.59(39)  & & 3 & 10.8/4 \\
         & 8.147(19)  & 2.63(91)  & & 4 & 8.9/3 \\ 
         & 8.950(25)  & 3.06(18)  & & 2 & 9.6/5 \\
         & 9.087(20)  & 3.25(23)  & & 2 & 6.3/5 \\ \hline
$0^{-+}$  & 8.998(28)  & 2.53(21)  & & 3 & 4.7/4 \\
         & 10.101(33)  & 3.89(24)  & & 2 & 4.1/5 \\ \hline
$0^{--}$  & 6.060(9)  & 2.88(21)  & & 3 & 6.0/4 \\
         & 7.759(15)  & 1.92(37)  & & 3 & 1.5/4 \\
         & 9.110(35)  & 2.11(55)  & & 3 & 3.2/4 \\
         & 9.709(30)  & 1.7(1.5)  & & 4 & 6.0/3 \\ \hline
$0^{+-}$  & 10.133(37)  & 3.38(94)  & & 3 & 17.0/4 \\
         & 11.350(57)  &  1.70(3)  & & 3 &  3.3/4 \\ \hline
\end{tabular}
\caption{Glueballs with spin 0: large-$N$ limit, with coefficients $c_1$ of $1/N^2$ and (where applicable) $c_2$ of $1/N^4$ corrections.}
\label{table_mksuN_J0}
\end{center}
\end{table}

\begin{table}[h]
\begin{center}
\begin{tabular}{|c|lcccc|}\hline
\multicolumn{6}{|c|}{ $SU(N\to\infty)$ : J=2 } \\ \hline
$J^{PC}$ & $M/\sqrt{\sigma}$ & $c_1$ & $c_2$ & $N\geq$ & $\chi^2/n_{df}$ \\ \hline
$2^{++}$  & 6.914(13)  & 2.62(32)   & 3.1(1.2) & 2 & 0.5/4 \\
        & 8.423(15)  & 2.74(14)   & & 2 & 3.8/5 \\
        & [9.59(9)]  &    & &  &  \\ 
        & 9.866(34)  & 1.91(84)   & & 3 & 4.0/4 \\
        & [10.81(10)]  &    & &  &  \\ \hline
$2^{-+}$  & 6.930(13)  & 2.42(34)  & 4.20(1.25) & 2 & 5.1/4 \\
        & 8.488(21)  & 0.41(88)   & & 4 & 1.9/3 \\
        & 9.732(32)  & 0.19(1.25)  &  & 4 & 7.3/3 \\
        & 9.905(30)  & 2.83(24)   & & 2 & 10.3/5 \\ 
        & 10.914(39) & 0.26(26)   & & 2 &  5.4/5\\ \hline
$2^{--}$  & 8.223(18)  & 3.24(48)   & & 3 & 5.8/4 \\
        & 9.592(24)  & 2.72(56)   & & 3 & 3.9/4 \\
        & 10.715(41) & 3.2(1.4)   & & 4 & 4.6/3 \\
        & 11.34(6)   & 7.0(1.6)   & & 3 & 28.4/4 \\ \hline
$2^{+-}$  & 8.224(20)  & 3.92(40)   & & 3 & 6.1/4 \\
        & 9.629(25)  & 2.74(68)   & & 3 & 2.6/4 \\
        & 10.870(41) & 1.74(1.06)  & & 3 & 3.9/4 \\
        & 11.26(6)  & 1.37(2.38)  &  & 3 & 11.5/4 \\ \hline
\end{tabular}
\caption{Glueballs with spin 2: large-$N$ limit, with coefficients $c_1$ of $1/N^2$ and (where applicable) $c_2$ of $1/N^4$ corrections. Values in square brackets are from averaging $N=12,16$ rather than from an extrapolation.}
\label{table_mksuN_J2}
\end{center}
\end{table}

\begin{table}[h]
\begin{center}
\begin{tabular}{|c|lccc|}\hline
\multicolumn{5}{|c|}{ $SU(N\to\infty)$ : J=1 } \\ \hline
$J^{PC}$ & $M/\sqrt{\sigma}$ & $c_1$ & $N\geq$ & $\chi^2/n_{df}$ \\ \hline
$1^{++}$  & 9.912(26)  & 2.57(24)  & 2 & 10.5/5 \\
        & 9.984(32)  & 4.2(28)  & 2 & 10.6/5 \\ 
        & 11.040(43) & 3.77(35) & 2 & 2.5/5 \\
        & 11.157(80) & 1.65(2.48) & 4 & 2.7/3 \\ \hline
$1^{-+}$  & 9.886(27)  & 2.67(21) & 2 & 14.5/5 \\
        & 9.969(30)  & 4.12(25)  & 2 & 12.2/5 \\ 
        & 11.195(40) & 2.55(37)  & 2 & 15.9/5 \\ \hline
$1^{--}$  & 9.401(29)  & 5.88(65)  & 3 & 5.87/4 \\
        & 9.740(32)  & 5.33(80)  & 3 & 3.39/4 \\
        & 10.684(44) & 4.9(1.2)  & 3 & 3.04/4 \\
        & 10.98(6)   & 4.4(1.6)  & 4 & 1.35/3 \\ \hline
$1^{+-}$  & 9.380(31)  & 4.95(80)  & 3 & 3.61/4 \\
        & 9.828(36)  & 3.85(98)  & 3 & 5.77/4 \\
        & 10.712(46) & 4.6(1.3)  & 3 & 3.60/4 \\
        & 10.99(6)   & 3.6(2.1)  & 4 & 0.86/3 \\ \hline
\end{tabular}
\caption{Glueballs with spin 1: large-$N$ limit, with coefficient $c_1$ of $1/N^2$ correction.}
\label{table_mksuN_J1}
\end{center}
\end{table}

\end{appendix}

\clearpage

\begin{figure}[htb]
\begin	{center}
\leavevmode
\input	{plot_mJ2_cont_su12.tex}
\end	{center}
\caption{Lightest five $2^{++}$ ($\bullet$) and lightest four $2^{--}$ ($\circ$)
glueballs with continuum extrapolations. For $SU(12)$.}
\label{fig_mJ2_cont_su12}
\end{figure}

\begin{figure}[htb]
\begin	{center}
\leavevmode
\input	{plot_m2K_N.tex}
\end	{center}
\caption{Lightest five $2^{-+}$ ($\bullet$) and lightest three $2^{--}$ ($\circ$)
glueballs versus $1/N^2$, with large-$N$ extrapolations.}
\label{fig_m2K_N}
\end{figure}


\end{document}